\documentclass[11pt]{article}
\pdfoutput=1

\usepackage{a4wide}
\usepackage[fleqn]{amsmath}

\RequirePackage{amsmath,amssymb,amsxtra,amsthm}

\RequirePackage[T1]{fontenc}
\RequirePackage[utf8]{inputenc}

\RequirePackage{mathrsfs}
\RequirePackage{mathpazo}

\RequirePackage{setspace}
\RequirePackage{array}
\RequirePackage{booktabs}

\RequirePackage[pdftex,final]{graphicx}
\graphicspath{{Plots/}}

\usepackage{cite}

\numberwithin{equation}{section}
\numberwithin{table}{section}
\numberwithin{figure}{section}

\author{
  \begin{minipage}{0.97\linewidth}
    \vspace{1cm}
    \begin{center}
      \begin{small}
        \textbf{Carlo Angelantonj}$^{1}$, \textbf{Cezar Condeescu}$^{2}$\footnote{On leave of absence from the Institute of Mathematics `Simion Stoilow' of the Romanian Academy, P.O. Box 1-764, Bucharest, RO-70700}${}\ $, \textbf{Emilian Dudas}$^{2,3}$ and \textbf{Gianfranco Pradisi}$^{4}$
     \end{small}
    \end{center}
    \vspace{.3cm} \hspace{1.3cm}\begin{minipage}{.75\linewidth}
      {\it \begin{footnotesize}
          \begin{itemize}
          \item[${}^1$] Dipartimento di Fisica Teorica, Universit\`a di Torino, and INFN Sezione di Torino\\
            Via P. Giuria 1, 10125 Torino, Italy
          \item[${}^2$] Centre de Physique Th\'eorique, Ecole Polytechnique and CNRS\\
          91128 Palaiseau, France
          \item[${}^3$] LPT, Bat. 210, Univ. de Paris-Sud, 91405 Orsay, France
          \item[${}^4$] Dipartimento di Fisica and Sezione INFN, Universit\`a di Roma ``Tor Vergata''\\
          Via della Ricerca Scientifica 1, 00133 Roma, Italy
          \end{itemize}
        \end{footnotesize}}
    \end{minipage}
    \vspace{1cm}
  \end{minipage}
}

\date{}

\title{\vspace{3cm}
  \begin{huge}
    \textbf{Non-perturbative transitions among\\[.5ex] intersecting-brane vacua}
  \end{huge}
}

\begin{document}

\begin{titlepage}
  \maketitle
  \thispagestyle{empty}

  \vspace{-14cm}
  \begin{flushright}
    DFTT 9/2011\\
    CPHT-RR041.0511\\
    LPT-ORSAY 11-41\\
    ROM2F/2011/05
  \end{flushright}

  \vspace{11cm}

  \begin{center}
    \textsc{Abstract}\\
  \end{center}
We investigate the transmutation of D-branes into Abelian magnetic backgrounds on the world-volume of higher-dimensional  branes, within the framework of global models with compact internal dimensions. The phenomenon, T-dual to brane recombination in the inter\-secting-brane picture, shares some similarities to inverse small-instanton transitions in non-compact spaces, though in this case the Abelian magnetic background is a consequence of the compactness of the internal manifold, and is not ascribed to a zero-size non-Abelian instanton growing to maximal size.
We provide details of the transition in various supersymmetric orientifolds and non-supersymmetric tachyon-free vacua  with Brane Supersymmetry Breaking, both from brane recombination and from a field theory Higgs mechanism viewpoints.

\end{titlepage}

\setstretch{1.1}

\tableofcontents

\section{Introduction and summary of results}
\label{sec:intro}

It is well known that D5 branes  can be described as gauge instantons on the world-volume of D9 branes in the limit in which the instanton size, related to the vacuum expectation value of the D9-D5 states, is zero \cite{Witten:1995gx}. This suggests that, blowing up the small instanton size, it is possible to connect orientifold vacua with D9 and D5 branes to string vacua with internal magnetic field fluxes. The resulting phenomenon has been briefly discussed in the literature \cite{csu} in some supersymmetric examples \cite{Angelantonj:2000hi, as}.  From a string theory perspective this transition can be elegantly described, in the T-dual framework of intersecting branes, in terms of ``brane recombination'', namely in terms of a new configuration of homology cycles wrapped by the branes that respects the RR charge distribution.
 It has also been suggested that this phenomenon can be captured in the low-energy limit by a conventional Higgs mechanism, at least in some simple examples, where scalars living at the brane intersections condense. 
For a ${\rm D}p$-${\rm D}(p+4)$ configuration the corresponding {\em vev}'s were identified by Witten \cite{Witten:1995gx} to be related to the instanton size. As a result, in the type I set-up, the condensate interpolates between zero-size instantons, in the zero-{\em vev} limit, and  constant {\it non-Abelian} magnetic field, in the large-{\em vev} limit corresponding to a ``fat'' gauge instanton of maximal size.

In the present paper, we address the problem of possible non-perturbative transitions among different orientifold vacua, and our results extend the discussion in the literature in two respects. On the one hand, by considering the low-energy field equations for the Abelian theory living on a  D9 brane with a compact internal space, we show that  an {\it Abelian} constant magnetic field is generated by the solution, that is naturally related to the normalisable constant zero-mode of the Laplacian in the compact space. We show that this magnetic field  (and not the non-Abelian configuration generated by a standard ``fat'' instanton) is T-dual to the supersymmetric angle configuration in toroidal intersecting-brane models. 
We point out interesting analogies and differences between the two classical configurations. The Abelian configuration contains, in addition to the constant magnetic field, a singular part which exactly reproduces the singularity of a more conventional non-Abelian instanton configuration in the singular gauge and in the zero-size limit. 
This singularity is actually due to the zero-size limit, and it is blown-up by giving the instanton a finite-size, related to the {\em vev} of D9-D5 states. We analyse the corresponding singularity for the Abelian case by studying the effects of the {\em vev} and of the Dirac-Born-Infeld non-linearities. We could not find a simple way to resolve the singularity, although it is possible that higher-derivative corrections to the Dirac-Born-Infeld Action may provide the tool.

On the other hand, we extend the analysis of non-perturbative transitions to the class of non-supersymmetric vacua with Brane Supersymmetry Breaking. There is a large literature on non-supersymmetric string vacua built over the years, in the heterotic strings \cite{ss1,ss2,ss3,ss4,ss5,ss6,ss7}, type 0 orientifolds \cite{nonsusy0b1,nonsusy0b2}, type II orientifolds with supersymmetry broken by compactification \cite{ssi1,ssi2,ssi3,Angelantonj:1999gm}, by magnetic fields \cite{Angelantonj:2000hi,magnetic1a,magnetic1b,magnetic2a,magnetic2b,magnetic2c,magnetic2d,magnetic2e,Larosa:2003mz}, at the string scale \cite{bsb1a,bsb1b,bsb1c,bsb1d,Angelantonj:1999ms} or by a combination of these effects \cite{Angelantonj:2003hr, Angelantonj:2005hs}. Although most of the non-supersymmetric vacua are manifestly unstable at the classical level due to the tachyonic states appearing in some regions of the moduli space some, most notably the models with ``Brane Supersymmetry Breaking'' \cite{bsb1a,bsb1b,bsb1c,bsb1d,Angelantonj:1999ms} have no manifest classical instabilities. At tree-level, these models exhibit supersymmetry in the closed-string sector, while in the open-string sector supersymmetry is explicitly broken or, more precisely, non-linearly realised on stacks of anti-branes \cite{jihad,pr}. Technically, the breaking of supersymmetry is ascribed to the simultaneous presence of  $\overline{{\rm D}p}$ branes and ${\rm O} p_{+}$ planes, where ${\rm O}p_{+}$ are exotic orientifold planes with positive tension and positive RR-charge. This system breaks supersymmetry in a way similar to  ${\rm D}p-\overline{{\rm D}p}$ brane-antibrane pairs, but with the notable difference of not introducing any tachyonic instability.

It is widely believed that all non-supersymmetric string vacua are unstable towards decaying into supersymmetric ones. When classical instabilities are present,
the decay is clearly due to tachyon condensation \cite{Sen:1998sm, Sen:2004nf} that indeed interpolates between different unstable non-supersymmetric orientifold vacua \cite{Angelantonj:2000xf}.  In the Brane Supersymmetry Breaking case, however, it can only occur through non-perturbative effects. The least understood and most intriguing cases are the ones in which the candidate supersymmetric vacua have completely different geometry, O-plane contents and  different twisted sectors when compared to their non-supersymmetric parents. The only known proposal in this direction \cite{carloemilian} makes explicit use of S-duality, and thus the real dynamics of the transition cannot be followed within a perturbative CFT set-up. We have nothing to add about these cases in the present work. There are however other examples in which the non-supersymmetric classically stable model and the candidate supersymmetric vacuum have the same geometry and differ only in the configuration of branes. The fate of such vacua is part of the subject of the present paper. We argue that the {\em magnetic transmutation}, T-dual to brane recombination discussed in the literature, is responsible for the transition among non-supersymmetric vacua as well. Whereas in the supersymmetric case the states acquiring {\em vev}'s are flat directions, in non-supersymmetric examples we argue that positive masses are generated for these states, that makes the transition truly non-perturbative in nature. We present explicit examples  built upon the $\mathbb{Z}_2$ orientifold of the type IIB \cite{Angelantonj:1999ms} in six dimensions and upon the $\mathbb{Z}_2 \times \mathbb{Z}_2$ orientifolds of  type IIA superstring in four dimensions, and discuss the relation of supersymmetric vacua or of configurations with Brane Supersymmetry Breaking and discrete torsion \cite{Angelantonj:1999ms} with magnetised vacua in six \cite{Angelantonj:2000hi} and four  \cite{Dudas:2005jx, Blumenhagen:2005tn, ms, zavala, honecker} dimensions.

The paper is organised as follows.  In section \ref{sec:recombination} we discuss the transition among vacua as a brane recombination process, for models in six dimensions built upon the  $T^4/\mathbb{Z}_2$ orbifold, and in four dimensions built upon the type IIB  $T^6/\mathbb{Z}_2\times \mathbb{Z}_2$ orientifold with discrete torsion. We address the problem both for supersymmetric and Brane Supersymmetry Breaking vacua. The description of the transition in terms of brane recombination is shown to work nicely in all examples.
In section \ref{higgsing} we analyse the same problem from a low-energy viewpoint, in terms of a field theory Higgs mechanism triggered by non-vanishing {\em vev}'s of open-string states living at the intersection of different stacks of branes. We perform a case analysis for six-dimensional examples and show that when supersymmetry is preserved the Higgs mechanism exactly captures the dynamics by yielding the correct massless spectrum of the corresponding intersecting brane models. On the contrary, when supersymmetry is broken some states ought to get masses from Yukawa (for fermions) and quartic-order (for scalars) couplings that we argue to exist. In four dimensions, however, the models we study involve $\overline{{\rm D} 5}_h$ anti-branes that are not standard  instantons of the gauge theory living on the D9 branes, rather they should be interpreted as new stringy non-perturbative configurations. As a consequence, at low energy the Higgs mechanism  does not accurately captures the dynamics of the transition, whereas brane recombination does, as shown in section \ref{sec:recombination}. 
In section \ref{sec:stability}  we perform an analysis of the quantum stability of the Brane Supersymmetry Breaking models, by computing the vacuum energy as a function of the anti-brane positions and/or Wilson lines and by computing the quantum induced masses for the states whose {\em vev}'s trigger the recombination process. We argue that a positive mass for ${\rm D}7-\overline{{\rm D}7 '}$ states is generated, that suggests a non-perturbative transition associated to a real  tunnelling phenomenon through a potential barrier. Section \ref{fieldtheory}  contains a detailed field-theory analysis in a superfield formalism which shows explicitly the connection between the {\em vev} of the
Higgs field  and the emergence of the Abelian magnetic field. 
We also compare properties of our Abelian classical solution in a compact space with the conventional SU(2) instanton solution. Appendix \ref{app:6d} collects some partition functions that are used in the paper, while in appendix \ref{app:higgs} more examples of six dimensional models are reported, both supersymmetric and with Brane Supersymmetry Breaking.


\section{Brane recombination}
\label{sec:recombination}

In this section we study the possible connections among different string vacua, supersymmetric and not, as a result of brane recombination. In particular, we shall show that all string vacua, at least in a given orbifold construction, live in the same moduli space and are connected one to the other by the process of brane recombination. For simplicity we shall concentrate on some interesting prototype examples in $d=6$ and $d=4$ based on $T^4/\mathbb{Z}_2$ and $T^6/\mathbb{Z}_2\times \mathbb{Z}_2$ orientifolds, with and without supersymmetry. The complete one-loop amplitudes are given in the appendices, where additional examples are also discussed. In the following we shall discuss in detail few cases, and for simplicity we shall adopt a more geometrical language, in terms of homology cycles wrapped by the branes.

In general, an orientifold of type II superstrings compactified on suitable Calabi-Yau manifolds ${\mathcal X}_d$ involves an action of the world-sheet parity $\varOmega$, possibly combined with some (order-two) automorphism ${\mathcal R}$ of the internal manifold. In the following, we shall consider the anti-holomorphic involution ${\mathcal R}:\ z_i \to \bar z_i$, $i=1,\ldots , d$, so that the effective orientifold projection involves $\tilde\varOmega = \varOmega\, (-1)^{F_{\rm L}}\, {\mathcal R}$, where the left-handed space-time fermionic index $(-1)^{F_{\rm L}}$ is needed in order that $\tilde\varOmega$ be order-two on the whole string spectrum.

As is well known, the fixed locus of $\tilde\varOmega$ defines the location of the orientifold planes, that extend through the whole non-compact Minkowski space and wrap additional $d$-cycles ${\boldsymbol \pi}_{\rm O}$ of the internal Calabi-Yau space. Given our choice of the anti-holomorphic involution, the fixed locus identifies a special Lagrangian submanifold of ${\mathcal X}_d$, and thus their introduction preserves some amount of supersymmetries. As usual, the consistency of the construction, {\it i.e.} R-R tadpole cancellation, requires the introduction of suitable number of D-branes that, aside from expanding in the non-compact dimensions, wrap suitable internal $d$-cycles
${\boldsymbol \varPi}_a$.
The open-string excitations thus comprise unitary gauge groups and chiral matter in (anti-)symmetric and bi-fundamental representations, with multiplicities given by the intersection numbers ${\mathcal I}_{ab}= {\boldsymbol \varPi}_a \circ {\boldsymbol \varPi}_b$ of pairs of branes. We shall not review here the details of the construction, but refer the interested reader to the vast literature on the subject \cite{Blumenhagen:2002wn}. In the following we shall study specific examples related to six and four dimensional orbifold compactifications.

\subsection{Six-dimensional models}
\label{subsec:6d}

The six-dimensional models we want to study are all based on the $T^4 /\mathbb{Z}_2$ orbifold compactification of type IIB superstring.
For simplicity we shall assume that the $T^4$ factorises into the product $T^2_1 \times T^2_2$, both with purely imaginary complex structure $U^{(i)} = i \, U_2^{(i)}$, $i=1,2$. The analysis can be easily extended to the case where the complex structure of the $T^2$'s has a quantised non-vanishing real part. 
We denote by $(z_1 , z_2)$ the complex coordinate on the $T^4$, where each entry $z_i$ identifies the position on the $i$-th $T^2$ factor. The ${\mathbb Z}_2$ is generated by the single element $g$ that reverts all coordinates: $(z_1 , z_2) \to - (z_1 , z_2 )$.

To define the homology of $T^4/{\mathbb Z}_2$ is it convenient to start from that of the covering $T^4$. If one denotes by ${\boldsymbol a}_i$ and ${\boldsymbol b}_i$ the canonical horizontal and vertical one-cycles of each rectangular $T^2_i$, with intersection form given by
\begin{equation}
{\boldsymbol a}_i \circ {\boldsymbol a}_j =0\,, \quad
{\boldsymbol b}_i \circ {\boldsymbol b}_j =0\,, \quad
{\boldsymbol a}_i \circ {\boldsymbol b}_j =\delta_{ij}\,,
\end{equation}
the homology of $T^4$ is clearly given by suitable combinations of the ${\boldsymbol a}$ and ${\boldsymbol b}$ cycles. In par\-ti\-cular $H_2 (T^4 , {\mathbb Z})$ is spanned by
\begin{equation}
\bar{\boldsymbol \pi}_i = ({\boldsymbol a}_1 \otimes {\boldsymbol a}_2,\, {\boldsymbol b}_1 \otimes {\boldsymbol b}_2,\,
{\boldsymbol a}_1 \otimes {\boldsymbol b}_2,\, {\boldsymbol b}_1 \otimes {\boldsymbol a}_2,\, {\boldsymbol a}_1 \otimes {\boldsymbol b}_1,\, {\boldsymbol a}_2 \otimes {\boldsymbol b}_2 )\,,
\end{equation}
with intersection form
\begin{equation}
\bar I = \sigma_1 \otimes {\rm diag} (1,-1,-1)\,.
\end{equation}

Modding-out the torus  by the $\mathbb{Z}_2$ action implies that the $\bar {\boldsymbol \pi}_i$ cycles are not invariant under the ${\mathbb Z}_2$. Therefore, one is bound to construct new cycles that correspond to invariant orbits of the orbifold group. In the case at hand one finds that the invariant two-cycles are
\begin{equation}
{\boldsymbol \pi}_i = (1+g) \,\bar {\boldsymbol \pi}_i\,.
\end{equation}
Moreover, when computing the intersection form, one has to take into account that the $T^4$ is a double cover of the orbifold, and therefore
\begin{equation}
I_{ij} = \tfrac{1}{2} \,{\boldsymbol \pi}_i \circ {\boldsymbol \pi}_j = 2 \,\bar I_{ij}\,.
\end{equation}
These cycles, however, do not span the lattice $H_2 ( T^4/\mathbb{Z}_2 , \mathbb{Z})$ since the resolved orbifold singularities introduce new two-cycles ${\boldsymbol e}_{xy}$, $x,y = 1, \ldots , 4$, called exceptional or collapsed cycles, associated to the $\mathbb{P}^1$'s localised at the 16 fixed points with coordinates $(z_1^x, z_2^y)$, where $\{z_j\}^x \in (0,\frac{1}{2} , \frac{U^{(j)}}{2} , \frac{1}{2} (1+U^{(j)}))$. Their intersection form
\begin{equation}
I_{xy,vw} = - 2 \, \delta_{xv}\, \delta_{yw}\,,
\end{equation}
is given by the Cartan matrix of the Lie algebra $(A_1 )^{16}$, while ${\boldsymbol \pi}_i \circ {\boldsymbol e}_{xy} =0$.
Although the Kummer basis $({\boldsymbol \pi}_i ,\, {\boldsymbol e}_{xy})$ does not span entirely $H_2 ( T^4/\mathbb{Z}_2 , \mathbb{Z})$, it is actually a convenient choice to discuss the orbifold $T^4 /\mathbb{Z}_2$. A generic two-cycle, can thus be written as
\begin{equation}
{\boldsymbol \varPi}_a = \sum_{i=1}^6 c_a^i \, {\boldsymbol\pi}_i + \sum_{x,y=1}^4 \epsilon_a^{xy} \, {\boldsymbol e}_{xy}\,,
\end{equation}
where the $c^i_a$ and the $e^{xy}_a$ are suitable coefficients. It is convenient to refer to
\begin{equation}
\hat {\boldsymbol \varPi}_a = \sum_{i=1}^6 c_a^i \, {\boldsymbol \pi}_i
\end{equation}
as the bulk cycle, since it is inherited from the covering $T^4$ and exists at a generic point on the orbifold. Notice that in the literature it is customary to refer to bulk cycles if the coefficients $c_a^i$ are suitable integers. We find, however, more appropriate to use this term for the component in the homology inherited by the covering torus, independently of the coefficients $c_a^i$.

\subsubsection{Supersymmetric vacua}
\label{sec:brsusy}

We have now all the ingredients to discuss the orientifold of type IIB superstring compactified on the $T^4 /\mathbb{Z}_2$ orbifold.

Let us start discussing the supersymmetric case, where the orientifold projection employed is precisely the $\tilde\varOmega$ previously introduced. The fixed locus of $\tilde\varOmega\,  (1+g)$ is given by the ${\boldsymbol \pi}_1$ and
${\boldsymbol \pi}_2$ two-cycles that, taking into account a correct normalisation of the RR charge, identify the cycles
\begin{equation}
{\boldsymbol\varPi}_{\rm O} = 2\, {\boldsymbol\pi}_1 \,, \qquad
{\boldsymbol\varPi}_{{\rm O}'} =2\, {\boldsymbol \pi}_2\,,
\label{6doplanes}
\end{equation}
wrapped by the O7 and ${\rm O}7'$ planes, respectively.

D7 branes are free to wrap a generic two-cycle ${\boldsymbol\varPi}_a$ on the $T^4/\mathbb{Z}_2$. The Fourier coefficients $c$'s and $\epsilon$'s have now a physical interpretation in terms of charges of untwisted and twisted RR forms, respectively. In order to make contact with the perturbative CFT construction reported in appendix
\ref{app:6d}, it is useful to limit our attention to factorisable bulk two-cycles, that are obtained by combining two one-cycles, one on each $T^2$, and to express then Fourier coefficients in terms of the number of times the branes wraps these one-cycles. To be specific, we write
\begin{equation}
\hat {\boldsymbol\varPi}_a = \sum_{i=1}^4 c_a^i\, {\boldsymbol \pi}_i = \bigotimes_{i=1}^2 \left( m_a^i \, {\boldsymbol a}_i + n_a^i \, {\boldsymbol b}_i \right)\,,
\end{equation}
with $m_a^i$ and $n_a^i$ co-prime integers. As a result, a bulk cycle is entirely specified by the integers $(m_a^i , n_a^i)$, and
\begin{equation}
(c_a^1 ,\, c_a^2 ,\, c_a^3 ,\, c_a^4  ) = \tfrac{1}{2} (m^1_a \, m_a^2 ,\, n^1_a \, n^2_a ,\, m_a^1 \, n^2_a ,\, n^1_a \, m^2_a )\,.
\end{equation}
The Fourier coefficients $\epsilon_a^{xy}$ associated to the collapsed cycles identify the fixed point $(z_1^x ,\, z^y_2)$ crossed by the cycle associated to the D7 brane, and are related to the orbifold action on the Chan-Paton coefficients. With a suitable normalisation
\begin{equation}
\epsilon^{xy}_a =
\begin{cases}
 \pm\tfrac{1}{2} \,  i& {\rm if} \quad  (z_1^x ,\, z^y_2)\in \hat{\boldsymbol\varPi}_a\,,
 \\
 0 & {\rm otherwise}\,,
\end{cases}
\end{equation}
and define the charge $Q_a^{xy}=\epsilon_a^{xy}$ with respect to the twisted RR six-form potentials.

Although, by construction, ${\boldsymbol\varPi}_a$ is invariant under the orbifold action, generically it will not be so under $\tilde\varOmega$, and thus on the orientifold one has to introduce also image branes wrapping the cycle ${\boldsymbol \varPi}_{\bar a} = \tilde\varOmega \cdot {\boldsymbol \varPi}_a$. The Fourier coefficients of ${\boldsymbol \varPi}_{\bar a}$ are completely determined once we define the action of $\tilde\varOmega$ on the basis two-cycles:
\begin{equation}
\begin{split}
& \tilde\varOmega \cdot ({\boldsymbol\pi}_1 ,\, {\boldsymbol\pi}_2 ,\, {\boldsymbol\pi}_3,\, {\boldsymbol\pi}_4 ,\, {\boldsymbol\pi}_5 ,\, {\boldsymbol\pi}_6 )  =  ({\boldsymbol\pi}_1 ,\, {\boldsymbol\pi}_2 ,\, -{\boldsymbol\pi}_3,\, -{\boldsymbol\pi}_4 ,\, - {\boldsymbol\pi}_5 ,\,-  {\boldsymbol\pi}_6 ) \,,
\\
& \tilde\varOmega \cdot {\boldsymbol e}_{xy} = - {\boldsymbol e}_{xy}\,.
\end{split}
\end{equation}
As a result, on the $T^4 /\mathbb{Z}_2$ orbifold, the invariant configuration ${\boldsymbol \varPi}_a + {\boldsymbol\varPi}_{\bar a}$ only wraps bulk cycles, and this is consistent with the perturbative string description where the twisted tadpoles are identically vanishing. Moreover, ${\boldsymbol \varPi}_a + {\boldsymbol\varPi}_{\bar a}$ has only components along ${\boldsymbol\pi}_1$ and ${\boldsymbol\pi}_2$, consistently with the geometry of the orientifold planes.

At this point, one has all the ingredients to derive the light (chiral) spectrum of a vacuum involving a given number of D7 branes. A vacuum configuration is then fully
determined by specifying the wrapping numbers $(m_a^i , n_a^i)$ of the $a$-th  stack  of $N_a$ D7 branes together with the coordinate of (at least) one fixed point crossed by them.
In the following we shall assume that all branes pass through the fixed point at the origin of the $T^4$.

As a result, each D7 brane  supports a unitary gauge group so that
\begin{equation}
G_{\rm CP} = \prod_a {\rm U} (N_a )\,.
\end{equation}
Open strings living at the intersection of a ${\rm D}7_a$ and ${\rm D}7_b$ brane are in the bi-fundamental representation $(N_a , \bar N_b)$ and come in ${\mathcal I}_{ab}$ families, where now
\begin{equation}
{\mathcal I}_{ab} = {\boldsymbol\varPi}_a \circ {\boldsymbol\varPi}_b = \frac{1}{2} \,\prod_{i=1}^2 \left( m_a^i n_b^i - m_b^i n_a^i \right) - 2\, \sum_{x,y} \epsilon_a^{xy} \epsilon_b^{xy}\,.
\label{intersectionone}
\end{equation}
Open strings living at the intersection of a ${\rm D}7_a$ and ${\rm D}7_{\bar b}$ brane are in the bi-fundamental representation $(N_a , N_b)$ and come in ${\mathcal I}_{a\bar b}$ families, where now
\begin{equation}
{\mathcal I}_{a\bar b} = {\boldsymbol\varPi}_a \circ {\boldsymbol\varPi}_{\bar b} = \frac{1}{2} \,\prod_{i=1}^2 \left( m_a^i n_b^i + m_b^i n_a^i \right) + 2\, \sum_{x,y} \epsilon_a^{xy} \epsilon_b^{xy}\,.
\label{intersectiontwo}
\end{equation}
Finally, any time a brane intersects its $\tilde\varOmega$ image one has chiral matter in the symmetric and
anti-symmetric representations. In particular, the number of (anti-)symmetric representations is
\begin{equation}
\begin{split}
\frac{1}{2} \left[ {\mathcal I}_{a\bar a} \mp( {\mathcal I}_{aO} + {\mathcal I}_{a O'})\right] &=
\frac{1}{2} \left[ {\boldsymbol \varPi}_a \circ {\boldsymbol \varPi}_{\bar a} \mp {\boldsymbol \varPi}_a \circ ( {\boldsymbol\varPi}_{\rm O} + {\boldsymbol\varPi}_{{\rm O}'} )\right]
\\
&= \frac{1}{4} \left[ 4 \, \prod_{i=1}^2 m^i_a \, n_a^i - 4 \mp 4\, \left( \prod_{i=1}^2 m_a^i + \prod_{i=1}^2 n_a^i \right) \right]\,,
\end{split}
\end{equation}
where we have used the property that, for any D7 brane,
\begin{equation}
\sum_{x,y} 2\,\epsilon_a^{xy} = 0 \ {\rm mod}\ 4i\,, \quad {\rm and} \qquad 4\, \sum_{x,y} \epsilon_a^{xy}\, \epsilon_a^{xy} =-4\,.
\label{intersectionthree}
\end{equation}
Although a generic configuration of D7 branes is not supersymmetric, if their relative angles with respect to the O planes satisfy the familiar condition $\varphi_a^1 + \varphi_a^2 =0$, for each stack of ${\rm D}7_a$ branes, then some amount of supersymmetry is preserved. The condition on the angles is equivalent to the requirement that the two-cycles wrapped by the branes be actually a special Lagrangian submanifold calibrated by ${\rm Re} (\varOmega_d )$, where here $\varOmega_d$ is the holomorphic $d$-form of the internal manifold. In the T-dual version in terms of magnetised background, the supersymmetry conditions translates into the self-duality of the internal flux, $H_1 = H_2$ \cite{Angelantonj:2000hi}.

If we compare the expressions (\ref{intersectionone}), (\ref{intersectiontwo}) and (\ref{intersectionthree}) with the spectrum worked-out in appendix \ref{app:6d} and summarised in table \ref{tab:6dsusyspectrum}, we can identify
\begin{equation}
S_{ab} = - 4\, \sum_{x,y} \epsilon^{xy}_a \, \epsilon^{xy}_b\,,
\end{equation}
{\it i.e.} the number of mutual intersections between branes $a$ and $b$ coinciding with the orbifold fixed points is precisely given by minus four times the inner product of the Fourier coefficients relative to the collapsed cycles.

As an example, the original ${\rm U} (16) \times {\rm U} (16)$ vacuum of \cite{Bianchi:1990yu, Gimon:1996rq} involves two stacks of 16 branes each, with wrapping numbers
\begin{equation}
\hat{\boldsymbol\varPi}_{{\rm D}7} \sim (m^1,n^1;m^2,n^2)=(1,0;1,0)\,,
\qquad
\hat{\boldsymbol\varPi}_{{\rm D}7'} \sim (m^1,n^1;m^2,n^2)=(0,1;0,1)\,.
\end{equation}

Pairs of stacks of D7 branes can recombine on the $T^4 /\mathbb{Z}_2$. Although the dynamics of this process cannot be described by the underlying perturbative  CFT, the rules for analysing the recombination process are well established, and essentially amount to RR charge conservation \cite{Cremades:2002cs}. For the $\tilde\varOmega$ orientifold this is particularly simple, since the invariant combinations of branes carry a charge only with respect to the ${\boldsymbol \pi}_1$ and ${\boldsymbol \pi}_2$ cycles. In fact, one finds that the recombination of $N_a$ ${\rm D7}_a$ and $N_b$ ${\rm D7}_b$ brane yields $N_c$ ${\rm D}7_c$ branes
\begin{equation}
N_c \, ({\boldsymbol\varPi}_c +{\boldsymbol\varPi}_{\bar c}) \equiv \left[ N_a\, ({\boldsymbol\varPi}_a +{\boldsymbol\varPi}_{\bar a} )\right] \cup \left[ N_b\, ({\boldsymbol\varPi}_b +{\boldsymbol\varPi}_{\bar b} ) \right] \,,
\end{equation}
with $N_c = {\rm GCD} (N_a , N_b)$, and wrapping numbers determined by the conditions
\begin{equation}
\begin{split}
m_c^1 m_c^2 &= \frac{N_a}{N_c}\, m_a^1 m_a^2 + \frac{N_b}{N_c}\, m_b^1 m_b^2 \,,
\\
n_c^1 n_c^2 &= \frac{N_a}{N_c}\, n_a^1 n_a^2 + \frac{N_b}{N_c}\, n_b^1 n_b^2\,.
\end{split}
\label{branerec}
\end{equation}
The amazing result is that {\em all} supersymmetric $T^4 /\mathbb{Z}_2$ orientifold vacua, with the same closed-string spectrum, are {\it all} in the same moduli space, and thus are connected to the ${\rm U} (16) \times {\rm U} (16)$ model of \cite{Bianchi:1990yu, Gimon:1996rq} via the recombination of suitable numbers of branes. 
This connection is expected to hold also in the case of compactifications on orbifolds of skew tori (with quantised real component of the complex structure \cite{Angelantonj:1999xf}), though we have not analysed it in detail.

Let us consider in fact the complete recombination of the D7 and ${\rm D}7'$ branes. The resulting configuration involves a single stack of 16 branes with wrapping numbers $(1,1;1,1)$ thus yielding a ${\rm U} (16)$ gauge group together with 4 hypermultiplets in the anti-symmetric $120$ representation. Similarly, one could recombine only 4 D7 branes together with the 16 ${\rm D}7'$ ones, so that the resulting configuration contains for instance two different stacks of branes of the type
\begin{equation}
\begin{split}
12 \  {\rm D}7 \ {\rm branes \ with:}&\quad  (m^1,n^1;m^2,n^2)=(1,0;1,0)\,,
\\
4 \ {\rm D}7 ''\ {\rm branes\ with:}& \quad (m^1,n^1;m^2,n^2)=(1,2;1,2)\,.
\end{split}
\end{equation}
The gauge group is now ${\rm U} (12)\times {\rm U} (4)$ with hypermultiplets in the representations $2\, (66,1) + 10 \, (1,6) + 4 (12,4)$. This is precisely the magnetised vacuum {\it without D5 branes}  found in \cite{Angelantonj:2000hi}.

In the T-dual picture in terms of magnetised branes the recombination process has the suggestive interpretation of brane transmutation, {\it i.e.} some D5 branes dilute into a constant magnetic background on the world-volume of the D9 branes. Alternatively, the process is somewhat dual to the small instanton transitions of \cite{Witten:1995gx}, where now the instanton becomes fatter and fatter and invades the whole compact space. We shall give evidence to this picture in section 5.

In the field theory limit, the recombination of branes $a$ and $b$ is described in terms of the Higgsing of massless scalars in the bi-fundamental $(N_a , N_b)$, and we shall review it in the next section.

\subsubsection{Brane Supersymmetry Breaking vacua}
\label{susec:bsb}

Brane Supersymmetry Breaking \cite{bsb1b} is an interesting deformation of the supersymmetric $T^4 /\mathbb{Z}_2$ orientifold, where the world-sheet parity is further dressed with an involution of the internal manifold $\sigma$ that affects the projection of the twisted sector, $\hat\varOmega = \tilde\varOmega \, \sigma$. In this way, the Klein-bottle projection (anti-)symmetrises the (NS-NS) R-R twisted sector and therefore the closed-string spectrum, still ${\mathcal N}=(1,0)$ supersymmetric, comprises 17 tensor multiplets together with 4 hypermultiplets aside, of course, of the gravitational supermultiplet. From a geometrical perspective, the involution $\sigma$ affects the nature of the orientifold planes that now are ${\rm O}7_-$ and ${\rm O}7^\prime_+$, {\it i.e.} the O-planes wrapping the ${\boldsymbol\pi}_2$ cycle have positive NS-NS tension and positive R-R charge. As a result, they have a different orientation with respect to the more conventional ${\rm O}7^\prime_-$, so that ${\boldsymbol\varPi}_{{\rm O}^\prime_+} = - 2 \, {\boldsymbol\pi}_2$, while ${\boldsymbol\varPi}_{{\rm O}_-} = + 2 \, {\boldsymbol\pi}_1$ as before.

An additional difference with respect to the previous supersymmetric case, is that the presence of $\sigma$ also affects the components of a D7 brane with respect to the collapsed cycles, and the behaviour of the latters under $\hat\varOmega$:
\begin{equation}
\epsilon^{xy}_a =
\begin{cases}
\pm \tfrac{1}{2} & {\rm if} \quad  (z_1^x ,\, z^y_2)\in \hat{\boldsymbol\varPi}_a\,,
 \\
 0 & {\rm otherwise}\,,
\end{cases}
\label{twistedcarges}
\end{equation}
and
\begin{equation}
\begin{split}
& \hat\varOmega \cdot ({\boldsymbol\pi}_1 ,\, {\boldsymbol\pi}_2 ,\, {\boldsymbol\pi}_3,\, {\boldsymbol\pi}_4 ,\, {\boldsymbol\pi}_5 ,\, {\boldsymbol\pi}_6 )  =  ({\boldsymbol\pi}_1 ,\, {\boldsymbol\pi}_2 ,\, -{\boldsymbol\pi}_3,\, -{\boldsymbol\pi}_4 ,\, - {\boldsymbol\pi}_5 ,\,-  {\boldsymbol\pi}_6 ) \,,
\\
& \hat\varOmega \cdot {\boldsymbol e}_{xy} = + {\boldsymbol e}_{xy}\,.
\end{split}
\label{OactionBSB}
\end{equation}
The choice of the coefficients $\epsilon^{xy}_a$ reflects the orbifold action on the Chan-Paton labels, that is now real \cite{bsb1b}. As a result of eq.
(\ref{OactionBSB}), the $\hat\varOmega$ invariant combination
\begin{equation}
{\boldsymbol\varPi}_a + {\boldsymbol\varPi}_{\bar a} = 2 \, c_a^1 \, {\boldsymbol\pi}_1 + 2 \, c_a^2 \, {\boldsymbol\pi}_2 + 2\, \sum_{x,y} \epsilon^{xy}_a\, {\boldsymbol e}_{xy}
\end{equation}
has now components not only along the bulk ${\boldsymbol\pi}_1$ and ${\boldsymbol \pi}_2$ cycles, but also along the exceptional ones. This implies that additional conditions have to be imposed on a consistent configuration of D7 branes, resulting in additional tadpoles for the twisted RR six-form potentials. Of course, this is in agreement with the perturbative CFT description, that is summarised in  appendix \ref{app:6d}.

The chiral open-string spectrum is still captured by the intersection form of the two-cycles wrapped by the branes. A generic ${\rm D}7_a$ brane still supports a unitary gauge group ${\rm U} (N_a)$, while open strings stretched between intersecting ${\rm D}7_a$ and ${\rm D}7_b$ branes  come in bi-fundamental representations $(N_a , \bar N_b)$ with a degeneracy given by
\begin{equation}
{\mathcal I}_{ab} = {\boldsymbol\varPi}_a \circ {\boldsymbol\varPi}_b = \frac{1}{2} \left[ \prod_{i=1}^2 \left( m_a^i \, n_b^i - m_b^i \, n_a^i \right)- 4 \sum_{x,y} \epsilon^{xy}_a \, \epsilon^{xy}_b \right]\,,
\end{equation}
open strings stretched between intersecting ${\rm D}7_a$ and ${\rm D}7_{\bar b}$ branes  come in bi-fundamental representations $(N_a , N_b)$ with a degeneracy given by
\begin{equation}
{\mathcal I}_{a\bar b} = {\boldsymbol\varPi}_a \circ {\boldsymbol\varPi}_{\bar b} = \frac{1}{2} \left[ \prod_{i=1}^2 \left( m_a^i \, n_b^i + m_b^i \, n_a^i \right)
- 4 \sum_{x,y} \epsilon^{xy}_a \, \epsilon^{xy}_b \right]\,,
\end{equation}
and, finally, any time a brane intersect its $\hat\varOmega$ image chiral fermions carry a (anti-)symmetric representation with a degeneracy given by
\begin{equation}
\begin{split}
\frac{1}{2} \left[ {\mathcal I}_{a\bar a} \mp ({\mathcal I}_{a O_-} + {\mathcal I}_{aO^\prime_+})\right] &= \frac{1}{2} \left[ {\boldsymbol\varPi}_a \circ {\boldsymbol\varPi}_{\bar a} \mp \left(
{\boldsymbol\varPi}_a \circ {\boldsymbol\varPi}_{O_-} + {\boldsymbol\varPi}_a \circ {\boldsymbol\varPi}_{O^\prime_+} \right)\right]
\\
&=\frac{1}{4} \left[ 4\, \prod_{i=1}^2 m_a^i\, n_a^i -4 \mp 4\, \left( \prod_{i=1}^2 m_a^i - \prod_{i=1}^2 n_a^i\right)\right]\,.
\end{split}
\end{equation}
Comparison with table 2 allows us to identify
\begin{equation}
\epsilon_a \, \epsilon_b \, S_{ab} = -4 \, \sum_{x,y} \epsilon^{xy}_a \, \epsilon_{b}^{xy}\,,
\end{equation}
where, in the CFT analysis, $S_{ab}$ counts, as usual, the number of mutual intersections between branes $a$ and $b$ coinciding with the orbifold fixed points, while $\epsilon_a$ takes into account the action of the orbifold on the Chan-Paton degeneracies. Note also that, since the $\epsilon^{xy}_a$ are now real, $4\,\sum_{x,y} \epsilon^{xy}_a \, \epsilon^{xy}_a=4$.
Special care is needed when a D7 brane wraps the same cycle as an orientifold plane, since now gauge groups become orthogonal or symplectic depending on whether the
O-plane is an ${\rm O}7_-$ or an ${\rm O}7_+^\prime$ one.

Since supersymmetry is explicitly broken in the open-string sector, charged scalars do not carry the same representations as their ``would be'' fermionic superpartners. Although the geometrical description adopted in this section is not suited for determining their spectrum, the perturbative CFT analysis of appendix \ref{app:6d} gives complete information about their representations and multiplicities, and the complete spectrum is summarised in table \ref{non-susy spectrum}.

Also in this case with Brane Supersymmetry Breaking we conjecture that {\em all} $T^4 /\mathbb{Z}_2$ vacua, with the same closed-string spectrum, are connected to the original ${\rm SO} (16)^2 \times {\rm USp} (16)^2$ model of \cite{bsb1b} via the recombination of suitable numbers of branes, though they might be separated by an energy barrier as argued in section 4.

Before we show the connection by working out some explicit examples, one should note that the rules for brane recombination  (\ref{branerec}) are now changed since they have to take into account also the conservation of the twisted charges, that are now non-vanishing. Let us suppose we recombine one ${\rm D}7_a$ and one ${\rm D}7_b$ brane
\begin{equation}
\left( {\boldsymbol \varPi}_a + {\boldsymbol \varPi}_{\bar a} \right) \cup  \left( {\boldsymbol \varPi}_b + {\boldsymbol \varPi}_{\bar b} \right) =
(m_a^1 m_a^2 + m_b^1 m_b^2 ) {\boldsymbol \pi}_1 + (n_a^1 n_a^2 + n_b^1 n_b^2 ){\boldsymbol\pi}_2 +2\,  \sum_{x,y} (\epsilon^{xy}_a + \epsilon^{xy}_b ) \, {\boldsymbol e}_{xy}\,.
\end{equation}
This cannot yield automatically a factorisable ${\rm D}7_c$ brane since, in general
\begin{equation}
2 \sum_{x,y} \left( \epsilon^{xy}_a + \epsilon^{xy}_b \right) \not= 4 \,,
\end{equation}
that is to say the associated ${\boldsymbol\varPi}_c$ cycle crosses more or less fixed points than the canonical four touched upon by a factorisable cycle. As a result, in order to allow the recombination of two or more branes into a factorisable one the additional condition
\begin{equation}
2 \sum_{x,y} \sum_a  \epsilon^{xy}_a = 0\ {\rm mod}\ 4
\label{branerecBSB}
\end{equation}
must hold. In this case the bulk cycle of the recombined branes is identified by the wrapping numbers $(m_c^i , n_c^i)$, with
\begin{equation}
m_c^1 \, m_c^2 = \sum_a m_a^1 \, m_a^2\,, \qquad n_c^1 \, n_c^2 = \sum_a n_a^1 \, n_a^2\,.
\end{equation}

Let us discuss now some simple examples where the various recombinations of branes in the original model \cite{bsb1b} yield new and old vacua with Brane Supersymmetry Breaking. The ${\rm SO} (16)^2 \times {\rm USp} (16)^2$ model involves eight copies of physical branes wrapping the following cycles:
\begin{equation}
\begin{split}
{\boldsymbol\varPi}^\pm_{{\rm D}7} &= \hat {\boldsymbol \varPi}_{{\rm D}7} \pm \tfrac{1}{2} \left( {\boldsymbol e}_{11} +  {\boldsymbol e}_{12} +  {\boldsymbol e}_{21} +  {\boldsymbol e}_{22} \right) \,,
\\
{\boldsymbol\varPi}^\pm_{\overline{{\rm D} 7} '} &= \hat {\boldsymbol \varPi}_{\overline{\rm D}7'} \pm \tfrac{1}{2} \left({\boldsymbol e}_{11} +  {\boldsymbol e}_{13} +  {\boldsymbol e}_{31} +  {\boldsymbol e}_{33} \right) \,,
\end{split}
\end{equation}
where $\hat {\boldsymbol \varPi}_{{\rm D}7} = \frac{1}{2} \, {\boldsymbol\pi}_1$ and $\hat {\boldsymbol \varPi}_{\overline{\rm D}7'} = - \frac{1}{2}\, {\boldsymbol\pi}_2$, with the minus sign implying that the ${\rm D}7'$ branes are actually anti-branes, and thus  their cycle has an opposite orientation. The sign in front of the exceptional cycles reflects, as already stated, the different orientifold action on the associated Chan-Paton labels, and thus the various twisted RR charges.

In this case, one needs to recombine at least three different stacks of branes in order to get a factorisable one. For instance, one could recombine\footnote{The additional factor 2 takes into account the fact that the ${\rm D}7$ and $\overline{{\rm D}7}'$ branes wrap the same cycle as their orientifold images. Here and in the following we shall omit to indicate the orientifold images, though they are tacitly present in the identification of the invariant cycle.}
\begin{equation}
(4\times 2\, {\boldsymbol\varPi}^-_{{\rm D}7}) \cup (8\times 2\, {\boldsymbol\varPi}^+_{\overline{{\rm D} 7} '} )\cup (8\times 2\, {\boldsymbol\varPi}^-_{\overline{{\rm D} 7} '} )
\end{equation}
to yield a new vacuum with eight horizontal D7 branes of type ${\boldsymbol\varPi}^+_{{\rm D}7}$, four horizontal D7 branes of type ${\boldsymbol\varPi}^-_{{\rm D}7}$
 and four oblique ${\rm D}7_o$ branes wrapping the cycle
\begin{equation}
{\boldsymbol\varPi}_{{\rm D}7_o}  = \tfrac{1}{2} \left( {\boldsymbol\pi}_1 - 4 \, {\boldsymbol\pi}_2 + 2 \, {\boldsymbol\pi}_3 - 2\, {\boldsymbol\pi}_4 \right) - \tfrac{1}{2} \left(
 {\boldsymbol e}_{11} +  {\boldsymbol e}_{12} +  {\boldsymbol e}_{21} +  {\boldsymbol e}_{22} \right) \,.
\end{equation}
This configuration clearly satisfies both the untwisted and the twisted tadpole conditions, and reproduces the Brane Supersymmetry Breaking vacuum {\it without D5 antibranes} of \cite{Angelantonj:2000hi} and with anti-self-dual magnetic background, with gauge group ${\rm SO} (16) \times {\rm SO} (8) \times {\rm U} (4)$.

Another possibility would be to recombine all branes of the original model. This would identify the two-cycle ${\boldsymbol \pi}_1 - {\boldsymbol\pi}_2$, without any leg along the collapsed cycles. What kind of brane could this correspond to? Notice that in the original configurations all branes were crossing the orbifold fixed points, and in particular the fixed point at the origin of the $T^4$ with $(z_1,z_2) =(0,0)$. As a result, also the recombined brane(s) should cross the origin, since the recombination process does not involve any deformation associated to brane displacements and/or Wilson lines. Moreover, in this vacuum all branes passing through the origin cross exactly four fixed points, and thus wrap four collapsed cycles. As a result, the complete recombination of all branes in the original model must result in, at least, two different stacks of branes wrapping the same bulk cycle ${\boldsymbol\pi}_1 - {\boldsymbol\pi}_2$ but with opposite twisted charges. Given the Fourier coefficients $c^i = (\frac{1}{2},-\frac{1}{2},0,0)$ one has the solution
\begin{equation}
(8\times 2\, {\boldsymbol\varPi}^+_{{\rm D}7}) \cup (8\times 2\, {\boldsymbol\varPi}^-_{{\rm D}7}) \cup (8\times 2\, {\boldsymbol\varPi}^+_{\overline{{\rm D} 7} '} )\cup (8\times 2\, {\boldsymbol\varPi}^-_{\overline{{\rm D} 7} '} )
=8 ( {\boldsymbol\varPi}_{d} + {\boldsymbol\varPi}_{d'} )\,,
\end{equation}
with
\begin{equation}
\begin{split}
{\boldsymbol\varPi}_{d} &= \tfrac{1}{2} \left( {\boldsymbol \pi}_1 - {\boldsymbol \pi}_2 + {\boldsymbol \pi}_3 - {\boldsymbol \pi}_4 \right) + \tfrac{1}{2} \left( {\boldsymbol e}_{11}+{\boldsymbol e}_{14}+{\boldsymbol e}_{41}+ {\boldsymbol e}_{44} \right)\,,
\\
{\boldsymbol\varPi}_{d'} &=\tfrac{1}{2} \left( {\boldsymbol \pi}_1 - {\boldsymbol \pi}_2 + {\boldsymbol \pi}_3 - {\boldsymbol \pi}_4 \right)- \tfrac{1}{2} \left( {\boldsymbol e}_{11}+{\boldsymbol e}_{14}+{\boldsymbol e}_{41}+ {\boldsymbol e}_{44} \right)\,.
\end{split}
\end{equation}
The Chan-Paton gauge group is now $G_{\rm CP} = {\rm U} (8) \times {\rm U} (8)$ with left-handed  fermions in the adjoint representation and in four copies of the $(28,1)+(1,28)$, and right handed fermions in two copies of the bi-fundamental representation $(8,\bar 8)$. From the CFT analysis of appendix \ref{app:6d} one also finds 16 real scalars in the $(8,8)$ together with 8 real scalars in the $(8,\bar 8)$, as summarised in table \ref{u8u8}.

More examples can of course be studied, though we leave a detailed analysis to the interested reader.

\begin{table}
\centering
\begin{tabular}{cccc}
\toprule
Multiplicity & Representation & Multiplet/Field & Chirality\\
\toprule
$4$ & $(28,1)+(1,28)$ & Weyl Fermion & Left-handed
\\
\midrule
$2$ & $(8,\bar8)$ & Hypermultiplet & Right-handed
\\
\midrule
$16$ & $(8,8)$ & Scalars & -
\\
\midrule
$1$ & $(64,1)+(1,64)$ & Gauge Multiplet & Left-handed\\
\bottomrule
\end{tabular}
\caption{Massless spectrum of the ${\rm U}(8)\times {\rm U} (8)$ model. Left-handed (right-handed) fermions correspond to $C_4$ ($S_4$) characters.}
\label{u8u8}
\end{table}
\subsection{Four-dimensional models}
\label{subsec:4d}

In four dimensions we focus our attention on a specific class of models based on the $T^6 /\mathbb{Z}_2 \times \mathbb{Z}_2$ orbifold with discrete torsion. The naive orientifold construction with orthogonal D-branes is a generalisation of the Brane Supersymmetry Breaking construction with supersymmetry broken in the open-string sector \cite{Angelantonj:1999ms}. However, it was shown in \cite{ms, Dudas:2005jx, Blumenhagen:2005tn} that supersymmetric vacua exist also for this compactification and involve non-trivial angles and/or magnetic backgrounds. The natural question is then whether the two constructions are related by some (non-perturbative) effect, like brane recombination.

The $\mathbb{Z}_2 \times \mathbb{Z}_2$ group is generated by the elements
\begin{equation}
g= (+,-,-)\,, \qquad h = (-,-,+)\,,
\end{equation}
while, as usual,  $f = gh = (-,+,-)$, and the $i$-th entry denotes the action of the orbifold on the $i$-th factor in the factorisable $T^6 = T^2 \times T^2 \times T^2$. The presence of discrete torsion, {\it i.e.} a relative sign $\epsilon=-1$ in the independent ${\rm SL} (2,\mathbb{Z})$ orbit in the twisted sector, exchanges the $h_{2,1}$ and $h_{1,1}$ Hodge numbers of the smooth Calabi-Yau manifold \cite{Vafa:1994rv}, and  acts as mirror symmetry in the type II compactification.

The description of the homology of the ${\mathcal X} = T^6 /\mathbb{Z}_2 \times \mathbb{Z}_2$ space is similar to the six-dimensional case previously studied. $H_3 ({\mathcal X}, \mathbb{Z})$ is generated by the eight bulk cycles
\begin{equation}
\begin{split}
\{ \bar{\boldsymbol\pi} \}_i = & \left(
{\boldsymbol a}_1 \otimes {\boldsymbol a}_2 \otimes {\boldsymbol a}_3 ,\,
{\boldsymbol b}_1 \otimes {\boldsymbol b}_2 \otimes {\boldsymbol b}_3 ,\,
{\boldsymbol b}_1 \otimes {\boldsymbol b}_2 \otimes {\boldsymbol a}_3 ,\,
{\boldsymbol a}_1 \otimes {\boldsymbol a}_2 \otimes {\boldsymbol b}_3 ,\, \right.
\\
& \left. \ \
{\boldsymbol b}_1 \otimes {\boldsymbol a}_2 \otimes {\boldsymbol b}_3 ,\,
{\boldsymbol a}_1 \otimes {\boldsymbol b}_2 \otimes {\boldsymbol a}_3 ,\,
{\boldsymbol a}_1 \otimes {\boldsymbol b}_2 \otimes {\boldsymbol b}_3 ,\,
{\boldsymbol b}_1 \otimes {\boldsymbol a}_2 \otimes {\boldsymbol a}_3
\right)
\end{split}
\end{equation}
inherited from the $T^6$. These three-cycles are common both to the orbifolds with and without discrete torsion. In the former case, however, one has also to consider the exceptional cycles that are built by tensoring a collapsed two-cycle ${\boldsymbol e}^i_{xy}$, localised at the fixed point $z^i_{xy}$ of the $i$-th element of the orbifold group, with a one-cycle of the spectator $T^2$:
\begin{equation}
\begin{split}
{\boldsymbol \alpha}^g_{xy} &=2\, {\boldsymbol a}_1 \otimes {\boldsymbol e}^g_{xy} \,,
\\
{\boldsymbol \beta}^g_{xy} &=2\,  {\boldsymbol b}_1 \otimes {\boldsymbol e}^g_{xy} \,,
\end{split}
\qquad
\begin{split}
{\boldsymbol \alpha}^f_{xy} &=2\,  {\boldsymbol a}_2 \otimes {\boldsymbol e}^f_{xy} \,,
\\
{\boldsymbol \beta}^f_{xy} &=2\,  {\boldsymbol b}_2 \otimes {\boldsymbol e}^f_{xy} \,,
\end{split}
\qquad
\begin{split}
{\boldsymbol \alpha}^h_{xy} &=2\,  {\boldsymbol a}_3 \otimes {\boldsymbol e}^h_{xy} \,,
\\
{\boldsymbol \beta}^h_{xy} &=2\,  {\boldsymbol b}_3 \otimes {\boldsymbol e}^h_{xy} \,.
\end{split}
\end{equation}
Altogether, these generate $H_3 ({\mathcal X} , \mathbb{Z})$ whose dimension is indeed $b_3=104$. Here and in the following we use the index $i$ to label both the $i$-th element of the orbifold group $(g,f,h)$ and the $i$-th $T^2$ component of the factorisable $T^6$. These are in fact related since, for instance, $g$ leaves the first $T^2$ fixed, while $h$ is inert on the second one, and $f$ on the third $T^2$.

The intersection form of these cycles can be straightforwardly derived from the one of the covering $T^6$ and from the intersection form of the exceptional cycles in $T^4 /\mathbb{Z}_2$. In particular, noting that under the $\mathbb{Z}_2 \times \mathbb{Z}_2$ action each bulk three-cycle has exactly three orbifold images
\begin{equation}
{\boldsymbol\pi}_i = (1+g)\,(1+h)\, \bar{\boldsymbol\pi}_i\,,
\end{equation}
and that collapsed cycles of different twisted sectors do not intersect, one finds the following non-vanishing entries of the intersection form:
\begin{equation}
I_{ij}= \tfrac{1}{4}\, {\boldsymbol \pi}_i \circ {\boldsymbol \pi}_j = - 4\, i\, \sigma_2 \otimes {\bf 1}_4\,,
\end{equation}
and
\begin{equation}
I_{xy,vw}^{\lambda\kappa}= \tfrac{1}{2}\, {\boldsymbol \alpha}^\lambda_{xy} \circ {\boldsymbol\beta}^\kappa_{vw} =- 4\, \delta^{\lambda\kappa}\, \delta_{xv}\, \delta_{yw}\,,
\end{equation}
where $\sigma_2$ is the Pauli matrix, ${\bf 1}_4$ is a $4\times 4$ identity matrix, and we have used the fact that, for a given twisted sector, ${\boldsymbol e}_{xy} \circ {\boldsymbol e}_{vw} = -2 \, \delta_{xv}\, \delta_{yw}$, while  ${\boldsymbol a}_i \circ {\boldsymbol b}_j =  \delta_{ij}$. A generic three-cycle of ${\mathcal X}$ can thus be written as
\begin{equation}
{\boldsymbol\varPi}_a =\sum_{i=1}^8 c_a^i \, {\boldsymbol\pi}_i +\sum_{\lambda=g,f,h} \sum_{x,y=1}^4 \left( \mu^{xy}_{a,\lambda}\, {\boldsymbol\alpha}^\lambda_{xy} +\nu^{xy}_{a,\lambda}\, {\boldsymbol\beta}^\lambda_{xy} \right) \,.
\end{equation}
Also in this case it is customary to refer to
\begin{equation}
\hat {\boldsymbol\varPi}^a = \sum_{i=1}^8 c^i_a \, {\boldsymbol\pi}_i
\end{equation}
as the bulk cycle.

Modding-out the $T^6 /\mathbb{Z}_2 \times \mathbb{Z}_2$ compactification by  $\tilde\varOmega = \varOmega \, {\mathcal R}\, (-1)^{F_{\rm L}}$ implies that four different types of O6 planes are introduced at the fixed locus of
$\tilde\varOmega $, $\tilde\varOmega \, g$, $\tilde\varOmega \, f$ and $\tilde\varOmega \, h$. In particular, given the action of the orbifold on the three-cycles and the fact that $\tilde\varOmega \cdot {\boldsymbol a}_i = +{\boldsymbol a}_i$ and $\tilde\varOmega \cdot {\boldsymbol b}_i = -{\boldsymbol b}_i$, one obtains the following geometry of
${{\rm O}6^o_{\epsilon_o}}$, ${{\rm O}6^g_{\epsilon_g}}$, ${{\rm O}6^f_{\epsilon_f}}$ and ${{\rm O}6^h_{\epsilon_h}}$ planes with associated three-cycles
\begin{equation}
{\boldsymbol\varPi}_{o,{\epsilon_o}} = 2\, \epsilon_o\,  {\boldsymbol \pi}_1\,, \quad
{\boldsymbol\varPi}_{g,{\epsilon_g}} = - 2\,\epsilon_g\,  {\boldsymbol \pi}_7\,, \quad
{\boldsymbol\varPi}_{f,{\epsilon_f}} =  -2 \,\epsilon_f\,  {\boldsymbol \pi}_5\,, \quad
{\boldsymbol\varPi}_{h,{\epsilon_h}} = -2 \,\epsilon_h\,  {\boldsymbol \pi}_3\,, 
\label{oplanesigns}
\end{equation}
where we have properly normalised the cycles as pertained to O6 planes. The signs $\epsilon_\alpha =\pm 1$  determine the type of orientifold plane $\alpha = o,g,f,h$, with positive $\epsilon_\alpha$ associated to a conventional O-plane with negative tension and charge, while $\epsilon_\alpha =-1$ refers to an {\it exotic} plane with positive tension and charge. These signs are actually not arbitrary, but have to be correlated to the presence/absence of discrete torsion $\epsilon$ as \cite{Angelantonj:1999ms}
\begin{equation}
\epsilon = \epsilon_o \, \epsilon_g \,  \epsilon_f \,\epsilon_h \,.
\end{equation}
In the following, we shall make the definite choice $\epsilon=-1$ and $(\epsilon_o , \epsilon_g , \epsilon_f , \epsilon_h ) = (+,+,+,-)$ as in \cite{Angelantonj:1999ms, Dudas:2005jx, Angelantonj:2009yj}, while the other options can be discussed in a similar fashion and require only minor modifications.

A generic D6-brane on the $T^4/\mathbb{Z}_2 \times \mathbb{Z}_2$ orbifold is thus identified by the Fourier coefficients of  its associated three-cycle ${\boldsymbol\varPi}_a$ with respect to the basis in $H_3$ previously introduced. In particular, as in the six-dimensional case, these coefficients are uniquely specified by the wrapping numbers $(m_a^i , n_a^i)$ associated to the canonical one-cycles of the $i$-th $T^2$.
As a result,\footnote{The signs in eq. (\ref{oddsigns}) can be determined by the consistency with the CFT description of the vacuum configuration, and differ from those of \cite{Blumenhagen:2005tn} the similar minus signs appear in the definition of the cycle wrapped by the O-planes. However, from a direct comparison with the transverse-channel Klein-bottle partition function it seems to us that the definitions (\ref{oplanesigns}) and (\ref{oddsigns}) are more appropriate.}
\begin{equation}
\{ c_a \}^i = \tfrac{1}{4}\, \left(  m^a_1 m^a_2 m^a_3\, ,\,  n^a_1 n^a_2 n^a_3 \,,\, n^a_1 n^a_2 m^a_3 \,,\,  m^a_1 m^a_2 n^a_3 \,,\,  n^a_1 m^a_2 n^a_3\,,\, m^a_1 n^a_2 m^a_3 \,,\,    m^a_1 n^a_2 n^a_3 \,,\,  n^a_1 m^a_2 m^a_3
\right)\,,
\label{oddsigns}
\end{equation}
while
\begin{equation}
\mu^{xy}_{a,\lambda} = \tfrac{1}{2} \, m^a_\lambda \, \epsilon^{xy}_{a,\lambda}
\,,
\qquad
\nu^{xy}_{a,\lambda} = \tfrac{1}{2}\, n^a_\lambda \, \epsilon^{xy}_{a,\lambda} \,.
\end{equation}
The Fourier coefficients $\epsilon^{xy}_{a,\lambda}$ associated to the collapsed cycles identify the fixed points crossed by the bulk cycle associated to the D6 brane, and are related to the orbifold action on the Chan-Paton coefficients. So in the case at hand with $(\epsilon_o , \epsilon_g, \epsilon_f , \epsilon_h)= (+,+,+,-)$, one has
\begin{equation}
\epsilon^{xy}_{a,g } =
\begin{cases}
\pm \tfrac{1}{2}i & {\rm if} \quad  (\bullet ,\, z_2^x ,\, z^y_3)\in \hat{\boldsymbol\varPi}_a\,,
 \\
 0 & {\rm otherwise}\,,
\end{cases}
\qquad
\epsilon^{xy}_{a,f} =
\begin{cases}
\pm \tfrac{1}{2}i & {\rm if} \quad  ( z_1^x ,\,\bullet ,\, z^y_3)\in \hat{\boldsymbol\varPi}_a\,,
 \\
 0 & {\rm otherwise}\,,
\end{cases}
\label{twistcharge4d1}
\end{equation}
while
\begin{equation}
\epsilon^{xy}_{a,h} =
\begin{cases}
\pm \tfrac{1}{2} & {\rm if} \quad  (z_1^x ,\, z^y_2,\, \bullet)\in \hat{\boldsymbol\varPi}_a\,,
 \\
 0 & {\rm otherwise}\,.
\end{cases}
\label{twistcharge4d2}
\end{equation}

Generically, the cycle ${\boldsymbol\varPi}_a$ is not left invariant by $\tilde\varOmega$, and thus one needs also introduce its image ${\boldsymbol\varPi}_{\bar a}$ in order to build an invariant configuration. To this end, one has to use the previous action of $\tilde\varOmega$ on the canonical one-cycles of each $T^2$, as well as its action on the collapsed ones
\begin{equation}
\tilde\varOmega \cdot {\boldsymbol\alpha}^\lambda_{xy} = - \epsilon_\lambda \, {\boldsymbol\alpha}^\lambda_{xy}\,,
\qquad
\tilde\varOmega \cdot {\boldsymbol\beta}^\lambda_{xy} = \epsilon_\lambda \, {\boldsymbol\beta}^\lambda_{xy}\,,
\end{equation}
where the signs $\epsilon_\lambda$ are precisely those that determine the type of O-planes associated to the $g$, $f$ and $h$ twists. Also in this case, it is useful to introduce the invariant combination $\tilde{\boldsymbol\varPi}_a =  {\boldsymbol\varPi}_a + {\boldsymbol\varPi}_{\bar a} $, in order to determine the effective cycle wrapped by an invariant combination of branes.

The chiral spectrum can be derived as usual by computing the intersection numbers between different stacks of branes and between branes and orientifold planes \cite{Blumenhagen:2005tn}, or by a direct CFT computation of the annulus and M\"obius-strip amplitudes \cite{Angelantonj:2009yj}. There are some subtleties here associated to the different nature of the orientifold planes. For instance, while D6 branes lying on ${\rm O}6_-$ planes yield unitary gauge groups, when lying on a ${\rm O}6_+$ the Chan-Paton coefficients become real and the gauge groups turns out to be symplectic. Moreover, when D6 branes  wrap the same cycle as an O6-plane the twisted charge assignments of eqs. (\ref{twistcharge4d1}) and (\ref{twistcharge4d2}) have to be modified so to reflect the correct orbifold action on the Chan-Paton labels, that can be found, for instance in \cite{Angelantonj:2002ct, Dudas:2000bn}. In the following, we shall not be concerned with the complete spectrum but will mainly focus our attention on the gauge group.

Also for this four-dimensional example, different vacua with common closed-string sector, {\it i.e.} different solutions of the tadpole conditions, can be connected via brane recombination. In particular, it is quite amusing that the simplest vacuum with Brane Supersymmetry Breaking \cite{Angelantonj:2009yj} is actually connected to the supersymmetric solutions of \cite{ Dudas:2005jx, Blumenhagen:2005tn, Camara:2010zm}. Let us show this connection in few illustrative examples.

The simplest solution of tadpole conditions is in terms of different stacks of D6 and $\overline{{\rm D}6}$ branes that are parallel to the various O-planes. Their wrapping numbers are
\begin{equation}
\begin{split}
{\rm D}6_o &= (1,0\,;\, 1,0\,;\, 1,0)\,,
\\
{\rm D}6_f &= (0,-1\,;\, 1,0 \,;\, 0,1)\,,
\end{split}
\qquad
\begin{split}
{\rm D}6_g &= (1,0\,;\, 0,1 \,;\, 0,-1)\,,
\\
{\rm D}6_h &= (0,1\,;\, 0,1\,;\, 1,0)\,.
\end{split}
\end{equation}
while the associated twisted charges identify different stacks. In particular, the consistency of the construction implies the following invariant cycles wrapped by the various D6 branes and their images
\begin{equation}
\begin{split}
\tilde{\boldsymbol\varPi}_o^\pm &= +\tfrac{1}{2} {\boldsymbol\pi}_1 \pm 2  \sum_{x,y=1}^4 \mu^{xy}_{a,h} \, {\boldsymbol\alpha}^h_{xy}
= +\tfrac{1}{2} {\boldsymbol\pi}_1 \pm \tfrac{1}{2} \, \left( {\boldsymbol \alpha}^h_{11} +  {\boldsymbol \alpha}^h_{12}
+  {\boldsymbol \alpha}^h_{21} + {\boldsymbol \alpha}^h_{22} \right)
\,,
\\
\tilde{\boldsymbol\varPi}_g^\pm &= - \tfrac{1}{2} {\boldsymbol\pi}_7 \pm  2 \sum_{x,y=1}^4 \nu^{xy}_{a,f} \, {\boldsymbol\beta}^f_{xy} =
- \tfrac{1}{2} {\boldsymbol\pi}_7 \pm \tfrac{1}{2} \,\left({\boldsymbol\beta}^f_{11}+ {\boldsymbol\beta}^f_{13}+ {\boldsymbol\beta}^f_{31}+ {\boldsymbol\beta}^f_{33}
\right)
\,,
\\
\tilde{\boldsymbol\varPi}_f^\pm &= -\tfrac{1}{2} {\boldsymbol\pi}_5 \pm 2 \sum_{x,y=1}^4 \nu^{xy}_{a,g} \, {\boldsymbol\beta}^g_{xy} =
- \tfrac{1}{2} {\boldsymbol\pi}_5 \pm \tfrac{1}{2} \, \left(
{\boldsymbol\beta}^g_{11}+ {\boldsymbol\beta}^g_{13}+ {\boldsymbol\beta}^g_{31}+ {\boldsymbol\beta}^g_{33}
\right)
\,.
\end{split}
\end{equation}
As anticipated the Chan-Paton labels of the ${\rm D}6_h$ branes are now real and carry non-vanishing charges with respect to all twisted RR forms
\begin{equation}
\tilde{\boldsymbol\varPi}_h = +\tfrac{1}{2} {\boldsymbol\pi}_3 +  2\sum_{x,y=1}^4 \left(
\nu^{xy}_{g} \, {\boldsymbol\beta}_{xy}^g + \nu^{xy}_{f} \,{\boldsymbol\beta}_{xy}^f + \mu^{xy}_{h} \, {\boldsymbol\alpha}^h_{xy} \right) \,.
\end{equation}
In particular, one has to distinguish four independent three-cycles
\begin{equation}
\begin{split}
\tilde{\boldsymbol\varPi}_h^1 &= +\tfrac{1}{2} {\boldsymbol\pi}_3 +\tfrac{1}{2} \, \left(
{\boldsymbol\beta}^g_{11} + {\boldsymbol\beta}^g_{13} + {\boldsymbol\beta}^g_{31} + {\boldsymbol\beta}^g_{33} +
{\boldsymbol\beta}^f_{11} + {\boldsymbol\beta}^f_{13} + {\boldsymbol\beta}^f_{31} + {\boldsymbol\beta}^f_{33} +
{\boldsymbol\alpha}^h_{11} + {\boldsymbol\alpha}^h_{13} + {\boldsymbol\alpha}^h_{31} + {\boldsymbol\alpha}^h_{33} \right)
\\
\tilde{\boldsymbol\varPi}_h^2 &= + \tfrac{1}{2} {\boldsymbol\pi}_3 +\tfrac{1}{2} \, \left(
{\boldsymbol\beta}^g_{11} + {\boldsymbol\beta}^g_{13} + {\boldsymbol\beta}^g_{31} + {\boldsymbol\beta}^g_{33} -
{\boldsymbol\beta}^f_{11} - {\boldsymbol\beta}^f_{13} - {\boldsymbol\beta}^f_{31} - {\boldsymbol\beta}^f_{33}  -
{\boldsymbol\alpha}^h_{11} - {\boldsymbol\alpha}^h_{13} - {\boldsymbol\alpha}^h_{31} - {\boldsymbol\alpha}^h_{33}
\right)
\\
\tilde{\boldsymbol\varPi}_h^3 &= +\tfrac{1}{2} {\boldsymbol\pi}_3 -\tfrac{1}{2} \, \left(
{\boldsymbol\beta}^g_{11} + {\boldsymbol\beta}^g_{13} + {\boldsymbol\beta}^g_{31} + {\boldsymbol\beta}^g_{33} -
{\boldsymbol\beta}^f_{11} - {\boldsymbol\beta}^f_{13} - {\boldsymbol\beta}^f_{31} - {\boldsymbol\beta}^f_{33} +
{\boldsymbol\alpha}^h_{11} + {\boldsymbol\alpha}^h_{13} + {\boldsymbol\alpha}^h_{31} + {\boldsymbol\alpha}^h_{33}
\right)
\\
\tilde{\boldsymbol\varPi}_h^4 &= +\tfrac{1}{2} {\boldsymbol\pi}_3 -\tfrac{1}{2} \, \left(
{\boldsymbol\beta}^g_{11} + {\boldsymbol\beta}^g_{13} + {\boldsymbol\beta}^g_{31} + {\boldsymbol\beta}^g_{33} +
{\boldsymbol\beta}^f_{11} + {\boldsymbol\beta}^f_{13} + {\boldsymbol\beta}^f_{31} + {\boldsymbol\beta}^f_{33} -
{\boldsymbol\alpha}^h_{11} - {\boldsymbol\alpha}^h_{13} - {\boldsymbol\alpha}^h_{31} - {\boldsymbol\alpha}^h_{33}
\right)
\end{split}
\end{equation}
all wrapping the same bulk cycle ${\boldsymbol \pi}_3$ but with different $\epsilon$ coefficients, and thus with different twisted RR charges. The resulting gauge group is ${\rm U} (8)^2 \times {\rm U}(8)^2 \times {\rm U}(8)^2 \times {\rm USp} (8)^4$ and the spectrum is not supersymmetric due to the presence of the ${\rm D}6_h$ antibranes, needed to cancel the charge of the exotic ${\rm O}6^h$ planes.

Actually, it was long believed that this $T^6 /\mathbb{Z}_2 \times \mathbb{Z}_2$ orientifold with discrete torsion could admit only non-supersymmetric vacua, due to the positive tension of the exotic O-planes. However, it was noticed in \cite{ms, Dudas:2005jx,Blumenhagen:2005tn} that this is not really correct, since rotated (or magnetised) branes
can actually ``mimic'' negative tension objects and thus allow for four-dimensional ${\mathcal N}=1$ supersymmetric vacua. The simplest instance involves four stacks of four D6 branes each, wrapping the bulk cycles associated to the wrapping numbers $(1,1;1,1;1,-1)$. The resulting gauge group is ${\rm U} (4)^4$ while (supersymmetric)  matter comes in
all possible (non-chiral) bi-fundamental representations and in chiral anti-symmetric representations for each factor \cite{Angelantonj:2009yj}.

Actually, these two vacua are not disconnected since the supersymmetric one can be obtained by fully recombining the (orthogonal) D6 branes of \cite{Angelantonj:1999ms}.
As in the six-dimensional configurations with Brane Supersymmetry Breaking, some care is needed to study the recombination process, since twisted tadpoles imply that different stacks of branes wrap the same bulk cycle, though they have different $\epsilon$ coefficients. In details, the recombination of the bulk cycles clearly yields
\begin{equation}
[{\boldsymbol\pi}_1]\cup [ {\boldsymbol\pi}_3]\cup [-  {\boldsymbol\pi}_5] \cup [ - {\boldsymbol\pi}_7] = \hat{\boldsymbol\varPi}_{\rm susy} =
{\boldsymbol\pi}_1 + {\boldsymbol\pi}_3 - {\boldsymbol\pi}_5 - {\boldsymbol\pi}_7\,,
\end{equation}
with
\begin{equation}
m_1 m_2 m_3 = 1\,, \quad
n_1 n_2 m_3 = 1\,, \quad
n_1 m_2 n_3 =-1\,,\quad
m_1 n_2 n_3 =-1\,,
\end{equation}
whose solution corresponds, for instance, to the wrapping numbers $(1,1;1,1;1,-1)$. The identification of the correct collapsed components requires some care, and finally one gets the following
three-cycles  associated to the four (invariant) supersymmetric stacks
\begin{equation}
\begin{split}
\tilde {\boldsymbol\varPi}^1_{\rm susy} &= \tfrac{1}{2} \, \hat{\boldsymbol\varPi}_{\rm susy}
+ \tfrac{1}{2} \sum_{x,y=1,4} \left[ i\,  {\boldsymbol\beta}_{xy}^g +i \,  {\boldsymbol\beta}_{xy}^f +  {\boldsymbol\alpha}_{xy}^h \right] \,,
\\
\tilde {\boldsymbol\varPi}^2_{\rm susy} &= \tfrac{1}{2} \, \hat{\boldsymbol\varPi}_{\rm susy}
+ \tfrac{1}{2} \sum_{x,y=1,4} \left[ i\,  {\boldsymbol\beta}_{xy}^g -i \,  {\boldsymbol\beta}_{xy}^f -  {\boldsymbol\alpha}_{xy}^h \right] \,,
\\
\tilde {\boldsymbol\varPi}^3_{\rm susy} &= \tfrac{1}{2} \, \hat{\boldsymbol\varPi}_{\rm susy}
- \tfrac{1}{2} \sum_{x,y=1,4} \left[ i\,  {\boldsymbol\beta}_{xy}^g -i \,  {\boldsymbol\beta}_{xy}^f +  {\boldsymbol\alpha}_{xy}^h \right] \,,
\\
\tilde {\boldsymbol\varPi}^4_{\rm susy} &= \tfrac{1}{2} \, \hat{\boldsymbol\varPi}_{\rm susy}
- \tfrac{1}{2} \sum_{x,y=1,4} \left[ i\,  {\boldsymbol\beta}_{xy}^g +i \,  {\boldsymbol\beta}_{xy}^f -  {\boldsymbol\alpha}_{xy}^h \right] \,.
\end{split}
\end{equation}

Another way to solve the tadpole conditions, while preserving ${\mathcal N}=1$ supersymmetry, is to introduce four different stacks of D6 branes with wrapping numbers and multiplicities
\begin{equation}
\begin{split}
{\rm D}6_1\ : &\quad N_1 = 2 \quad {\rm and\ wrapping\ numbers}\quad (1\, ,\,-1\, ;\,1\, ,\,-4\, ;\,2\,,\,1)\,,
\\
{\rm D}6_2\ : &\quad N_2 = 2 \quad {\rm and\ wrapping\ numbers}\quad (1\,,\,-1\,;\,1\,,\,-4\,;\,2\,,\,1)\,,
\\
{\rm D}6_3\ : &\quad N_3 = 1 \quad {\rm and\ wrapping\ numbers}\quad (1\,,\,-2\,;\,3\,,\,-2\,;\,2\,,\,1)\,,
\\
{\rm D}6_4\ : &\quad N_4 = 1 \quad {\rm and\ wrapping\ numbers}\quad (1\,,\,6\,;\,1\,,\,-2\,;\,2\,,\,-1)\,.
\end{split}
\end{equation}
yielding a ${\rm U}(2)^2 \times {\rm U} (1)^2$ gauge group  \cite{Camara:2010zm}.

It is interesting to see that this vacuum can not be obtained from the (minimal) non-supersymmetric one with orthogonal D6 branes, after suitable brane recombinations. In fact, the (invariant) bulk cycles wrapped by the supersymmetric D6 branes are
\begin{equation}
\begin{split}
\tilde{\boldsymbol\varPi}_{1,2} &=\tfrac{1}{2}\left( 2\, {\boldsymbol\pi}_1
+ 8 \, {\boldsymbol\pi}_3 - {\boldsymbol\pi}_5 - 4\, {\boldsymbol\pi}_7 \, \right) -
\tfrac{1}{2}\left(
i \, \sum_{\genfrac{}{}{0pt}{}{x=1,2}{y=1,3}} {\boldsymbol\beta}_{xy}^g
\pm 4i \, \sum_{\genfrac{}{}{0pt}{}{x=1,4}{y=1,3}} {\boldsymbol\beta}_{xy}^f 
\mp 2 \,  \sum_{\genfrac{}{}{0pt}{}{x=1,4}{y=1,2}} {\boldsymbol\alpha}_{xy}^h\right),
\\
\tilde{\boldsymbol\varPi}_{3} &=\tfrac{1}{2}\left( 6\, {\boldsymbol\pi}_1 +
8\, {\boldsymbol\pi}_3 - 6\, {\boldsymbol\pi}_5 - 2\, {\boldsymbol\pi}_7\, \right) -
\tfrac{1}{2}\left(
i \, \sum_{\genfrac{}{}{0pt}{}{x=1,2}{y=1,3}} {\boldsymbol\beta}_{xy}^g 
+ 2i \, \sum_{\genfrac{}{}{0pt}{}{x=1,2}{y=1,3}} {\boldsymbol\beta}_{xy}^f 
- 2 \, \sum_{\genfrac{}{}{0pt}{}{x=1,2}{y=1,2}} {\boldsymbol\alpha}_{xy}^h\right),
\\
\tilde{\boldsymbol\varPi}_{4} &=\tfrac{1}{2}\left( 2\, {\boldsymbol\pi}_1
- 24 \, {\boldsymbol\pi}_3 -6\, {\boldsymbol\pi}_5 + 2\, {\boldsymbol\pi}_7\, \right) +
\tfrac{1}{2}\left(
6i \, \sum_{\genfrac{}{}{0pt}{}{x=1,2}{y=1,2}} {\boldsymbol\beta}_{xy}^g
- 2i \, \sum_{\genfrac{}{}{0pt}{}{x=1,3}{y=1,2}} {\boldsymbol\beta}_{xy}^f 
+ 2 \, \sum_{\genfrac{}{}{0pt}{}{x=1,3}{y=1,2}} {\boldsymbol\alpha}_{xy}^h\right),
\end{split}
\end{equation}
 and thus there are not enough ${\rm D}6$ branes to reproduce the previous cycles. A possible solution would be to add 24 brane-antibrane pairs of type ${\rm D}6_h$ and 2 brane-antibrane pairs of type ${\rm D}6_g$ in the original non-supersymmetric vacuum.


\section{The low-energy Higgs mechanism as brane recombination}
\label{higgsing}

Brane recombination is a purely stringy mechanism where some branes dilute into others and get transmuted in a (new) constant magnetic field background. In a sense, the process of brane recombination is the inverse of the small instanton transition of \cite{Witten:1995gx}. Although this interpretation is at first hand correct, as we shall argue in  section \ref{fieldtheory}, a deeper analysis suggests that in the case of brane recombination the process is subtler and involves non-perturbative effects associated to Abelian Born-Infeld configurations. 

As observed by Witten \cite{Witten:1995gx}, in the small instanton transition the hypermultiplet moduli space develops
a conical singularity at a finite distance.  In correspondence, when the instanton shrinks to zero size, the
target space exhibits an infinite tube where the dilaton grows, breaking the CFT perturbative description, signalling the presence of a solitonic brane.  The net result on the dynamics is just an enhancement of the gauge symmetry, namely a conventional (inverse) Higgs effect, where the size of the instanton is related to the {\em vev}'s
of scalar fields originating from open strings stretched between ${\rm D}p$ and ${\rm D} (p+4)$ branes.
Following similar/inverse reasonings, it is quite natural to associate the process of brane recombination to a conventional field theoretic  Higgs mechanism. These two phenomena share many common features, as for instance, the reduction of the rank of the gauge group and of the number of chiral fermions. However, despite these similarities, a proof of the equivalence is far from being under control since the process of brane recombination not only involves the condensation of the light (or tachyonic) Higgs fields, but the whole tower of string excitations take a non-trivial part in the process, that again cannot be described in CFT and requires some additional non-perturbative analysis.

Although, the description of brane recombination as a low-energy Higgs effect does not require the presence of some amount of supersymmetry, by analysing a series of simple six-dimensional examples, we shall show that whenever supersymmetry is absent, the presence of some (non-renormalisable) higher-order couplings in the low-energy Lagrangian has to be postulated in order to lift the extra states and to correctly recover the charged light string spectrum. More examples will be discussed in appendix \ref{app:higgs}.

\subsection{Higgs mechanism in six-dimensional supersymmetric vacua}

\subsubsection{A model with simple recombination}

The simplest example of a Higgs mechanism is the ${\rm U} (16) \times {\rm U} (16) \to {\rm U} (16)$ breaking in six-dimensional supersymmetric vacua, whose brane recombination process was already studied in section \ref{sec:brsusy}. The original spectrum of \cite{Bianchi:1990yu, Gimon:1996rq} comprises hypermultiplets in the anti-symmetric representations of the gauge group and in the bi-fundamental $(16,16)$. If  the bi-fundamental hypermultiplet gets a non-vanishing {\em vev} the gauge bosons associated to the anti-diagonal combination of the two ${\rm U}(16)$ become massive, and thus only a diagonal ${\rm U}(16)$ survives at low energies together with four hypermultiplets in the 120 representation, exactly the same spectrum pertaining to a single stack of (recombined) branes wrapping the diagonal of the two $T^2$'s.

\subsubsection{A model with multiple recombinations}

A more involved Higgs mechanism is required to obtain the light states of the ${\rm U} (12) \times {\rm U} (4)$ model of \cite{Angelantonj:2000hi}. This vacuum was reproduced in section \ref{sec:brsusy} via the recombination of four horizontal D7 branes and sixteen vertical ${\rm D} 7'$ ones. Actually, as a matter of facts, any time one partially recombines the original branes the process cannot be described by simply assigning non-trivial {\em vev}'s to bi-fundamental scalars. In fact, in the case at hand, a  {\em vev} for the $(16,16)$ hypermultiplets that would naively break
${\rm U} (16) \times {\rm U} (16)$ to ${\rm U} (12) \times {\rm U} (4)$ can always be rotated within the colour indices to reproduce the vacuum with the diagonal ${\rm U} (16)$. As a result, one needs to assign non-vanishing {\em vev}'s to diverse hypermultiplets transforming in different representations of the gauge group.

To reproduce the ${\rm U} (12) \times {\rm U} (4)$ model via a field theoretic Higgs mechanism one then needs to give suitable discrete and continuous {\em vev}'s to all hypermultiplets. In particular, one can describe the whole process in two steps. First, break
\begin{equation}
{\rm U} (16) \times {\rm U} (16) \to {\rm U} (12) \times {\rm U} (4) \times {\rm U} (4)^4
\end{equation}
by turning on discrete Wilson lines and/or brane displacements in such a way to split the sixteen D7 branes in two groups of twelve and four sitting at different fixed points, and the sixteen ${\rm D} 7 '$ in four groups of four branes each sitting at different fixed points. This breaking induces the following decomposition
\begin{equation}
\begin{split}
(120,1) \to &\  (66,1;1,1,1,1)+(1,6;1,1,1,1)+{\bf(12,4;1,1,1,1)}\,,
\\
(1,120) \to &\  (1,1;6,1,1,1)+(1,1;1,6,1,1)+(1,1;1,1,6,1)+(1,1;1,1,1,6)
\\
& + {\bf (1,1;4,4,1,1) + permutations}\,,
\\
(16,\overline{16}) \to &\  (12,1;\bar 4,1,1,1) + {\rm permutations}
\\
&+(1,4;\bar 4,1,1,1) + {\rm permutations} \,,
\\
(256,1) \to &\  (144,1;1,1,1,1) + (1,16;1,1,1,1) + {\bf (12,\bar 4;1,1,1,1)}\,,
\\
(1,256) \to &\  (1,1;16,1,1,1) +{\rm permutations}
\\
& + {\bf (1,1;4,\bar 4,1,1) + permutations}\,,
\end{split}
\end{equation}
where we have highlighted the states that take part in the six-dimensional ${\mathcal N}=(1,0)$ {\em discrete} Higgs mechanism, and decouple from the light spectrum. The second step then consists in assigning {\em vev}'s to the hypermultiplets in the representations
\begin{equation}
(1,4;\bar 4 , 1 , 1 ,1) + (1,4;1,\bar 4 , 1 ,1) + (1,4;1,1,\bar 4  ,1) + (1,4;1,1,1,\bar 4 ) \,.
\end{equation}
As a result, the ${\rm U} (4)^5$ is broken to the diagonal ${\rm U} (4)$ while the ${\rm U} (12)$ gauge group stays massless, according to the decomposition
\begin{equation}
\begin{split}
& (66,1;1,1,1,1) \to  (66,1)\,,
\\
& (1,6;1,1,1,1) + (1,1;6,1,1,1)+{\rm permutations} \to 5\times (1,6)\,,
\\
&(12,1;\bar 4,1,1,1)+{\rm permutations} \to 4\times (12,\bar 4)\,,
\\
&(1,4;\bar 4,1,1,1)+{\rm permutations} \to 4\times {\bf (1,16)}\,,
\\
&(144,1;1,1,1,1) \to (144,1)\,,
\\
&(1,16;1,1,1,1)+(1,1;16,1,1,1)+{\rm permutations}\to (1,16)+4\times {\bf (1,16)}\,,
\end{split}
\end{equation}
where again we have highlighted the states that become massive after the Higgsing. The massless spectrum thus obtained precisely reproduces the one of \cite{Angelantonj:2000hi} and of section \ref{sec:brsusy}.

Although, we have achieved the ${\rm U} (12) \times {\rm U} (4)$ vacuum via a two-steps low-energy process as corresponding in string theory to a rank-preserving breaking via (discrete) brane displacements and/or Wilson lines, followed by a recombination of multiple stacks, it seems plausible that the final result could be obtained in one stroke by turning on suitable discrete and continuous {\it vev}'s for the physical and un-physical scalars, as suggested by the simpler brane recombination process described in section \ref{sec:brsusy}.

\subsection{Higgs mechanism in six-dimensional Brane Supersymmetry Breaking vacua}

\subsubsection{Recombination with fractional branes}

We want to reproduce here the non-supersymmetric vacuum with all branes recombined to yield two independent stacks of D7 branes wrapping the same bulk cycle (corresponding to the diagonal of both $T^2$'s) but with opposite twisted charges. The final gauge group is ${\rm U} (8) \times {\rm U} (8)$ and the light spectrum is summarised in table \ref{u8u8}. In the low-energy effective field theory this vacuum can be obtained via a Higgs mechanism of the ${\rm SO} (16)^2 \times {\rm USp} (16)^2$ model of \cite{bsb1b}, after the scalars in the bi-fundamental representations
\begin{equation}
(16,1;1,16) + (1,16;16,1) \label{higgs1}
\end{equation}
acquire non-trivial {\em vev}'s. This scalars however, not only give masses to the gauge bosons associated to the broken generators, but must also contribute to tree-level masses for extra scalars and fermions, in oder to reproduce the light spectrum of table \ref{u8u8}. Let us analyse this phenomenon in some detail by decomposing the various
${\rm SO} (16)^2 \times {\rm USp} (16)^2$ representations of the Brane Supersymmetry Breaking model \cite{bsb1b} in terms of ${\rm U} (8) \times {\rm U} (8)$ ones, where each ${\rm U} (8)$ factor correspond to the diagonal breaking of ${\rm SO} (16) \times {\rm USp} (16)$. For the bi-fundamental scalars one finds
\begin{equation}
\begin{split}
(16,1;1,16) &\to {\bf (28+\overline{28},1)+(36+\overline{36},1)+ (64,1)}+(64,1)\,,
\\
(1,16;16,1) &\to {\bf (1, 28+\overline{28})+(1,36+\overline{36})+ (1,64)}+(1,64)\,,
\end{split}
\label{pippo}
\end{equation}
while for the adjoint vector bosons
\begin{equation}
\begin{split}
(120,1;1,1)+(1,120;1,1) &\to {\bf (28+\overline{28},1)+(64,1)+(1,28+\overline{28})}+(1,64)\,,
\\
(1,1;136,1)+(1,1;1,136) &\to {\bf (36+\overline{36},1)+(1,64)+(1,36+\overline{36})}+(64,1)\,,
\end{split}
\end{equation}
where again the highlighted states become massive\footnote{Here we have assumed that in the Brane Supersymmetry Breaking vacua the Higgs mechanism involves a doublet of scalars.}. The remaining massless spectrum would thus include four scalars and one right-handed fermion in the representations
\begin{equation}
(16,16;1,1)+(1,1;16,16)\to 2\times \left[ (8,8)+(\bar 8 , \bar 8)+(8,\bar 8)+(\bar 8, 8) \right]\,,
\end{equation}
one left-handed fermion in the representations
\begin{equation}
(120,1;1,1)+{\rm permutations} \to 2\times \left[ (28+\overline{28},1)+(64,1)+(1,64)+(1+28+\overline{28})\right]\,,
\end{equation}
and half left-handed fermion in the representations
\begin{equation}
(16,1;16,1)+(1,16;,1,16)\to 2\times \left[ (8,8)+(\bar 8 , \bar 8)+(8,\bar 8)+(\bar 8, 8) \right]\,.
\end{equation}
These are too many states if one wants to match the spectrum in table \ref{u8u8}. Therefore, for the Higgs mechanism to offer a field theory description of brane recombination, the Higgs fields ought to couple to un-matched matter fields to make them massive. A simple glimpse at the various quantum numbers, shows that, in the original model, the following allowed third-order ${\it fermion} \otimes {\it fermion} \otimes {\it scalar}$ Yukawa couplings
\begin{equation}
\begin{split}
& (16,16;1,1)\otimes (16,1;16,1)\otimes (1,16;16,1)\,,
\\
& (16,16;1,1)\otimes (1,16 ;1,16)\otimes (16,1;1,16)\,,
\\
& (1,1;16,16)\otimes (16,1;16,1)\otimes (16,1;1,16)\,,
\\
& (1,1;16,16)\otimes (1,16 ;1,16)\otimes (1,16;16,1)\,,
\end{split}
\end{equation}
would suffice to give mass to the extra fermions, while for the extra bosons one should invoke fourth-order scalar couplings of the form
\begin{equation}
(16,16;1,1)\otimes (16,1;1,16)\otimes (1,1;16,16)\otimes (1,16;16,1)\,,
\end{equation}
consistent with the fact the scalar masses $m^2_\phi \sim v^2$  should be proportional to the {\em vev}-squared.
Finally, there are two copies of Weyl fermions in the adjoint representation $(64,1)+(1,64)$, and one combination of them should get a mass in order
to match the spectrum of the ${\rm U} (8) \times {\rm U} (8)$ model. This is indeed possible through couplings to the adjoint scalars (\ref{pippo})
which trigger the Higgs mechanism.
\bigskip

\subsubsection{Recombination in the bulk}

Also in vacua with Brane Supersymmetry Breaking one has the possibility to move branes in the bulk. Of course, charge conservation implies that this deformation is consistent only if one moves pairs of fractional branes  with opposite twisted charges, so to preserve the twisted tadpole conditions. Then the compound of fractional branes recombines into bulk ones with reduced gauge group. Clearly, this operation is allowed both for orthogonal branes and for branes at arbitrary angles. In the former case, one obtains the vacuum described in appendix \ref{wl}, whose excitations are listed in table \ref{bsbbulk}. In the latter case, one has to properly deform the amplitudes as explained in appendix \ref{wl} to obtain the massless excitations listed in table \ref{bsbbulkrotated}.

\begin{table}
\centering
\begin{tabular}{cccc}
  \toprule
  Multiplicity & Representation & Field/Multiplet & Chirality
  \\
 \toprule
  $1$ & $64$ & Gauge Multiplet & Left-handed
  \\
  $1$ & $64$ & Hypermultiplet & Right-handed
  \\
  \midrule
  $8$ & $28$ & Weyl Fermion & Left-handed
  \\
  $16$ & $28$ & Scalars & -
  \\
  $16$ & $36$ & Scalars & -
  \\
  \bottomrule
\end{tabular}
\caption{Massless spectrum of the ${\rm U}(8)$ model containing one stack of diagonal D7 branes in the bulk.}
\label{u8}
\end{table}

As an example, a possible solution of tadpole conditions introduces a single stack of rotated branes wrapping the shifted diagonal of both $T^2$'s. According to table \ref{bsbbulkrotated}, the light excitations comprise a full vector multiplet in the adjoint of the ${\rm U} (8)$ gauge group and charged matter as in table \ref{u8}.  It is interesting to observe that, as expected, this vacuum can be obtained from the ${\rm U} (8) \times {\rm U} (8)$ one with fractional branes described in section \ref{susec:bsb}, by giving a non-trivial {\em vev} to the hypermultiplet in the $(8,\bar 8)$ representation. A super-Higgs mechanism seems to be at work also in this non-supersymmetric vacuum\footnote{in the sense, that the gauge bosons eat four scalars to become massive.}, and the resulting spectrum matches indeed the one of table \ref{u8}, since under the breaking ${\rm U} (8) \times {\rm U} (8) \to {\rm U} (8)$
\begin{equation}
\begin{split}
& 4\times [ (28,1)+(1,28) ] \to 8\times 28 \quad ({\rm Weyl\ fermions}) \,,
\\
& 16\times (8,8) \to 16\times [28 + 36 ]\quad ({\rm scalars})\,.
\end{split}
\end{equation}

It is interesting to check, whether the same vacuum could be obtained by recombining the orthogonal bulk branes in the ${\rm SO} (16) \times {\rm USp} (16)$ model. After all, the mental process one has followed to obtained the ${\rm U} (8)$ vacuum is to start from the Brane Supersymmetry Breaking model of \cite{bsb1b}, recombine all branes into the ${\rm U} (8) \times {\rm U} (8) $ vacuum and then move the branes in the bulk. It is conceivable, however, to follow a different path, where one first moves the orthogonal branes in the bulk to break the gauge group to ${\rm SO} (16) \times {\rm USp} (16)$ and then recombine them in the diagonal branes. Although in string theory these two ways to proceed yield the same result, it is not the case if one tries to describe the whole process in field theory by use of the Higgs mechanism. In fact, the breaking of the ${\rm SO} (16) \times {\rm USp} (16)$ gauge group in the diagonal ${\rm U} (8) $ can be achieved in field theory by assigning a non-vanishing {\em vev} to the hypermultiplet in the $(16,16)$ representation. By decomposing the various representations of table \ref{bsbbulk} in terms of those of the final gauge group one obtains\footnote{As usual, we have highlighted the states that become massive after the breaking of the gauge group.}
\begin{equation}
(16,16) \to {\bf 28+\overline{28}+36+\overline{36}+64} +64
\end{equation}
for the hypermultiplets including four scalars and one left-handed spinor,
\begin{equation}
\begin{split}
(120,1) &\to {\bf 28 +\overline{28}+64}\,,
\\
(1,136)&\to {\bf 36+\overline{36}} + 64
\end{split}
\end{equation}
for the gauge bosons, and
\begin{equation}
\begin{split}
(120,1) &\to 28+\overline{28}+64 \quad {\rm Weyl\ fermions\ (C)} \,,
\\
(1,120) & \to 28+\overline{28}+64 \quad {\rm Weyl\ fermions\ (C)} \,,
\\
(136,1) & \to 36+\overline{36} +64 \quad {\rm Hypermultiplets\ (S)} \,,
\\
4\times(1,120) & \to  4\times[28+\overline{28}+64]\quad {\rm Scalars}\,,
\\
(1,136) & \to 36+\overline{36}+64\quad {\rm Weyl\ fermion\ (S)} \,.
\end{split}
\end{equation}
Surprisingly, from the recombination of the bulk branes one gets at most six left-handed ({\it i.e.} $C$-type) fermions in the anti-symmetric 28 representation, while to reproduce table \ref{u8} one needs eight such fermions. Does it mean that the two paths are not {\it commuting}?

One possible explanation of this mismatch could be that, in vacua with Brane Supersymmetry Breaking, displacements and Wilson lines associated to orthogonal branes are not flat directions. At one loop, an effective potential is generated that tends to attract the bulk branes towards the orientifold planes and orbifold fixed points, as we show in the next section. On the other hand, Brane Supersymmetry Breaking vacua with intersecting branes do not induce any quantum potential, at least at one-loop order. As a result, while the transition ${\rm U} (8) \times {\rm U} (8) \to {\rm U} (8)$ is in principle allowed, the ${\rm SO} (16) \times {\rm USp} (16)$ is energetically disfavoured, and thus the branes are attracted towards the orientifold planes to yield the original vacuum \cite{bsb1b} rather than to recombine in the bulk.


\subsection{Remarks on the Higgs mechanism in the $T^6/\mathbb{Z}_2\times\mathbb{Z}_2$ orientifold}

What about the equivalence of brane recombination and the Higgs mechanism for four-dimensional vacua? Although the equivalence has been already used in supersymmetric constructions \cite{Cremades:2002cs}, it is far from being obvious and universal. As we saw in the previous pages, already within six-dimensional vacua it was non-trivial to show that a Higgs mechanism in the effective field theory could describe the recombination of branes when supersymmetry is not present. It is not so surprising that additional subtleties appear in the $T^6/\mathbb{Z}_2\times \mathbb{Z}_2$ orientifold with discrete torsion.

To start with, it is well established that the original chiral non-supersymmetric model with  ${\rm U} (8)^2\times {\rm U} (8)^2 \times {\rm U} (8)^2 \times {\rm USp} (8)^4$ gauge group can be connected to chiral supersymmetric vacua, and this transition can be correctly explained in terms of brane recombinations, as we have shown in section \ref{subsec:4d}. However, it is doubtful that one can capture this transitions by a simple standard Higgs mechanism in the effective field theory. One argument would be that the supersymmetry condition for intersecting branes
\begin{equation}
\theta_1 + \theta_2 + \theta_3 = 0
\end{equation}
becomes, in the T-dual picture, a non-linear relation among the magnetic fluxes
\begin{equation}
H_1  +  H_2  +  H_3  =  (\alpha')^2 \, H_1 \, H_2 \, H_3 \,, \label{susy4d}
\end{equation}
and depends non-trivially on $\alpha'$, at least for dimensional reasons. This has to be contrasted to six dimensions, where the supersymmetry condition $\theta_1 \pm \theta_2 =0$ translates into a (anti-)self-duality condition $H_1 = \pm H_2$, that does not involve $\alpha '$.
As a result, while in six-dimension the low-energy analysis is expected to be fully captured by the Yang-Mills Action, as we shall see explicitly in section \ref{fieldtheory}, in four dimensions eq. (\ref{susy4d}) can only  be captured by a full Dirac-Born-Infeld Action, and a complete  superfield analysis is  not yet available\footnote{See \cite{tsyetlin} for a review on the subject. }. The original argument \cite{Witten:1995gx} relating {\em vev}'s of $95$ scalars to Yang-Mills instanton size is also certainly modified, in particular since in this
case, not all D5 branes can be interpreted as gauge instantons on the parent D9 branes. Indeed, whereas in the six-dimensional models the gauge group associated to the D5 branes matches the dual gauge group for the theory on the D9 brane (in both the supersymmetric and non-supersymmetric cases), this is no longer the case for the $T^6/\mathbb{Z}_2\times \mathbb{Z}_2$ with discrete torsion due to the fact that $\overline{{\rm D}5}_h$ branes support a symplectic gauge group, whereas the D9 gauge groups are unitary. Dissolving all $\overline{{\rm D}5}_h$ is certainly essential in obtaining a supersymmetric model, but the corresponding phenomenon is not
a usual small instanton transition. Notice also that the magnetised branes in these models have three instanton-like charges, one of them having
negative mass and charge. Again, this cannot correspond to a simple field-theory instanton.

As a result, given the previous arguments, it is reasonable to assume that the transition towards a supersymmetric vacuum might correspond  to a new stringy non-perturbative configuration, which is not fully captured by a Yang-Mills Higgs mechanism. Brane recombination, however,  is perfectly applicable and gives important hints on the non-perturbative transition between the non-supersymmetric and the supersymmetric vacua.


\section{Stability in six dimensions}
\label{sec:stability}

A natural question that arises when studying string models with broken supersymmetry is that of stability. In general, loop corrections to physical quantities are not protected in non-supersymmetric vacua, and most of the times can wildly destabilise the tree-level vacuum. We shall address here this problem within the context of six-dimensional vacua with Brane Supersymmetry Breaking.  Although tachyon free at the tree level, one-loop corrections can in principle yield non-vanishing mass-terms for the charged scalars and thus it is crucial to determine their sign. We shall do this by computing the induced potential for the ${\rm D}7$-${\rm D 7}$ and $\overline{{\rm D}7}'$-$\overline{{\rm D 7}}'$ open strings and the one-loop two-point function for the ${\rm D}7$-$\overline{{\rm D}7}'$ scalars.

\subsection{The one-loop effective potential}
\label{energetics}

To compute the one-loop effective potential for the open-string states we employ the back\-ground-field method. Hence, we turn on non-trivial, constant, {\it vev}'s for the brane positions and Wilson lines $a$ and $b$, for the D7 and $\overline{{\rm D} 7}'$ branes, and compute the one-loop free energy $\varLambda (a,b)$. In principle, $\varLambda (a,b)$ receives contributions both from the annulus and M\"obius strip amplitudes. However, for the Brane Supersymmetry Breaking model the annulus diagram is identically vanishing and only ${\mathcal M}$ actually contributes. Moreover, from the fact that D7-O7 interactions are effectively supersymmetric, as can be seen by the explicit expression of the amplitudes in  appendix
\ref{wl}, it is evident that the vacuum energy does not depend on the position and Wilson lines of the D7 branes and thus these charged scalars stay massless also at the one-loop level
\begin{equation}
m_{77}^2=0 \,.
\end{equation}
On the contrary, an effective potential for the positions and Wilson lines of the $\overline{{\rm D}7}'$ branes, collectively denoted by $b_i$, is generated.
Assuming, for simplicity, that all anti-branes have been displaced the same amount and carry the same Wilson lines, the potential reads
\begin{equation}
\varLambda (b) \equiv - {\mathcal M} = - \frac{N_{\bar 7 '}}{4} \int_0^\infty \frac{dt}{t^4} \, \frac{ \hat V_4\, \hat O_4 + \hat O_4 \, \hat V_4 + \hat S_4 \, \hat S_4 + \hat C_4 \, \hat C_4}{\hat\eta^8} \, \left( \varGamma_{+2b} + \varGamma_{-2b} \right) \,.
\end{equation}
Here, $N_{\bar 7 '}$ counts the total number of antibranes, the {\em hatted} characters include suitable phases as described in \cite{Angelantonj:2002ct, Dudas:2000bn}, 
while $\varGamma_{\pm 2b}$ encodes the contribution of the Kaluza-Klein and winding zero-modes
\begin{equation}
\varGamma_{\pm 2b} = \sum_{m\in\mathbb{Z}^4} q^{(m\pm 2b)^T\, A^{-1} \, (m\pm 2b)}\,,
\end{equation}
with
\begin{equation}
A = {\rm diag} \left( \frac{\alpha'}{R_1^2}\,,\, \frac{R_2^2}{\alpha'}\,,\,\frac{\alpha'}{R_3^2}\,,\, \frac{R_4^2}{\alpha'} \right)\,,
\end{equation}
Here $R_i$ denotes the radius of the $i$-th compact direction, while $b= (b_1 , b_2 , b_3 , b_4)$ encodes the information about brane positions along the first and third compact coordinate and Wilson lines along the second and fourth compact directions.
In order to study the $b$ dependence of the potential it is useful to perform the $P=T S T^2 S$ transformation
\begin{equation}
\frac{it}{2} + \frac{1}{2} \to \frac{i}{2t}+\frac{1}{2} \equiv i\ell +\frac{1}{2}
\end{equation}
so that
\begin{equation}
\varLambda (b) = - {N_{\bar 7 '}}{2}\, \frac{R_2\, R_4}{R_1\, R_3} \, \int_0^\infty d\ell\, \frac{ \hat V_4 \hat O_4 + \hat O_4 \hat V_4 + \hat S_4  \hat S_4 + \hat C_4  \hat C_4}{\hat\eta^8} \, \sum_{m\in\mathbb{Z}^4} \, \cos (4 \pi m^T\,  b) \, e^{-2\pi \ell  m^T  A  m} \,,
\label{quantumpot1}
\end{equation}
and, after some algebra,
\begin{equation}
\begin{split}
\varLambda (b) =& - \frac{N_{\bar 7 '}}{2\pi} \, \frac{R_2 \, R_4 }{R_1\, R_3}\, d_0 \left( \int_0^\infty d\ell +
{\sum_{m}}^\prime \ \frac{\cos (4\pi m^T \, b )}{ m^T \, A\, m}
 \right)
\\
&-\frac{N_{\bar 7 '}}{4} \sum_{M=1}^\infty (-1)^M \, d_M
\sum_{\eta=\pm 1}
\sum_{m\in\mathbb{Z}^4} C_{M,m} (\eta\, b ) \, K_1 (2 \pi D_{M,m} (\eta \, b))
\,,
\end{split}
\label{quantumpot}
\end{equation}
where
\begin{equation}
\begin{split}
C_{M,m} (\eta\, b) &= \sqrt{\frac{M}{(m+2\eta b )^T\, A^{-1} \, (m+2 \eta b)}}\,,
\\
D_{M,m} (\eta\, b ) &= \sqrt{M\, (m+2\eta b)^T \, A^{-1} \, (m+2\eta b)}\,,
\end{split}
\end{equation}
$K_1 (x)$ is the modified Bessel function,
and we have used the fact that $\hat V_4 \, \hat O_4 + \hat O_4 \, \hat V_4$ and $\hat S_4 \, \hat S_4 + \hat C_4 \, \hat C_4$ admit the same $q$-power expansion
\begin{equation}
\frac{\hat S_4 \, \hat S_4 + \hat C_4 \, \hat C_4}{\hat\eta^8} = \sum_{M=0}^\infty (-1)^M\, d_M\, q^M\,.
\end{equation}

The potential $\varLambda (b)$ is highly ill defined, as can be seen from eqs. (\ref{quantumpot1})  and (\ref{quantumpot}). Not only it receives a divergent contribution from the tadpole (the first term in eq. (\ref{quantumpot}) proportional to $\int_0^\infty d\ell$), but it is not under control also for $b_i =0,\frac{1}{2}$, since for these choices
\begin{equation}
\varLambda (b_i =0,\tfrac{1}{2}) =  - \frac{N_{\bar 7 '}}{2\pi} \, \frac{R_2 \, R_4 }{R_1\, R_3}\,  \left(d_0 \int_0^\infty d\ell + {\sum_{M,m}}^\prime \frac{(-1)^M \, d_M}{\pi \, (M+ m^T A m)} \right)\,,
\end{equation}
and the convergence of the sum over the string states is not {\it a priory} guaranteed since the degeneracies $d_M$ grow exponentially with $\sqrt{M}$.

In the field theory limit, where $R_{1,3} \gg \sqrt{\alpha '}$ and $R_{2 ,4} \ll \sqrt{\alpha '}$, the second line in eq. (\ref{quantumpot}) is clearly exponentially suppressed and thus, for the choice $(b_1,b_2,b_3,b_4) = (b,0,0,0)$,
\begin{equation}
\varLambda (b) \simeq {\rm tadpole} - \frac{N_{\bar 7 '}}{2\pi} \, \frac{R_2 \, R_4}{R_1 \, R_3} \, d_0\, {\sum_{m}}^\prime \, \frac{\cos (4\pi m_1 b)}{m^{T}\, A \, m}
\end{equation}
This expression has an equivalent interpretation as the interaction of the branes with the orientifold planes and thus it is not surprising that it is proportional to the Green's function in four (compact) dimensions, that blows up for $b=0$ or $b=\frac{1}{2}$ when the branes sit on top of the orientifold planes, and thus the field theory description is not any more reliable. One may then conclude that the points $b=0,\frac{1}{2}$ are (the only) minima of $\varLambda (b)$, though an analytic proof is not available.

\begin{figure}[h!]
\begin{center}
  \includegraphics[width=6cm]{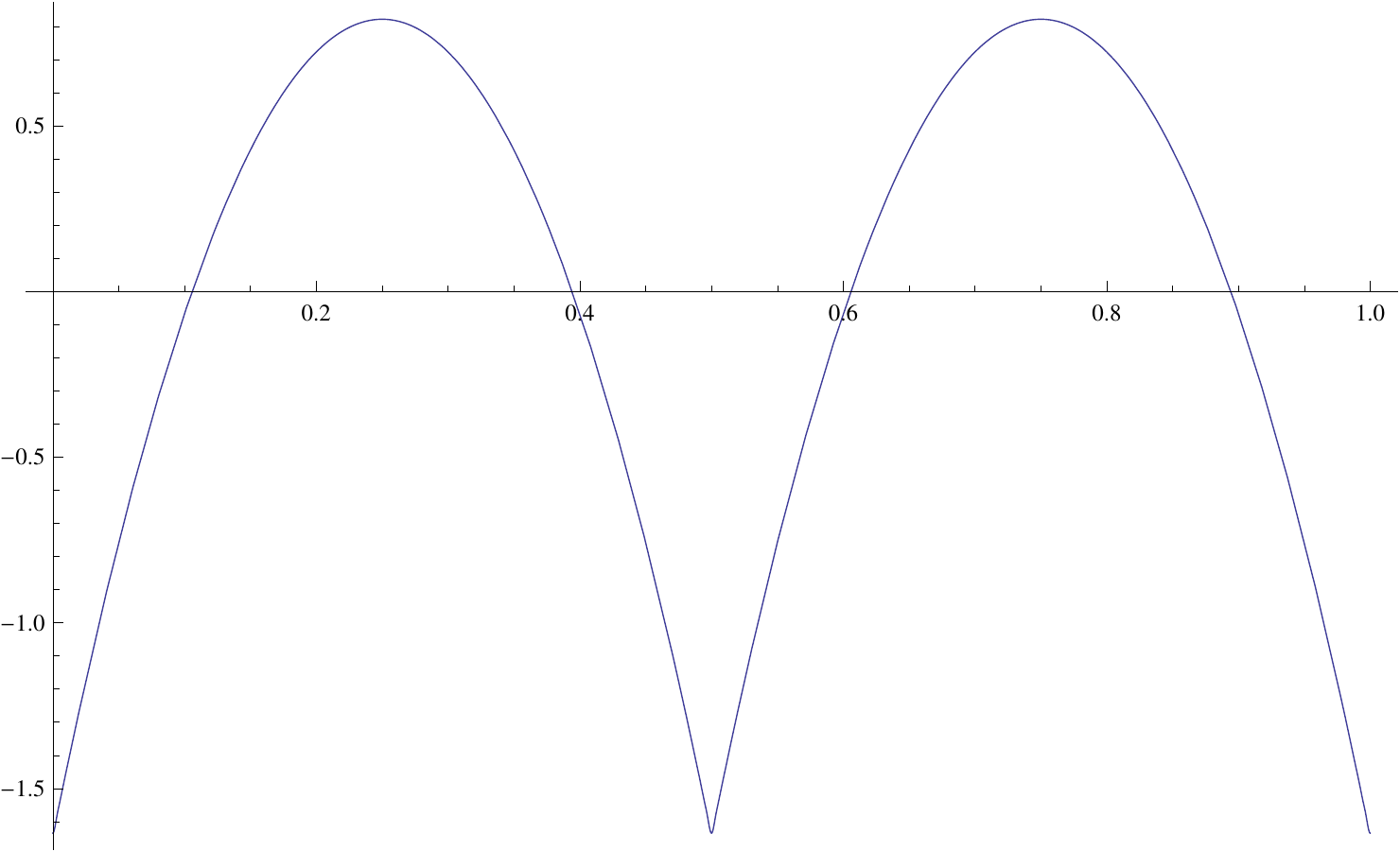}\\
  \caption{The effective potential  $\varLambda (b)$  in the region $R_1 \gg \sqrt{\alpha '}$ and  $R_{2,3,4}\sim \sqrt{\alpha '}$. }\label{plot}
  \end{center}
\end{figure}

In the extreme situation where only $R_1$ is in the field theory regime, and $R_{2,3,4}\sim \sqrt{\alpha '}$, the result simplifies further
\begin{equation}
\varLambda (b) \simeq {\rm tadpole} - \frac{N_{\bar 7 '}}{2\pi} \, \frac{R_2 \, R_4}{R_1 \, R_3} \, d_0\, \sum_{m_1 \not=0}  \frac{R_1^2\, \cos (4\pi m_1 b)}{\alpha'\, \, m_1^2}\,,
\end{equation}
and the finite contribution is plotted in fig. \ref{plot}, where the presence of the minima is manifest. Given these results, the field theory intuition and the fact that the second line in eq. (\ref{quantumpot}) is sizeable for $b=0,\frac{1}{2}$, we are tempted to conclude that the minima survive also for generic values of $R_i$, and the $\overline{D7} '$ are attracted towards the fixed points and there get a positive one-loop mass, $m^2_{\bar 7 '\bar 7 '} \simeq \frac{\partial^2 \varLambda }{\partial b_i \, \partial b_j} >0$.

\bigskip

\subsection{Masses for the $7\, \bar 7'$ scalars}
\label{95}

To estimate the mass of the $7\, \bar 7'$ scalars $\chi$ one cannot employ the background-field method, since open strings with mixed Neumann-Dirichlet boundary conditions do not couple to the world-sheet sigma model, at least in a simple way. Therefore, one should compute one-loop two-point functions on the annulus and M\"obius strip amplitudes with the insertion at the same boundary of the (twisted) vertex operators for the $\chi$ fields. Actually, also in this case the annulus diagram is expected to yield a vanishing contribution because the oriented open-string spectrum is still supersymmetric.

\begin{figure}[h!]
\begin{center}
  \includegraphics[width=12cm]{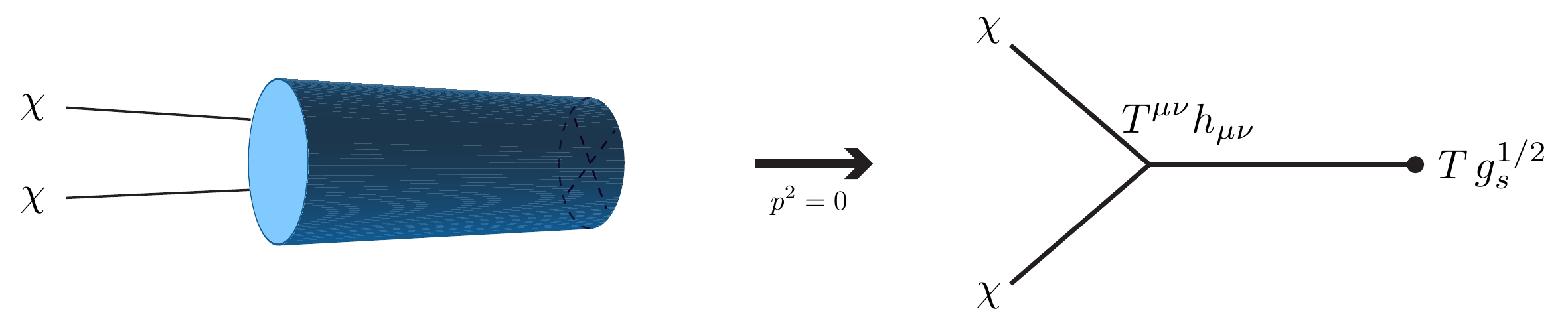}\\
  \caption{The relevant string theory diagram for estimating the one-loop masses of the $7\bar 7 '$ scalars should give in the limit $p^2=0$ the same result as the field theory diagram where only the massless states from the gravitational sector (namely the graviton) give a non-zero contribution. }\label{graph}
  \end{center}
\end{figure}

Although a full fledged string theory calculation of the amplitude can be performed, we shall content ourselves with a field theory analysis, since in the closed-string channel
for the M\"obius diagram massless closed string states will give the dominant contribution, while massive states will yield exponentially suppressed terms, as sketched in Fig. \ref{graph}.
\bigskip

The relevant part of the effective field theory Action (in the string frame) is given by
\begin{equation}
{\cal S}=-\int d^6x \sqrt{-g}\, e^{-\phi} \, \left( |\partial \chi|^2+T \right)+\int d^{10}x \sqrt{-g} \, e^{-2\phi} \, R + \ldots \,,
\end{equation}
where $\phi$ is the dilaton, $R$ is the Ricci scalar and $T$ is the tadpole coefficient, and indeed the $7\, \bar 7 '$ states are localised in six dimensions. In the Einstein frame
\begin{equation}
{\cal S}_E=\int d^{10}x\sqrt{-g_E}R-\int d^6x\sqrt{-g_E}\left( |\partial\chi|^2+T\, e^{\phi/2} \right) \,,
\end{equation}
and, after expanding around the flat Minkowski metric $g_{E,\mu\nu}\sim\eta_{\mu\nu}+h_{\mu\nu}$ and $\phi \sim \phi_0 +\varphi$, one obtains the relevant interaction
\begin{equation}
-\int d^6 x\, T^{\mu\nu}h_{\mu\nu}=-\frac{1}{2} \int d^6 x\,
\left[\partial^\mu \chi \, \partial^\nu \chi - \tfrac{1}{2}\, \eta^{\mu\nu}\, (\partial\chi)^2\right]h_{\mu\nu} \,.
\end{equation}
Since the dilaton does not couple directly to the $\chi$ states the only diagram contributing to the mass is the one in Fig. \ref{graph}, and taking into account the appropriate factors yields
\begin{equation}
\begin{split}
-im_{\chi}^2 & = - i (p^\mu p^\nu-\eta_{\mu\nu}p^2)\, \frac{-i}{p^2}\, \left(\eta_{\mu\rho}\eta_{\nu\sigma}+\eta_{\mu\sigma}\eta_{\nu\rho}-\tfrac{1}{4}\eta_{\mu\nu}\eta_{\rho\sigma}\right)\eta^{\rho\sigma} (-iT\, g_s^{1/2})
\\
& = i ( p^\mu p^\nu-\eta_{\mu\nu}p^2)\, \frac{1}{p^2}\, \left(2\eta_{\mu\nu}-\tfrac{3}{2}\eta_{\mu\nu}\right)\, T \, g_s^{1/2}
\\
& = - \frac{5\, i}{2} \, T \, g_s^{1/2} \,,
\end{split}
\end{equation}
with, as usual, $g_s = e^{\phi_0}$.
Note that inserting the coupling to massive (string and Kaluza-Klein) states in the above diagram would yield a zero contribution in the limit $p^2\rightarrow0$. As a result, we expect that the full-fledged string theory calculation should give the same result for the $7 \, \bar 7 '$ scalar masses (where the same limit is taken).
Therefore, we obtain a positive contribution to the mass proportional to the tadpole $T$
\begin{equation}
m_{\chi}^2=\tfrac{5}{2}\, T \, g_s^{1/2}\,, \label{mass77p}
\end{equation}
so that, the one-loop masses for the six-dimensional Brane Supersymmetry Breaking model remain non-tachyonic, and the vacuum is stable even quantum mechanically. When mapped back to the string frame, the $g_s$ dependence associates the contribution (\ref{mass77p}) to a disk diagram. This peculiar result can actually be ascribed to a breaking of perturbation theory in the presence of a tadpole.

It is debatable if a result proportional to the tadpole has a physical meaning\footnote{For a recent discussion, see \cite{karim}.}, in fact as it was shown in \cite{Dudas:2004nd} the standard perturbative relation between loops and principal powers is lost when expanding around a wrong vacuum, and to get meaningful results one should sum diagrams with an arbitrary number of tadpole insertions.
However, performing a similar calculation for the masses of the gauge fields, following the previous steps and taking into account the results in \cite{Dudas:2004nd}, one finds a vanishing result, in agreement with the physical intuition that the only effect of the tadpole is to stabilise the position of the ${\rm D}7 '$ branes. Clearly, a more detailed analysis is needed to really compute the scalar masses.


\section{Field theory analysis of D5 branes {\em vs} magnetised D9 transition.}
\label{fieldtheory}

In this section we perform an analysis of the classical solutions of the D9-D5 system by using the superfield formalism of \cite{superfield1,superfield2}. We take into account the disappearance  of D5 branes into the wordvolume of D9 branes by searching the classical D9 configuration generated by the {\em vev}'s of $95$ scalars, modeled by a FI term $\xi$ in the effective Action.  While it seems to be very hard to control the dynamics, we shall provide evidence that  an Abelian magnetic field background is actually generated, where the ${\rm U} (1)$ field is chosen inside the Cartan sub-algebra of the non-Abelian gauge group.  We shall show that  gauge field configurations in a compact space are quite different from the ``standard'' (Abelian and non-Abelian) instanton configurations. In the present section we shall use the setting of D9-D5 systems and magnetised D9's, that is T-dual to the setting of branes at angles employed until now. 


\subsection{The Abelian case}

In the presence of an Abelian gauge field, the general formula for the scalar potential of a (global) supersymmetric theory is
\begin{equation}
V = \tfrac{1}{2}D^2+|F|^2 \,. 
\label{d1}
\end{equation}
It is instructive to consider first the case of a single magnetic field along a two-torus.  It corresponds in the field theory limit to having a six-dimensional Super-Yang-Mills theory compactified on a $T^2$ with $A_4,A_5$ the components of the gauge field along the two-torus where the non-trivial {\em vev}'s are localised. We use complex coordinates and  a complex scalar field defined by
\begin{equation}
z = \tfrac{1}{2}(x_4+ix_5) \,, \qquad \varPhi  = A_5 + i A_4 \,,  
\label{d2}
\end{equation}
in such a way that $\partial = \partial_4 - i \partial_5$ and $\bar\partial = \partial_4 + i \partial_5$.
In order to parameterise a {\em vev} for the $59$ hypermultiplets, we add a localised FI term in the Action, generalising to six dimensions an example of ref. \cite{superfield2}.  The important point is that in our case the internal space is compact.  In terms of the complex scalar field and of the FI term, the $F$ and $D$ terms are thus given by
\begin{equation}
\begin{split}
D &=  -  \tfrac{1}{2}(\partial\bar\varPhi+\bar\partial\varPhi)+\xi\,\delta^{(2)}(z) \,, 
\\
F & = 0 \,.
\end{split} 
\label{d3}
\end{equation}
In order to find solutions to the equations of motion that extremise $V$
\begin{equation}
\partial D = \bar \partial D  =  0 \,, 
\label{d4}
\end{equation}
one can try the following Ansatz
\begin{equation}
\varPhi  =  \alpha \, \partial \, G_2 \, , 
\label{d5}
\end{equation}
with $G_2$ being the Green's function of the Laplacian on the two-torus
\begin{equation}
\partial \bar\partial \, G_2 \ =  \delta^{(2)} -  \frac{1}{V_2} \, , \label{d6}
\end{equation}
and $\alpha$ an arbitrary constant to be fixed. The constant term on the RHS of eq. (\ref{d6}) inversely proportional to the volume $V_2$ is necessary to take into account the compactness of the two-torus.  As we shall see, it is instrumental for our proposal of relating the non-perturbative transition to the generation of a non-trivial magnetic background. 
To neutralise the source term, being
\begin{equation}
D  =  -  \alpha  \left(\delta^{(2)}-\frac{1}{V_2}\right)+\xi\,\delta^{(2)} \,, \label{d7}
\end{equation}
we must choose $\alpha=\xi$. Notice that, as expected, the $D$ term is indeed constant and reads
\begin{equation}
D  \equiv -  F_{45}  =  \frac{\xi}{V_2} \,. \label{d8}
\end{equation}
It should be stressed here that the FI term exactly generates a constant magnetic field on the defect, an indication
of the transition from a ``fat'' instanton to a magnetised brane wrapping the two-torus.
More explicitly, we can write the gauge potential by using the Green's function on a two-torus $T^2$ of complex structure $\tau $ and volume $V_2 $ as
\begin{equation}
G_2 (z | \tau )  = - \tfrac{1}{4} \ln  \left| \frac{\theta_1 (z | \tau)}{\theta_1^\prime (0|\tau )} \right|^2 +
\frac{\pi ({\rm Im \ z})^2}{2 {\rm Im} \tau} \, . \label{d08}
\end{equation}
The gauge potential (\ref{d5}) becomes then
\begin{equation}
\varPhi (z)  =  - \frac{ \alpha \, \theta_1^\prime  (z | \tau )}{4 \, \theta_1 (z | \tau )}  - 
\frac{\pi \alpha\, (z- {\bar z})}{4 \, {\rm Im}\, \tau } \, .  
\label{d081}
\end{equation}
The last term in (\ref{d081}), as observed, describes the constant magnetic field as can be deduced from eq. (\ref{d8}).  It
is not periodic along the $\tau$ cycle of the torus, but the non-periodicity exactly compensates the source term.

Given this warm-up exercise in six-dimensions, we can move to consider the eight-dimensional  Super-Yang-Mills theory compactified on a factorisable four-torus $T^4$ with complex coordinates $z_1=\frac{1}{2}(x_4+ix_5),\ z_2=\frac{1}{2}(x_6+ix_7)$. This set-up can be associated to our six-dimensional vacua described in section \ref{subsec:6d}.
On the factorised $T^4$ it is useful to introduce a pair of complex scalars encoding the internal components of the gauge field
\begin{equation}
\begin{split}
\varPhi_1 &= A_5+iA_4 \, ,
\\
\varPhi_2 &= A_7+iA_6  \,,
\end{split} \label{d9}
\end{equation}
so that the $F$ and $D$ terms are given by
\begin{equation}
\begin{split}
\bar F &= -\frac{1}{\sqrt{2}}\, (\partial_1\varPhi_2-\partial_2\varPhi_1) \,, 
\\
D&=-\tfrac{1}{2}\, \sum_{i=1}^2(\partial_i\bar \varPhi^i+\bar\partial^i\varPhi_i)+\xi \,\delta^{(4)}\, , 
\end{split} \label{d10}
\end{equation}
with the constant $\xi$ parameterising the {\em vev} of the $95$ scalars responsible for the small instanton transition \cite{Witten:1995gx}.
The equations of motion from the stationary points  of $V$ can be expressed as
\begin{equation}
\bar\partial^i D - \sqrt{2}\, \epsilon^{ij}\, \partial_j F=0 \,.
\label{d11}
\end{equation}
Note that in the absence of the source terms, the equations $D=\bar{F}=0$ are equivalent to the anti-self-duality condition for the gauge field in the internal $T^4$
\begin{equation}
F_{\mu\nu}  =  - \tilde F_{\mu\nu}\equiv - \tfrac{1}{2} \epsilon_{\mu\nu\rho\sigma} \ F_{\rho\sigma} \,. \label{d12}
\end{equation}
Using the $T^2$ example previously studied, one can make the following educated guess
\begin{equation}
\varPhi_1 =  k_1 \, \partial_1 \, G_4 \, , \qquad \varPhi_2  =  k_2 \, \partial_2 \, G_4 \, ,
\label{solution}
\end{equation}
for the solution,  where $k_i$ are constants and $G_4$ is the Green's function of the four-dimensional Laplacian on a $T^4$ with volume $V_4$,
\begin{equation}
(\partial_1\bar \partial^1 + \partial_2 \bar\partial^2)\, G_4  =  \delta^{(4)}  -  \frac{1}{V_4} \,.
\label{green}
\end{equation}
In terms of the Ansatz (\ref{solution}), the $F$ term reads 
\begin{equation}
\bar F = \frac{1}{\sqrt{2}}\,  (k_1-k_2) \partial_1 \partial_2\, G_4 \, , 
\label{d12bis}
\end{equation}
so that the condition $\bar{F}=0$ implies $k_1=k_2\equiv k$. The $D$-term then becomes 
\begin{equation}
D  = -  k\, (\partial_1\bar\partial^1+\partial_2\bar\partial^2) \, G_4 + \xi \, \delta^{(4)} 
 = -  k \left(\delta^{(4)}-\frac{1}{V_4}\right) + \xi\, \delta^{(4)} \,,
\label{d13}
\end{equation}
that, in turns, relates the constant $k$ to the FI term, $k=\xi$. Consequently
\begin{equation}
D \equiv -  (F_{45}+F_{67})  =  \frac{\xi}{V_4}   
\label{d14}
\end{equation}
is constant. The anti-self-duality condition $F_{\mu\nu} = - \tilde F_{\mu\nu}$, valid in the infinite
volume case, is violated on a compact torus.
There are no solutions to the field equations if $k_1 \neq k_2$, a case in which a different Ansatz seems to be necessary\footnote{Of interest for the Brane Supersymmetry Breaking models would be the case $k_1=-k_2$. In the Brane Supersymmetry Breaking models, however, supersymmetry is non-linearly realised; it would be interesting to understand how to find solutions of the field equations in that framework.}.

A few comments are in order here about our solution  (\ref{solution}). One can write it in the following form
\begin{equation}
\left(
  \begin{array}{c}
    A_4 \\
    A_5 \\
    A_6 \\
    A_7 \\
  \end{array}
\right) = k\, \left(
                      \begin{array}{cccc}
                        0 & -1 & 0 & 0 \\
                        1 & 0 & 0 & 0 \\
                        0 & 0 & 0 & -1 \\
                        0 & 0 & 1 & 0 \\
                      \end{array}
                    \right)\left(
                             \begin{array}{c}
                               \partial_4 \\
                               \partial_5 \\
                               \partial_6 \\
                               \partial_7 \\
                             \end{array}
                           \right) \, G_4  \, .
\label{d16}
\end{equation}
Since the matrix in  eq. (\ref{d16}) is the generator of a rotation in the factorised torus $\bigotimes_{i=1}^2 T^2_i$, the gauge field takes the elegant form
\begin{equation}
A_\mu  =  k \, J_{\mu\nu} \, \partial_{\nu} G_4  =  M_{\mu\nu} \, \partial_{\nu} G_4   
\label{d17}
\end{equation}
with a field strength
\begin{equation}
F_{\mu\nu}  =  M_{\nu\rho} \, \partial_\mu\partial_\rho G_4  -  M_{\mu\rho} \, \partial_\nu\partial_\rho G_4 \, , 
\label{d18}
\end{equation}
where $M_{\mu\nu}$ is a self-dual matrix ($M_{\mu\nu}=\frac{1}{2}\epsilon_{\mu\nu\rho\sigma}M_{\rho\sigma}$).
The dual field strength results
\begin{equation}
\tilde F_{\rho\sigma}= - F_{\rho\sigma}- M_{\rho\sigma} \, \Delta G_4 \, .
\label{d20}
\end{equation}
Again, $F_{\mu\nu}$ is not anti-self-dual due to the non harmonicity of the Green's function on the torus. An anti-self-duality condition  can only be approximatively recovered in the large volume limit.

Away from the source at $r=0$ (where $r \equiv |x|=(x_4^2+x_5^2+x_6^2+x_7^2)^{1/2} $) 
the Green's function $G_n (x_i)$ of the Laplacian on a $T^n$ torus admit a local expansion\footnote{$r^{2-n}$ becomes $\ln (r)$ for $n=2$. For a global definition on the $T^n$, this expression needs to be periodised
by adding images, as in the two-dimensional case of eq. (\ref{d08}).}
\begin{equation}
G_n  =  c_1  + \frac{c_2}{r^{n-2}}  -  \frac{r^2}{2 n V_n} \, , \label{d021}
\end{equation}
and thus the gauge field on the $T^4$ admits the local expansion
\begin{equation}
A_{\mu}  =   - \tfrac{1}{4}\, M_{\mu\nu}\, x_\nu \left(  V_4^{-1} + 8 \, r^{-4} \right) \,.
 \label{d21}
\end{equation}

The first term in eq.(\ref{d21}) corresponds to a constant magnetic field with identical components on the
$T^2$'s in our factorized Ansatz for the Abelian generator. The last term in eq. (\ref{d21}) resembles
very closely the form of the ${\rm SU} (2)$ instanton solution in the singular gauge.  This is not surprising since,
defining $f(r)=G_4^\prime (r)/r$, the gauge field becomes
\begin{equation}
A_\mu =  M_{\mu\nu} \, x_\nu \, f(r) \, ,
\label{am}
\end{equation}
that is analogous to  the most general ${\rm U}(2)$-invariant expression.  Unfortunately, its
behaviour close to the origin is very singular, since in our case $f(r)=C r^{-4}$.  As a result, 
one obtains a divergent instanton number
\begin{equation}
\int d^4x \, F_{\mu\nu} \, \tilde F_{\mu\nu} = - 8\pi^2\int_{0}^\infty dr\, \frac{d}{dr} \left( r^4f^2 \right)= +\infty \,, 
\label{inst}
\end{equation}
where we have put $k=1$.
Strictly speaking, one cannot extend the integral up to infinity, but the contribution due to the
finite volume is only a constant term and cannot make finite the integral in eq. (\ref{inst}). We suggest that this
divergence should be interpreted in terms of the singularity of the small-instanton transition.  We are
considering here a ``fat'' instanton and we are formally taking the limit to zero-size.  However, since it cannot be shrunk to zero size, it should be considered as an instanton on a non-commutative torus in the limit of zero
$B$-field \cite{nekrasov,sw,Terashima,mmms}.  In that case, the behaviour close to the singularity is modified to be of the
form $f\sim r^{-2}$,  thus giving a finite instanton number, and equal to the number of $D5$-branes.

In our compact-space situation, the low-energy dynamics of a D-brane is described by the  Dirac-Born-Infeld Action\footnote{In what follows we take $\alpha'=\frac{1}{2 \pi}$.}
\begin{equation}
{\cal L}_{\rm DBI}  = \sqrt{\det(g+ \, F)} \, .
\label{dbilagrangian}
\end{equation}
Its non-linearity might then affect the solution and could  render the singularity milder. With an  Euclidian metric $g_{\mu\nu}=\delta_{\mu\nu}$, the Euler-Lagrange equations deriving from eq. (\ref{dbilagrangian}) are
\begin{equation}
\partial_\mu\left(F_{\mu\rho}\, (1-F^2)^{-1}_{\rho\nu}\, \sqrt{\det(1+ \, F)}\right)  =  0 \, ,
\label{dbi}
\end{equation}
and it can be actually shown that, in a non-compact space and away from the singularity, the gauge potential in eq. (\ref{d16}) is a solution of the full Dirac-Born-Infeld equations. This is essentially due to the fact that, away from the origin, we have an anti-self-dual field strength, as follows from eq. (\ref{d20}) with $\Delta G_4=0$. Using $\tilde F_{\mu\nu} +F_{\mu\nu} =0$, one can easily show that
\begin{equation}
(F^2)_{\mu\nu} =  -  \delta_{\mu\nu} \, \textrm{Pf}(F) \, ,
\qquad \det (1 +   F )   =  \left[ 1 +\,  \textrm{Pf}(F) \right]^2 \,,
\end{equation}
and  the Pfaffian can be written as $\textrm{Pf}(F)=F_{12}^2+F_{13}^2+F_{14}^2$.  Therefore, any anti-self-dual field strength is a solution of the full Dirac-Born-Infeld equations in a non-compact space.

In order to study the behaviour close to the singularity $r=0$, we can limit ourselves to the case in which the function $f(r \rightarrow 0) \gg 1$.  For $2f + r f' \not=0$, we expand
\begin{equation}
F_{\mu\rho}\, (1-F ^2)^{-1}_{\rho\nu} \,\sqrt{\det(1+  F)}  \simeq  \tfrac{1}{2} (2f + r f') M_{\mu \nu}
 +  \frac{f'(3f+r f')}{2rf} \, (M_{\mu\rho} x_{\rho} x_{\nu} - M_{\nu\rho} x_{\mu}  x_{\rho}) \,,  \label{dbi2}
\end{equation}
Eq.  (\ref{dbi})  can  be written at leading order as
\begin{equation}
\left[ \frac{r^3 f' (4f + rf')}{f} \right]' \ = \ 0 \ ,
\end{equation}
and thus 
\begin{equation}
r^3 \, f' \, (4f+ r f')  = \beta \, f \, , \label{dbi3}
\end{equation}
where $\beta$ is an integration constant.  Solutions exist only for $\beta=0$ and read
\begin{equation}
 f (r \rightarrow 0) \ = \ C_1 \quad  {\rm and} \quad f (r \rightarrow 0) \ = \ \frac{C_2}{r^4} \ , \label{dbi4}
\end{equation}
where $C_i$'s are (arbitrary) integration constants. Viceversa, the case $2f + r f' =0$ turns out to be incompatible with the Dirac-Born-Infeld field equation.

In the case of a compact space, however, the anti-self-duality of the field strength is  violated by the non
 harmonicity of the Green's function, since
\begin{equation}
F_{12} + F_{34}  = \Delta G_4 \, , \qquad F_{13} + F_{42}  =  0 \ {\rm and}  \qquad F_{14}  +  F_{23}  =  0 \, ,
\label{compact}
\end{equation}
where for simplicity we have taken $k=1$ in eq. (\ref{d16}). As a result, one cannot expect that the solution of the linearised equations be still a solution of the full Dirac-Born-Infeld Action. Expanding the equations of motion and keeping  the leading and sub-leading terms in a small field expansion, one can write
\begin{equation}
\partial_\mu\left(F_{\mu\nu}+  (F^3)_{\mu\nu}-\tfrac{1}{4}\,  F_{\mu\nu}\, {\rm Tr}(F^2)\right) =  0 \, .  \label{dbi1}
\end{equation}
The linear term reproduces the Maxwell's equations, while using eqs. (\ref{compact}) and (\ref{d17}) the sub-leading terms become
\begin{equation}
(F)^3_{\mu\nu}-\tfrac{1}{4}F_{\mu\nu}\, {\rm Tr}(F^2)=\Delta G_4\, {\rm Pf}(F)\, M_{\mu\nu}-\tfrac{1}{2}(\Delta G_4)^2 \, F_{\mu\nu} \,.
\label{fcube}
\end{equation}
We can then look for a perturbative solution
\begin{equation}
F_{\mu\nu}  =  F^{(0)}_{\mu\nu}  +  F^{(1)}_{\mu\nu} \, ,
\end{equation}
where $F^{(0)}_{\mu\nu}$ is the field strength of the potential in eq. (\ref{am}). Assuming the same functional dependence for the perturbation
\begin{equation}
A_{\mu}^{(1)}  =  M_{\mu\nu} \, x_{\nu} \, g(r) \,,
\end{equation}
and inserting the expansion in the linearised equation 
\begin{equation}
\partial_{\mu}\left[F^{(1)}_{\mu\nu}+(F^{(0)3})_{\mu\nu}-\tfrac{1}{4}F^{(0)}_{\mu\nu}{\rm Tr}\left(F^{(0)2}\right)\right] \ = \ 0 \ .
\label{forg}
\end{equation}
one finds
\begin{equation}
\frac{M_{\nu\mu}\, x_{\mu}}{r}\, \frac{d}{dr}\left(4\, g+r\, g'+V_4^{-1}\, {\rm Pf}(F^{(0)})\right)  =  0 \, ,
\end{equation}
whose solution is
\begin{equation}
g(r) \ =  -  V_4^{-1} \,  f^2(r) \, .
\end{equation}
Moreover, using ${\rm Pf}(F^{(0)}) = 2 (2f^2+rff')$ one also gets $4g+rg'+{\rm Pf}(F^{(0)})/V_4=0$. Hence the perturbative expansion of the gauge field is
\begin{equation}
A_{\mu}  =  M_{\mu\nu} \, x_{\nu} \, \left(f - V_4^{-1}\, f^2 + O(V_4^{-2}) \right) \, .
\label{secorder}
\end{equation}
The nature of the singularity is not  affected by the non-linear corrections, since eq. (\ref{secorder}) is only valid far from the singularity.  It is important to notice that the subleading correction, reliable only in the infrared, is in perfect agreement (at large volume) with the subleading correction to the profile of the Abelian part of a non-commutative ${\rm U}(2)$ instanton in the singular gauge, as found for instance in ref. \cite{ncinst}.

The previous discussion can also be rephrased in terms of a first-order formulation. Indeed, the equation (\ref{dbi}) is automatically satisfied if one imposes the generalised anti-self-duality condition
\begin{equation}
F_{\mu\rho}\, (1-F^2)^{-1}_{\rho\nu}\, \sqrt{\det(1+F)}= - \tilde F_{\mu\nu} ,
\label{self}
\end{equation}
that is compatible with the perturbative solution in eq. (\ref{secorder}).

Finally, let us mention that since the singular part of the solution has a fast variation near the singularity, the Dirac-Born-Infeld Action is in principle not a good approximation of the string effective Action. It is then plausible that higher-derivative corrections play an important role and change the behaviour of the solution near the singularity, leading eventually to a finite instanton charge.


\subsection{The non-Abelian case}

A constant background magnetic field actually emerges also in the zero-size limit of a non-Abelian gauge configuration, pertaining to the case where more D-branes are coincident. For simplicity, let us focus our attention to the case of an ${\rm SU}(2)$ instanton in a compact internal torus, with auxiliary fields given by
\begin{equation}
\begin{split}
D &= - \tfrac{1}{2}\,  ({\bar \partial}^i \varPhi_i + {\partial}_i {\bar \varPhi}^i) + [\varPhi_i , {\bar \varPhi}^i ] \, ,
\\
{\bar F} &= - \frac{1}{\sqrt{2}}\, \epsilon^{ij}\, {\cal D}_i \varPhi_j \,,
\end{split}
\label{d25}
\end{equation}
where ${\cal D}_i$ is the covariant derivative, acting on the various fields as
\begin{equation}
\begin{split}
{\bar {\cal D}}^i D &= {\bar {\partial}}^i D +2 [ {\bar \varPhi}^i ,D ] \, , 
\\
{\cal D}_j F  &=  \partial_j F - 2[\varPhi_j ,F ] \,,  
\\
{\cal D}_i \varPhi_j &=  \partial_i \varPhi_j -2 [\varPhi_i , \varPhi_j ] \, .
\end{split}
\label{d27}
\end{equation}
The equations of motions
\begin{equation}
{\bar {\cal D}}^i D  - \sqrt{2}\, \epsilon^{ij} \, {\cal D}_j F  =  0 \, ,  
\label{d26}
\end{equation}
are similar to eqs. (\ref{d11}) and can be solved adopting the standard Ansatz for the gauge connection
\begin{equation}
A_\mu =  i \, \sigma_{\mu\nu} \, \partial_\nu \log G_4 \, ,
\end{equation}
where $\sigma_{\mu\nu}$ are self-dual and each of its components is an anti-Hermitean matrix, in such a way that $A_\mu^{\dagger}=A_\mu$. Explicitly, they are given by
\begin{equation}
i\, \sigma_{\mu\nu}=\frac{1}{2}\left(
                              \begin{array}{cccc}
                                0 & -\sigma_3 & \sigma_2 & -\sigma_1 \\
                                \sigma_3 & 0 & -\sigma_1 & -\sigma_2 \\
                                -\sigma_2 & \sigma_1 & 0 & -\sigma_3 \\
                                \sigma_1 & \sigma_2 & \sigma_3 & 0 \\
                              \end{array}
                            \right) \,,
\end{equation}
where the $\sigma_i$ are the Pauli matrices, and the $\sigma_{\mu\nu}$ are related to the 't Hooft symbols. In complex formulation,
the fields can be written in the form
\begin{equation}
\begin{split}
\varPhi_i & = \tfrac{1}{4}(\sigma_3\, \partial_i-\sigma_-\, \epsilon_{ij}\,\bar\partial^j) \log G_4 \ ,
\\
 \bar\varPhi^i &=\tfrac{1}{4}(\sigma_3\, \bar\partial^i- \sigma_+ \,\epsilon^{ij}\, \partial_j) \log G_4 \ ,
 \end{split}
\end{equation}
with $\sigma_\pm=\sigma_1\pm i\sigma_2$, as usual.
In a non-compact space, the $F$ and $D$ flatness conditions imply that the gauge configuration be anti-self-dual since, when written in terms of the field strength they become
\begin{equation}
D  \equiv  -  ( F_{45}+F_{67}) =0 \, , \qquad
F  = \frac{1}{\sqrt{2}}\left[F_{47} + F_{56} + i (F_{46}-F_{57}) \right] =0 \, ,
\end{equation}
However, as in the Abelian case, the anti-self-duality condition fails to hold in a compact space, since now
\begin{equation}
\tilde F_{\mu\nu}=-\left(F_{\mu\nu}+i\, \sigma_{\mu\nu} \, G_4^{-1}\, \Delta G_4\right) \, ,
\end{equation}
and the Green's function on $T^4$ is not any more an harmonic function.

Some comments on the results of these last two sections are in order.  The constant internal magnetic field distributed
on the D-branes and related, by T-duality, to the rotation of D-branes, is associated to a diagonal ${\rm U}(1)$ inside the ${\rm U}(N)$ gauge group supported by $N$ magnetised D-branes.  It coincides with the constant magnetic field we find in the Abelian case or is embedded in the non-Abelian solution just shown, and it is a consequences of a non-vanishing FI term if the internal space is compact.  In the infrared, the behaviour is very reminiscent of the behaviour of the ${\rm SU}(2)$ BPST (anti-)instanton \cite{bpst} in the singular gauge.  In that case, however, the situation is quite different.  The constant magnetic field resulting when the maximum size equals the volume for the BPST instanton takes the form
\begin{equation}
F_{\mu \nu} \simeq  \frac{1}{g \, \rho^2} \, i \,  \bar \sigma_{\mu \nu} \, , \label{d24}
\end{equation}
where now $\bar \sigma_{\mu\nu}$ is  anti-self-dual, and $g$ is the gauge coupling constant.  In a form notation,
the magnetic field corresponds to the two form flux
\begin{equation}
F  =  H  \left[ \sigma_3\, (dz_1 \wedge d {\bar z}_1 - dz_2 \wedge d {\bar z}_2 ) + \sigma_{-} \, dz_1 \wedge d {\bar z}_2 +
 \sigma_{+} \, dz_2 \wedge d {\bar z}_1  \right] \, ,
 \label{d024}
\end{equation}
and it is not possible to bring this solution into the Cartan sub-algebra, by a gauge rotation. The ${\rm SU}(2)$ instanton is a genuinely non-Abelian solution and has no connection, for instance, with the Abelian configuration corresponding to two equal magnetic fields $H_1 = H_2$ in the factorised four-torus. The solutions we find are clearly not instantons.  The field strength is not anti-self-dual and it exhibits a real singularity at the origin, being not gauge-equivalent to a non-singular configuration. One possible interpretation of the solutions is related to non-commutative instantons in the limit of zero $B$-field and zero size, as mentioned in the previous section. However, we are convinced that an accurate analysis of the higher derivative and higher order corrections could shed new light on these solutions, that certainly deserve a much deeper investigation.

\vskip 1in

\subsection*{Acknowledgements}
We would like to thank K. Benakli, M. Bianchi, M. Bill\`o, P.G. Camara,M. Goodsell and A. Sagnotti for useful discussions. A.C. and G.P would like to thank CPHT at Ecole Polytechnique and the Theory Unit at CERN, C.C., E.D. and G.P.  are grateful to the Theory Department of the University of Torino for hospitality during various stages of this project. This work was partially supported by the European ERC Advanced Grants no. 226371 MassTeV and no. 226455 ``Supersymmetry, Quantum Gravity and Gauge Fields'' (SUPERFIELDS)., by the CNRS PICS Nos. 3747 and 4172, by the European Initial Training Network PITN-GA-2009-237920, by the ANR TAPDMS ANR-09-JCJC-0146 and by the grant CNCSIS `Idei' 454/2009, ID-44.


\appendix

\section{$T^4/{\mathbb{Z}}_2$ orientifolds with intersecting branes}
\label{app:6d}

We remind here the partition functions for the $T^4/{\mathbb Z}_2$ orientifold of type IIB superstring with D7 branes at angles, both for supersymmetric vacua and for configurations with Brane Supersymmetry Breaking. As we shall see, the two choices differ in some signs in the Klein-bottle amplitude and in the open-string sector.
To this end the standard world-sheet parity $\varOmega$ has to be dressed with the complex conjugation ${\mathcal R}:\ z_i \to \bar z_i$ on each complex coordinate of the $i$-th $T^2$ and, for consistency, with the operator $(-1)^{F_{\rm L}}$, $F_{\rm L}$ being the left-handed space-time fermion number. Moreover, one has the option to add some inner automorphism $\hat\sigma$ that flips the sign in the twisted sector of closed strings \cite{bsb1b,bsb1c,bsb1d}. 
As a result, the $\hat\varOmega = \varOmega \, {\mathcal R}\, (-1)^{F_{\rm L}}\, \hat\sigma $ orientifold will include ${\rm O}7_-$ and ${\rm O}7^\prime_\sigma$ planes, the former wrapping the horizontal ${\boldsymbol a}_i$ cycles of each $T^2$ and the latter wrapping the vertical ${\boldsymbol b}_i$ cycles, with an orientation determined by $\sigma$, of each $T^2$. Here $\sigma$ is the eigenvalue the operator $\hat\sigma$ has on the twisted sector and, as expected, select the kind of ${\rm O}7^\prime$ planes. 

The relevant one-loop amplitudes in closed-string sector thus read
\begin{equation*}
{\mathcal T} = \frac{1}{2} \left\{ |Q_o + Q_v |^2 \, \varLambda^{(4,4)} +  |Q_o - Q_v |^2 \, \left| \frac{2\eta}{\vartheta_2}\right|^4
 + 16\, |Q_s + Q_c |^2 \, \left| \frac{\eta}{\vartheta_4}\right|^4 + 16\, |Q_s - Q_c |^2 \, \left| \frac{\eta}{\vartheta_3}\right|^4\right\}
\end{equation*}
for the torus amplitude, and
\begin{equation*}
{\mathcal K} = \frac{1}{4} (Q_o + Q_v ) \, \left( \varLambda_0 + \varGamma_0 \right) + \frac{16}{2} \, \sigma\, (Q_s +Q_c ) \left( \frac{\eta}{\vartheta_4}\right)^2
\end{equation*}
for the Klein-bottle amplitude. 
Here we have introduced the notation 
\begin{equation}
\varLambda_{\pm 2a} = \sum_{m\in\mathbb{Z}^4} q^{(m\pm 2b)^T\, A \, (m\pm 2b)}\,, 
\qquad
\varGamma_{\pm 2b} = \sum_{m\in\mathbb{Z}^4} q^{(m\pm 2b)^T\, A^{-1} \, (m\pm 2b)}\,,
\label{gammalambda}
\end{equation}
for the contribution of the (shifted) Kaluza-Klein and winding zero modes
with 
\begin{equation}
A = {\rm diag} \left( \frac{\alpha'}{R_1^2}\,,\, \frac{R_2^2}{\alpha'}\,,\,\frac{\alpha'}{R_3^2}\,,\, \frac{R_4^2}{\alpha'} \right)\,.
\end{equation}
As usual, the choice $\sigma =+1$ corresponds to the conventional choice \cite{Bianchi:1990yu, Gimon:1996rq} and the massless excitations now comprise an ${\mathcal N}(1,0)$ supergravity multiplet coupled to one tensor multiplet and twenty hypermultiplets. Alternatively, the choice $\sigma =-1$ corresponds to vacua with Brane Supersymmetry Breaking \cite{bsb1b,bsb1c} and the massless excitations comprise an ${\mathcal N}(1,0)$ supergravity multiplet coupled now to seventeen tensor multiplets and four hypermultiplets.

The open-strings will involve D7 branes wrapping factorisable two-cycles on the $T^4=T^2 \times T^2$. As a result, each stack $a$ of branes is identified by four integers $(m^a_i , n^a_i)$, $i=1,2$ corresponding to the wrappings of the horizontal ${\boldsymbol a}_i$ and vertical ${\boldsymbol b}_i$ one-cycles of the $i$-th $T^2$. In writing the annunuls and M\"obius-strip amplitudes we do not include explicitly the contribution of horizontal and vertical D7 branes (that in the T-dual $\varOmega$ orientifold would correspond to D9 and D5 branes, respectively) since these simply correspond to the choices  $(1,\,0;\,1,\,0)$ and (0,\,1;\,0,\,1) for the wrapping numbers $(m^a_1,\, n^a_1;\, m^a_2,\, n^a_2) $. 
In the case of vacua with Brane Supersymmetry Breaking, however, some care has to be taken in the case D7 branes sit on orientifold planes, since in this configuration the Chan-Paton labels become real and the gauge groups orthogonal and/or symplectic. For completeness, in table \ref{non-susy spectrum} we report the correct spectrum also for these horizontal or vertical branes. A generic D7 brane, however, is not invariant under the $\hat\varOmega$ projection, and thus one is bound to introduce also its image under the orientifold action in order to get invariant configurations. Given the expression for $\hat\varOmega$, the image brane, in the following indicated with the label $\bar a$, has wrapping number $(m^a_1,\, -n^a_1;\, m^a_2,\, - n^a_2) $, if $(m^a_1,\, n^a_1;\, m^a_2,\, n^a_2) $ are those of the {\em fundamental} ${\rm D}7_a$ brane.
As a result, the annulus amplitude consists of the following contributions:
\begin{itemize}
\item ${\mathcal A}_{aa}$: open strings starting from and ending on the same stack of ${\rm D}7_a$ branes;
\item ${\mathcal A}_{a\bar a}$: open strings starting from a ${\rm D}7_a$ brane and ending on its ${\rm D}7_{\bar a}$ image;
\item ${\mathcal A}_{ab}$: open strings starting from a stack of ${\rm D}7_a$ branes and ending on a  stack of ${\rm D}7_b$ branes;
\item ${\mathcal A}_{a\bar b}$: open strings starting from a stack of ${\rm D}7_a$ branes and ending on a  stack of ${\rm D}7_{\bar b}$ image branes;
\end{itemize}
In equations,
\begin{equation*}
{\mathcal A}_{aa} = \frac{1}{2} \sum_a N_a\,\bar N_a \,\left[  (Q_o+Q_v) {\textstyle\left[\genfrac{}{}{0pt}{}{0}{0}\right]} \, P^{(1)} \, W^{(1)} \, P^{(2)}\, W^{(2)} + (Q_o - Q_v ) {\textstyle \left[
\genfrac{}{}{0pt}{}{0}{0}
\right]} \, \left(\frac{2\eta}{\vartheta_2}\right)^2 \right]\,,
\end{equation*}
\begin{equation*}
\begin{split}
{\mathcal A}_{a\bar a} =& \frac{1}{4} \sum_a \left[ N_a^2 \, (Q_o + Q_v )  {\textstyle\left[
\genfrac{}{}{0pt}{}{a\bar a}{ 0}
\right]}
+ \bar N_a^2 \,  (Q_o + Q_v )  {\textstyle\left[
\genfrac{}{}{0pt}{}{\bar a a}{0}
\right]} \right] \frac{I_{a\bar a}}{\varUpsilon_1 {\textstyle\left[
\genfrac{}{}{0pt}{}{a\bar a}{0}
 \right] }}
\\
&+\frac{1}{4} \sum_a \left[ \epsilon_a^2
\, N_a^2 \, (Q_o - Q_v )  {\textstyle\left[
\genfrac{}{}{0pt}{}{a\bar a}{0}\right]}
+ \bar\epsilon_a^2\, \bar N_a^2 \,  (Q_o - Q_v )  {\textstyle\left[
\genfrac{}{}{0pt}{}{\bar a a}{0}\right]} \right] \frac{S_{a\bar a}}{\varUpsilon_2 {\textstyle\left[
\genfrac{}{}{0pt}{}{a\bar a}{0} \right] }}\,,
\end{split}
\end{equation*}
\begin{equation*}
\begin{split}
{\mathcal A}_{ab} =& \frac{1}{2} \sum_{\genfrac{}{}{0pt}{}{a,b}{a>b}} \left[ N_a\,\bar N_b \, (Q_o + Q_v )  {\textstyle\left[ \genfrac{}{}{0pt}{}{ab}{0}\right]}
+ \bar N_a\, N_b \,  (Q_o + Q_v )  {\textstyle\left[
\genfrac{}{}{0pt}{}{\bar a \bar b}{0}\right]} \right] \frac{I_{ab}}{\varUpsilon_1 {\textstyle\left[
\genfrac{}{}{0pt}{}{ab}{0} \right] }}
\\
&+\frac{1}{2} \sum_{\genfrac{}{}{0pt}{}{a,b}{a>b}} \left[ \epsilon_a \bar\epsilon_b\, 
N_a\, \bar N_b \, (Q_o - Q_v )  {\textstyle\left[ \genfrac{}{}{0pt}{}{ab}{0}\right]}
+ \bar\epsilon_a \epsilon_b\, 
\bar N_a\, N_b \,  (Q_o - Q_v )  {\textstyle\left[ \genfrac{}{}{0pt}{}{\bar a \bar b}{0}\right]} \right] \frac{S_{ab}}{\varUpsilon_2 {\textstyle\left[ \genfrac{}{}{0pt}{}{ab}{0} \right] }}\,,
\end{split}
\end{equation*}
\begin{equation*}
\begin{split}
{\mathcal A}_{a\bar b} =& \frac{1}{2} \sum_{\genfrac{}{}{0pt}{}{a,b}{a>b}} \left[ N_a\, N_b \, (Q_o + Q_v )  {\textstyle\left[ \genfrac{}{}{0pt}{}{a\bar b}{0}\right]}
+ \bar N_a\, \bar N_b \,  (Q_o + Q_v )  {\textstyle\left[ \genfrac{}{}{0pt}{}{\bar a  b}{0}\right]} \right] \frac{I_{a\bar b}}{\varUpsilon_1 {\textstyle\left[ \genfrac{}{}{0pt}{}{a\bar b}{0} \right] }}
\\
&+\frac{1}{2} \sum_{\genfrac{}{}{0pt}{}{a,b}{a>b}} \left[ \epsilon_a \epsilon_b \,
N_a\,  N_b \, (Q_o - Q_v )  {\textstyle\left[ \genfrac{}{}{0pt}{}{a\bar b}{0}\right]}
+ 
\bar\epsilon_a \bar \epsilon_b\, 
\bar N_a\, \bar N_b \,  (Q_o - Q_v )  {\textstyle\left[ \genfrac{}{}{0pt}{}{\bar a  b}{0}\right]} \right] \frac{S_{a\bar b}}{\varUpsilon_2 {\textstyle\left[ \genfrac{}{}{0pt}{}{a\bar b}{0} \right] }}\,.
\end{split}
\end{equation*}

In writing these amplitudes we have followed the notation in \cite{Angelantonj:2005hs}. The momentum and winding lattice sum, depend clearly on the total length $L_\|$ of the D-brane and on the minimal separation $L_\perp$ of the multiple wrappings
\begin{equation*}
L^{a,(i)}_\| = \sqrt{ (m^a_i\, R_1^{(i)})^2 +  (n^a_i\, R_2^{(i)})^2}\,,\qquad
L^{a,(i)}_\perp = R^{(i)}_1\, R^{(i)}_2\, / L^{a,(i)}_\| \,,
\end{equation*}
with $R^{(i)}_{1,2}$ the sizes of the horizontal and vertical sides of the $i$-th $T^2$. The characters depend on the relative angle of branes $a$ and $b$ through
\begin{equation*}
\begin{split}
Q_o {\textstyle\left[ \genfrac{}{}{0pt}{}{ab}{cd}\right]} &= V_4 \, \left[ O_2 (\zeta_1) O_2 (\zeta_2 ) + V_2 (\zeta_1) V_2 (\zeta_2 ) \right] - C_4 \,\left[ S_2 (\zeta_1) C_2 (\zeta_2 ) + C_2 (\zeta_1) S_2 (\zeta_2 ) \right]\,,
\\
Q_v {\textstyle\left[ \genfrac{}{}{0pt}{}{ab}{cd}\right]} &= O_4 \, \left[ V_2 (\zeta_1) O_2 (\zeta_2 ) + O_2 (\zeta_1) V_2 (\zeta_2 ) \right] - S_4 \,\left[ S_2 (\zeta_1) S_2 (\zeta_2 ) + C_2 (\zeta_1) C_2 (\zeta_2 ) \right]\,,
\end{split}
\end{equation*}
where the internal SO(4) symmetry is clearly broken to ${\rm SO}(2) \times {\rm SO} (2)$ due to the non-trivial rotations of the branes, and
\begin{equation*}
\begin{split}
O_2 (\zeta ) &= \frac{e^{2i\pi\zeta}}{\eta (\tau)}\, \left[ \vartheta_3 (\zeta |\tau ) + \vartheta_4 (\zeta |\tau )\right]\,,
\\
V_2 (\zeta ) &= \frac{e^{2i\pi\zeta}}{\eta (\tau)}\, \left[ \vartheta_3 (\zeta |\tau ) - \vartheta_4 (\zeta |\tau )\right]\,,
\end{split}
\qquad
\begin{split}
S_2 (\zeta ) &= \frac{e^{2i\pi\zeta}}{\eta (\tau)}\, \left[ \vartheta_2 (\zeta |\tau ) + i\, \vartheta_1 (\zeta |\tau )\right]\,,
\\
C_2 (\zeta ) &= \frac{e^{2i\pi\zeta}}{\eta (\tau)}\, \left[ \vartheta_2 (\zeta |\tau ) - i\, \vartheta_1 (\zeta |\tau )\right]\,,
\end{split}
\end{equation*}
with $\zeta_i = [ (\phi_a^i - \phi_b^i )\tau + (\phi^i_c-\phi^i_d )]/\pi$. The contribution of the rotated world-sheet bosonic coordinates is encoded in the combinations
\begin{equation*}
\varUpsilon_{1} {\textstyle \left[ \genfrac{}{}{0pt}{}{ab}{cd} \right]} = \prod_{i=1,2} \frac{\vartheta_1 (\zeta_i |\tau) }{i\,\eta (\tau)}\, e^{2i\pi \zeta_i}\,,
\qquad
\varUpsilon_{2} {\textstyle \left[ \genfrac{}{}{0pt}{}{ab}{cd} \right]} = \prod_{i=1,2} \frac{\vartheta_2 (\zeta_i |\tau) }{\eta (\tau)}\, e^{2i\pi \zeta_i}\,.
\end{equation*}
Finally,
\begin{equation*}
I_{ab} = \prod_{i=1,2} \left( m^a_i\, n^b_i - m^b_i\, n^a_i \right)
\end{equation*}
counts the number of intersections branes $a$ and $b$ have on the $T^4$, while
\begin{equation*}
S_{ab} = \prod_{i=1,2} \left( 1 + \varPi (m^a_i + m^b_i) \, \varPi (n^a_i + n^b_i) \right) \,
\end{equation*}
with $\varPi (\mu) = (1+e^{i\pi \mu})/2$, $\mu \in \mathbb{Z}$, counts the number of mutual intersections that coincide with the fixed points of the $T^4/\mathbb{Z}_2$ orbifold, assuming that all branes cross the fixed point at the origin of the $T^4$. These are related to the wrapping of the collapsed cycles on the resolved orbifold.

The phases $\epsilon_a$ reflect the action of the $\mathbb{Z}_2$ orbifold group on the Chan-Paton labels of the D7 branes. These phases are of course correlated with the sign $\sigma =\pm 1$, since in the supersymmetric case the action of $\mathbb{Z}_2$ is complex while in vacua with Brane Supersymmetry Breaking it is real. In particular,
\begin{equation}
\epsilon_a^2 = -\sigma\,,
\end{equation}
$\bar\epsilon_a$ is the complex conjugate of $\epsilon_a$, and the phases have to be chosen to satisfy the twisted tadpole conditions
\begin{equation}
\sum_a \left( \epsilon_a \, N_a +\bar\epsilon_a \, \bar N_a \right) =0\,. \label{twistedtadpoles}
\end{equation}
Although in the supersymmetric $\sigma =+1$ case eq. (\ref{twistedtadpoles}) vanishes identically, since $N_a \equiv \bar N_a$, and thus all $\epsilon_a$ can be taken to be $\epsilon_a = i$, for the choice $\sigma =-1$ one has to have some stacks with positive $\epsilon_a$ and others with negative $\epsilon_b$.

The last one-loop amplitude is the M\"obius-strip one and gets contributions from the brane intersections with the orientifold planes. One has
\begin{equation*}
\begin{split}
{\mathcal M}_a =& - \frac{1}{4} \sum_a  \left[ N_a\, (\hat Q_o + \hat Q_v ) {\textstyle \left[
\genfrac{}{}{0pt}{}{a\bar a}{0} \right]} + \bar N_a \, (\hat Q_o + \hat Q_v ) {\textstyle \left[ \genfrac{}{}{0pt}{}{\bar a a}{0} \right]} \right] \frac{I_{a{\rm O}}}{\hat\varUpsilon_1 {\textstyle \left[ \genfrac{}{}{0pt}{}{a\bar a}{0} \right]}}
\\
&+ \frac{1}{4} \sum_a  \left[ N_a\, (\hat Q_o - \hat Q_v ) {\textstyle \left[ \genfrac{}{}{0pt}{}{a\bar a}{0} \right]} + \bar N_a \, (\hat Q_o - \hat Q_v ) {\textstyle \left[ \genfrac{}{}{0pt}{}{\bar a a}{0} \right]} \right] \frac{I^\epsilon_{a{\rm O}'}}{\hat\varUpsilon_2 {\textstyle \left[ \genfrac{}{}{0pt}{}{a\bar a}{0} \right]}} \,,
\end{split}
\end{equation*}
where, here and in the rest of the Appendix, the {\em hatted} characters include suitable phases as explained in \cite{Angelantonj:2002ct, Dudas:2000bn}, and
\begin{equation*}
I_{a{\rm O}}=4\, \prod_{i=1,2}  n^a_i\,, \qquad  I^\sigma_{a{\rm O}'}= 4\, \sigma\, \prod_{i=1,2}  m^a_i\,,
\end{equation*}
count respectively the intersections of the ${\rm D}7_a$ branes with the horizontal and vertical orientifold planes. The intersection number $I^\sigma_{a{\rm O}'}$ depends on the sign $\sigma$ since, while the horizontal ${\rm O}7$ planes the wrapping numbers can be taken to be $ (1,\, 0;\, 1,\, 0)$ in both cases, to the vertical ${\rm O}7 '$ planes one has to associate the wrapping numbers $ (0,\, 1;\, 0,\, \sigma)$, that reflect the RR charges of the ${\rm O}7 ^{\prime}_\sigma$. As a result, for a given stack of ${\rm D}7_a$ branes, the intersection $I^\sigma_{a{\rm O}'}$ flips the sign for the two choices $\sigma=\pm 1$.  Notice that on the $T^4/\mathbb{Z}_2$ orientifold there are exactly four copies of ${\rm O}7$ and ${\rm O}7 '$ planes, as reflected by the multiplicity in the intersection numbers.

Finally, for completeness we report the untwisted RR tadpole conditions
\begin{equation}
\sum_a m^a_1 \, m^a_2 \, N_a = 16\,,
\qquad
\sum_a n^a_1 \, n^a_2 \, N_a = 16\, \sigma\,.
\end{equation}

\begin{table}
\centering
\begin{tabular}{ccc}
\toprule
Multiplicity & Representation & Relevant Indices \\
\toprule
$\frac{1}{4}(I_{a\bar a}+I_{a{\rm O}}+I^+_{a{\rm O'}}+4)$ & $\frac{N_a(N_a-1)}{2}$ & $\forall a$
\\
\midrule
$\frac{1}{4}(I_{a\bar a}-I_{a{\rm O}}-I^+_{a{\rm O'}}+4)$ & $\frac{N_a(N_a+1)}{2}$ & $\forall a$
\\
\midrule
$\frac{1}{2}(I_{ab}+S_{ab})$ & $(N_a,\bar N_b)$ & $a < b$
\\
\midrule
$\frac{1}{2}(I_{a\bar b}-S_{ab})$& $(N_a,N_b)$& $a < b$
\\
\bottomrule
\end{tabular}
\caption{Representations and multiplicities of charged hypermultiplets living at the D7 brane intersections on a $T^4 /\mathbb{Z}_2$. The gauge group is $G=\prod_a {\rm U} (N_a)$.}
\label{tab:6dsusyspectrum}
\end{table}

\begin{table}
\centering
\begin{tabular}{cccc}
\toprule
Multiplicity & Representation & Multiplet/Fields & Chirality\\
\toprule
$1$ & $p_a\bar p_a+\frac{m_1(m_1-1)}{2}+\frac{m_2(m_2-1)}{2}$ & Gauge Multiplet & Left-handed
\\ [1ex]
$1$ & $\frac{d_1(d_1+1)}{2}+\frac{d_2(d_2+1)}{2}$ & Vector Boson & -
\\
\midrule
$\frac{1}{4}(-I_{a\bar a}-I_{aO}+I^-_{aO'}-4)$ & $\frac{p_a(p_a-1)}{2}$ & 4 scalars & -
\\[1ex]
$\frac{1}{4}(-I_{a\bar a}- I_{aO}-I^-_{aO'}+4)$ & $\frac{p_a(p_a-1)}{2}$ & Weyl fermion & Left-handed
\\[1ex]
$\frac{1}{4}(-I_{a\bar a}+I_{aO}-I^-_{aO'}-4)$ & $\frac{p_a(p_a+1)}{2}$ & 4 scalars & -
\\[1ex]
$\frac{1}{4}(-I_{a\bar a}+ I_{aO}+I^-_{aO'}+4)$ & $\frac{p_a(p_a+1)}{2}$ & Weyl fermion & Left-handed
\\
\midrule
$\frac{1}{2}(-I_{ab}+\epsilon_a\epsilon_bS_{ab})$ & $(p_a,\bar p_b)$ & Weyl fermion & Left-handed
\\[1ex]
$\frac{1}{2}(-I_{ab}-\epsilon_a\epsilon_bS_{ab})$ & $(p_a,\bar p_b)$ & 4 scalars & -
\\[1ex]
$\frac{1}{2}(-I_{a\bar b}+\epsilon_a\epsilon_bS_{ab})$ & $(p_a,p_b)$ & Weyl fermion & Left-handed
\\[1ex]
$\frac{1}{2}(-I_{a\bar b}-\epsilon_a\epsilon_bS_{ab})$ & $(p_a, p_b)$ & 4 scalars & -
\\
\midrule
$1$ & $(m_1,m_2)$ & hypermultiplet & Right-handed
\\[1ex]\hline
$1$ & $\frac{d_1(d_1-1)}{2}+\frac{d_2(d_2-1)}{2}$ & Weyl fermion & Left-handed
\\[1ex]
$1$ & $(d_1,d_2)$ & 4 scalars & -
\\[1ex]
$1$ & $(d_1,d_2)$ & Weyl fermion & Right-handed
\\
\midrule
$1$ & $(m_1,d_1)+(m_2,d_2)$ & 1/2 Weyl fermion & Left-handed
\\[1ex]
$1$ & $(m_1,d_2)+(m_2,d_1)$ & 2 scalars & -
\\
\midrule
$\frac{1}{2}(-I_{a7}+\epsilon_aS_{a7})$ & $(p_a,m_1)$ & Weyl fermion & Left-handed
\\[1ex]
$\frac{1}{2}(-I_{a7}-\epsilon_aS_{a7})$ & $(p_a,m_1)$ & 4 scalars & -
\\[1ex]
$\frac{1}{2}(-I_{a7}-\epsilon_aS_{a7})$ & $(p_a,m_2)$ & Weyl fermion & Left-handed
\\[1ex]
$\frac{1}{2}(-I_{a7}+\epsilon_aS_{a7})$ & $(p_a, m_2)$ & 4 scalars & -
\\
\midrule
$\frac{1}{2}(I_{a \bar 7 '}+\epsilon_aS_{a\bar 7 '})$ & $(p_a,d_1)$ & Weyl fermion & Left-handed
\\[1ex]
$\frac{1}{2}(I_{a\bar 7 '}-\epsilon_aS_{a\bar 7 '})$ & $(p_a,d_1)$ & 4 scalars & -
\\[1ex]
$\frac{1}{2}(I_{a\bar 7 '}-\epsilon_aS_{a\bar 7 '})$ & $(p_a,d_2)$ & Weyl fermion & Left-handed
\\[1ex]
$\frac{1}{2}(I_{a\bar 7 '}+\epsilon_aS_{a\bar 7 '})$ & $(p_a, d_2)$ & 4 scalars & -
\\
\bottomrule
\end{tabular}
\caption{Massless spectrum of the non-supersymmetric $T^4 /\mathbb{Z}_2$ orientifold for branes at angles. Left-handed (right-handed) fermions correspond to $C_4$ ($S_4$) characters. For completeness we also include a certain number of straight ${\rm D7}$, with Chan-Paton labels $m_1$ and $m_2$, and $\overline{{\rm D}7} '$ branes, with Chan-Paton labels $d_1$ and $d_2$, to stress the subtleties emerging when branes wrap the same cycle as orientifold planes.}
\label{non-susy spectrum}
\end{table}

\subsection{Partition functions of the Brane Supersymmetry Breaking model in the bulk}
\label{wl}
Finally, we consider here the possibility to move all branes in the bulk in the six-dimensional Brane Supersymmetry Breaking vacua.
The closed-string amplitudes are not affected by this deformation, while the annulus and M\"obius strip amplitudes depend now on the Wilson lines and brane displacements. Since branes wrap generic bulk cycles, the orbifold group does not act on the open-string spectrum. Denoting collectively by $a$ ($b$) the Wilson lines and positions of the D7 ($\overline{{\rm D}7} '$) branes one has:
\begin{equation}
\begin{split}
{\mathcal A}&=\tfrac{1}{2}\Bigl\{
(n^2\, \varGamma_0 + d^2\, \varLambda_0) ( V_4O_4-S_4S_4+O_4V_4-C_4C_4)
\\
&+4\, n d \, (O_4S_4-C_4O_4+V_4C_4-S_4V_4)\, \left(\frac{\eta}{\vartheta_4}\right)^2
\\
&+ \tfrac{1}{2}\,\left[ n^2 \, (\varGamma_{+2a} + \varGamma_{-2a} )
 + d^2\, ( \varLambda_{+2b} + \varLambda_{-2b} ) \right]\,(V_4O_4-S_4S_4+O_4V_4-C_4C_4)\Bigr\} \,,
\end{split}
\end{equation}
and
\begin{equation}
\begin{split}
{\mathcal M} &=\tfrac{1}{4}\Bigl[
n \,  (\varGamma_{+2a} + \varGamma_{-2a} )\, (\hat V_4\hat O_4-\hat S_4 \hat S_4+\hat O_4\hat V_4-\hat C_4\hat C_4)
\\
& + d\, ( \varLambda_{+2b} + \varLambda_{-2b} ) \, (-\hat V_4\hat O_4-\hat S_4 \hat S_4-\hat O_4\hat V_4-\hat C_4\hat C_4)
\\
&+ 2\, n \, (\hat V_4\hat O_4+\hat S_4 \hat S_4-\hat O_4\hat V_4-\hat C_4\hat C_4)\, \left(\frac{2\eta}{\vartheta_2}\right)^2
\\
&+2\, d\, (-\hat V_4\hat O_4+\hat S_4 \hat S_4+\hat O_4\hat V_4-\hat C_4\hat C_4)\, \left(\frac{2\eta}{\vartheta_2}\right)^2\Bigr]\,,
\end{split}
\end{equation}
where $\varGamma_\alpha$ and $\varLambda_\alpha$ were defined in eq. (\ref{gammalambda}).
As usual, we have denoted by $n$ ($d$) the Chan-Paton label of the D7 ($\overline{{\rm D}7} '$) branes.
The twisted tadpoles are now absent since bulk branes are not charged with respect to twisted forms, while the
(massless) untwisted ones are not affected by this deformation, though the total number of physical branes is now halved due to the separation between branes and their $\hat\varOmega$ images. The massless excitations are summarised in table \ref{bsbbulk}.

Also in this case, one has the possibility to rotate the bulk branes. The deformation is rather standard and can be deduced from intersecting-brane vacua compactified on a flat $T^4$. The only difference is in the M\"obius strip amplitude, since in this case D7 branes interact also with ${\rm O}7^\prime_+$ planes. The corresponding massless open-string spectrum is summarised in table \ref{bsbbulkrotated}.

\begin{table}
\centering
\begin{tabular}{cccc}
  \toprule
  Multiplicity & Representation & Field/Multiplet & Chirality
  \\
  \toprule
  $1$ & (120,1) & Gauge Multiplet & Left-handed
  \\
  $1$ & (1,136) & Vector Boson & -
  \\
  $1$ & (1,120) & Weyl Fermion & Left-handed
  \\
  $1$ & (136,1) & Hypermultiplet & Right-handed
  \\
  $4$ & (1,120) & Scalar & -
  \\
  $1$ & (1,136) & Weyl Fermion & Right-handed
  \\\
  $1$ & (16,16) & Hypermultiplet & Left-handed
  \\
  \bottomrule
\end{tabular}
\caption{Massless spectrum of the Brane Supersymmetry Breaking model with orthogonal branes in the bulk. The gauge group is ${\rm SO} (16)\times {\rm USp}(16)$.}
\label{bsbbulk}
\end{table}

\begin{table}
\centering
\begin{tabular}{cccc}
  \toprule
  Multiplicity & Representation & Field/Multiplet & Chirality \\
  \toprule
  $1$ & $p_a\bar p_a$ & Gauge Multiplet & Left-handed
  \\
  1 & $p_a\bar p_a$ & Hypermultiplet & Right-handed
   \\
   \midrule
  $-I_{a\bar a}-\frac{1}{2}(I_{aO}-I^-_{aO'})$ & $\frac{p_a(p_a-1)}{2}$ & $4$ Scalars & -
  \\
  $-I_{a\bar a}-\frac{1}{2}( I_{aO} +I^-_{aO'}) $ & $\frac{p_a(p_a-1)}{2}$ & Weyl fermion & Left-handed
  \\
  $-I_{a\bar a}+\frac{1}{2}(I_{aO}-I^-_{aO'}) $ & $\frac{p_a(p_a+1)}{2}$ & $4$ Scalars & -
  \\
  $-I_{a\bar a}+\frac{1}{2} (I_{aO}+I^-_{aO'})$ & $\frac{p_a(p_a+1)}{2}$ & Weyl fermion & Left-handed
  \\
  \midrule
  $I_{ab}$ & $(p_a,\bar p_b)$ & Hypermultiplet & Left-handed
  \\
  $I_{a\bar b}$ & $(p_a, p_b)$ & Hypermultiplet & Left-handed
  \\
  \bottomrule
\end{tabular}
\caption{Massless spectrum for Brane Supersymmetry Breaking vacua with intersecting branes in the bulk. The generic gauge group is $\prod_a {\rm U} (p_a)$.}
\label{bsbbulkrotated}
\end{table}

\section{Six-dimensional Higgsings: some additional examples}
\label{app:higgs}

In this appendix, we present the explicit field-theory Higgs procedure that interpolates between different six-dimensional vacua. These deformations have a stringy description in terms of brane recombination. We anticipate that in the supersymmetric examples the Higgs mechanism leads precisely to the massless spectrum of the intersecting brane models obtained after recombination. In the non-supersymmetric examples, instead, it is often the case that after the Higgs fields acquire a {\it vev}, some states have to get a mass from Yukawa (for fermions) or quartic couplings (for scalars) in order to match the spectrum of the corresponding intersecting brane models.

\subsection{A supersymmetric ${\rm U} (8)\times {\rm U} (8)$ model}
\label{susy}

From a stringy perspective, such a vacuum can be obtained after eight horizontal D7 branes are recombined with all vertical ${\rm D} 7 '$ ones. The resulting configuration consists of two stacks of eight branes each and with wrapping numbers
\begin{equation}
{\rm D}7_1 \ : \ \  (1,0;1,0)\,,
\qquad
{\rm D}7_2 \ : \ \ (1,2;1,1)\,.
\label{u8u8susy}
\end{equation}
The open-string light excitations comprise a vector multiplet with gauge group ${\rm U} (8) \times {\rm U} (8)$ and hypermultiplets in the representations $2\times (28,1) + 6 \times (1,28)+2\times (8,\bar 8)$.

Also in this case, the low-energy description of the gauge-symmetry breaking is in terms of a {\em two-steps} Higgs mechanism. 

\medskip
\noindent
{\bf Step one.} First, one has to break the original ${\rm U} (16) \times {\rm U} (16)$ symmetry to ${\rm U} (8)^2 \times {\rm U} (8)^2$, by displacing the branes on different fixed points, or by introducing suitable discrete Wilson lines. This discrete deformation can be ascribed in field theory to a {\em discrete} Higgs mechanism. The original spectrum decomposes then according to
\begin{equation}
\begin{split}
& 2 \times (120,1) \to  2 \times\left[(28,1,1,1)+(1,28,1,1) +  {\bf (8,8,1,1)}\, \right] \,,
\\
& 2 \times (1,120) \to 2 \times\left[(1,1,28,1)+(1,1,1,28)+{\bf (1,1,8,8)}\, \right] \,,
\\
& (16,\overline{16}) \to (8,1,\bar8,1)+(8,1,1,\bar8)+(1,8,\bar8,1)+(1,8,1,\bar8) \,,
\\
& (256,1) \to (64,1,1,1)+(1,64,1,1)+{\bf (8,\bar  8,1,1)}+{\bf (\bar8,8,1,1)} \,,
\\
& (1,256) \to (1,1,64,1)+(1,1,1,64)+{\bf (1,1,8,\bar8)}+{\bf (1,1,\bar8,8)} \,. 
\end{split}
\end{equation}
Here and in the following representations in bold face correspond to states that become massive as a result of the Higgs mechanism. 

\medskip
\noindent
{\bf Step two.} At this point, one can assign non-vanishing {\em vev}'s to states transforming in the bi-fundamentals $(1,8,\bar 8 , 1)+ (1,8 ,1,\bar 8)$ that breaks the last three ${\rm U} (8)$ factors to a diagonal one, while leaving untouched the first ${\rm U} (8)$. This process corresponds to the recombination of eight D7 and sixteen ${\rm D} 7 '$ branes. Decomposing the previous representations in terms of the final ${\rm U} (8) \times {\rm U} (8)$ gauge group, one finds 
\begin{equation}
\begin{split} 
& 2 \times (28,1,1,1) \to 2 \times (28,1) \,,
\\
& 2 \times [(1,28,1,1)+(1,1,28,1)+(1,1,1,28)] \to  6 \times (1,28) \,,
\\
& (8,1,\bar8,1)+(8,1,1,\bar8) \to 2 \times (8,\bar8) \,,
\\
& (1,8,\bar8,1)+(1,8,1,\bar8) \to 2 \times {\bf (1,64)} \,,
\\
& (64,1,1,1) \to  (64,1) \,,
\\
& (1,64,1,1)+(1,1,64,1)+(1,1,1,64) \to 2 \times {\bf (1,64)}+(1,64) \,. 
\end{split}
\end{equation}
After the highlighted states become massive and decouple from the light spectrum, one exactly recovers the massless excitations associated to the intersecting branes (\ref{u8u8susy}).

\subsection{A non-supersymmetric ${\rm SO} (16)\times {\rm SO} (8)\times  {\rm U} (4)$ model}

For completeness, we analyse here the Brane Supersymmetry Breaking vacuum with gauge group ${\rm SO} (16)\times {\rm SO} (8)\times  {\rm U} (4)$ and its low-energy construction in terms of a Higgs mechanism. It was shown in section \ref{susec:bsb} that 
this vacuum can be obtained by recombining four D7 branes together with the sixteen $\overline{{\rm D}7} '$. Twisted tadpole cancellation imposes tight constraints on the model, and thus the resulting configuration  involves three stacks of branes with wrapping numbers
\begin{equation}
{\rm D}7_+ \ : \ (1,0;1,0)\,,
\quad
{\rm D}7_- \ : \ (1,0;1,0)\,,
\quad 
{\rm D}7_o \ : \ (1,-2;1,2) \,,
\end{equation}
with associated two-cycles ${\boldsymbol\varPi}_{{\rm D}7}^\pm$ and ${\boldsymbol\varPi}_{{\rm D}7_o}$ as explained in section \ref{susec:bsb}. The spectrum of light excitations can be deduced from table \ref{non-susy spectrum} and is reported for completeness in table \ref{sosou}.

\begin{table}
\centering
\begin{tabular}{ccc}
\toprule
Multiplicity & Representation & Multiplet/Field \\
\toprule
$1$ & $(16,8,1)$ & Hypermultiplet
\\
\midrule
$8$ & $(16,1,4+\bar4)$ & Scalar
\\
\midrule
$2$ & $(1,8,4+\bar4)$ & Weyl Fermion
\\
\midrule
$24$ & $(1,1,6)$ & Scalar
\\
\midrule
$10$ & $(1,1,6)$ & Weyl Fermion
\\
\bottomrule
\end{tabular}
\caption{Massless spectrum of the ${\rm SO} (16)\times {\rm SO} (8)\times {\rm U}(4)$ model.}
\label{sosou}
\end{table}

The low-energy field theory description is rather complicated and involves now four steps. For brevity, in the following we shall focus our attention just on the gauge group breaking patterns giving a brief description of the massless spectrum at the end.

\medskip
\noindent
{\bf Step one.} One breaks the original ${\rm SO} (16)^2 \times {\rm USp} (16)^2$  gauge group to ${\rm SO} (16) \times {\rm SO} (8) \times {\rm U} (4) \times {\rm USp} (8) \times {\rm USp} (16)$ by assigning a {\em vev} to the complex scalar in the representation
\begin{equation}
\begin{split}
(1,16;16,1) \to&\ (1,8,1,8,1)+{\it (1,1,16,1,1)} +{\bf (1,8,4+\bar4,1,1)}+{\bf (1,1,4+\bar4,8,1)}
\\
&+{\bf (1,1,16,1,1)}+{\bf (1,1,10+6,1,1)}+{\bf (1,1,\overline{10}+\bar6,1,1)} \,,
\end{split}
\end{equation}
where the states in bold face become massive and we have written in italic the states that acquires the {\em vev} and triggers the gauge symmetry breaking.

\medskip
\noindent
{\bf Step two.} One further breaks the group to ${\rm SO} (16) \times {\rm SO} (8) \times {\rm U} (4) \times {\rm USp} (8)^2$ by assigning a {\em vev} to the quadruplet of scalars in the representation
\begin{equation}
\begin{split}
(1,1,4+\bar 4,1,16) \to&\ {\it (1,1,16,1,1)}+(1,1,6+\bar6,1,1)+{\bf (1,1,4+\bar4,1,8)}
\\
&+{\bf (1,1,16,1,1)}+{\bf (1,1,10 +\overline{10},1,1)}
\end{split}
\end{equation}
where, again, the states in bold face become massive and we have written in italic the states that acquires the {\em vev} and triggers the gauge symmetry breaking.

\medskip
\noindent
{\bf Step three.} One assigns a {\em vev} to the scalars in the representation
\begin{equation}
(1,1,1,8,8) \to {\it (1,1,1,28)} +{\bf (1,1,1,36)}\,,
\end{equation}
to break ${\rm USp} (8)^2$ to the diagonal ${\rm USp}(8)$, so that the residual symmetry is ${\rm SO} (16) \times {\rm SO} (8) \times {\rm U} (4) \times {\rm USp} (8)$.

\medskip
\noindent
{\bf Step four.} The final step consists in breaking the gauge symmetry to the desired ${\rm SO} (16) \times {\rm SO} (8) \times {\rm U} (4)$ group using a {\em vev} for the scalars in the representation
\begin{equation}
(1,1,4+\bar 4,8) \to {\it (1,1,16)} + (1,1,6+\bar 6) + {\bf (1,1,16) + (1,1,10+\overline{10})}\,.
\end{equation}

Although we omit the details on the light spectrum, let us conclude with some comments: 

\begin{itemize}
\item One obtains 24 scalars in the $(1,1,6)$ representation, that match the spectrum in table \ref{sosou}. 

\item There are in addition  four scalars in the representation $(16,8,1)$ and eight scalars in the $(16,1,4+\bar 4)$, that also appear in table \ref{sosou}.

\item Extra scalars in the $(1,1,16)$ and in the $(1,8,4+\bar 4)$ seem to appear in the low-energy description. However, these are expected to acquire a mass through quartic-order couplings with the Higgs fields. We have checked that these couplings are actually compatible with the various quantum numbers.

\item Similar considerations hold for the fermions. One recovers those in table \ref{sosou} plus additional states that however are expected to get a mass via Yukawa couplings. Also in this case we have checked that quantum numbers do not prevent these couplings, that then are expected to emerge.

\end{itemize}



\begin{thebibliography}{99}

\bibitem{Witten:1995gx}
  E.~Witten,
  ``Small instantons in string theory,''
  Nucl.\ Phys.\  {\bf B460 } (1996)  541-559.
  [hep-th/9511030].

  \bibitem{csu}
 M.~Cvetic, G.~Shiu, A.~M.~Uranga,
``Chiral four-dimensional N=1 supersymmetric type 2A orientifolds from intersecting D6 branes,''
  Nucl.\ Phys.\  {\bf B615 } (2001)  3-32.
  [hep-th/0107166].

\bibitem{Angelantonj:2000hi}
  C.~Angelantonj, I.~Antoniadis, E.~Dudas, A.~Sagnotti,
  ``Type I strings on magnetized orbifolds and brane transmutation,''
  Phys.\ Lett.\  {\bf B489 } (2000)  223-232
  [hep-th/0007090].
  
\bibitem{as}
    C.~Angelantonj, A.~Sagnotti,
  ``Type I vacua and brane transmutation,''
  [hep-th/0010279].

\bibitem{ss1}
R.~Rohm,
``Spontaneous Supersymmetry Breaking In Supersymmetric String Theories,''
Nucl.\ Phys.\ B {\bf 237} (1984) 553;

\bibitem{ss2}
C.~Kounnas and M.~Porrati,
``Spontaneous Supersymmetry Breaking In String Theory,''
Nucl.\ Phys.\ B {\bf 310} (1988) 355;

\bibitem{ss3}
I.~Antoniadis, C.~Bachas, D.~C.~Lewellen and T.~N.~Tomaras,
``On Supersymmetry Breaking In Superstrings,''
Phys.\ Lett.\ B {\bf 207} (1988) 441;

\bibitem{ss4}
S.~Ferrara, C.~Kounnas, M.~Porrati and F.~Zwirner,
``Superstrings With Spontaneously Broken Supersymmetry And Their Effective Theories,''
Nucl.\ Phys.\ B {\bf 318} (1989) 75;

\bibitem{ss5}
C.~Kounnas and B.~Rostand,
``Coordinate Dependent Compactifications And Discrete Symmetries,''
Nucl.\ Phys.\ B {\bf 341} (1990) 641;

\bibitem{ss6}
I.~Antoniadis,
``A Possible New Dimension At A Few Tev,''
Phys.\ Lett.\ B {\bf 246} (1990) 377;

\bibitem{ss7}
E.~Kiritsis and C.~Kounnas,
``Perturbative and non-perturbative partial supersymmetry breaking:  ${\mathcal N} = 4 \to  {\mathcal N} = 2 \to {\mathcal N} = 1$,''
Nucl.\ Phys.\ B {\bf 503} (1997) 117 [arXiv:hep-th/9703059].

\bibitem{nonsusy0b1}
A.~Sagnotti,
  ``Surprises in open string perturbation theory,''
  Nucl.\ Phys.\ Proc.\ Suppl.\  {\bf 56B } (1997)  332-343.
  [hep-th/9702093]~;

\bibitem{nonsusy0b2}
C.~Angelantonj,
  ``Nontachyonic open descendants of the 0B string theory,''
  Phys.\ Lett.\  {\bf B444 } (1998)  309-317.
  [hep-th/9810214].

\bibitem{ssi1}
J.~D.~Blum, K.~R.~Dienes,
  ``Strong/weak coupling duality relations for nonsupersymmetric string theories,''
  Nucl.\ Phys.\  {\bf B516 } (1998)  83-159
  [hep-th/9707160],  Phys.\ Lett.\  {\bf B414 } (1997)  260-268
  [hep-th/9707148].

\bibitem{ssi2}
I.~Antoniadis, E.~Dudas, A.~Sagnotti,
  ``Supersymmetry breaking, open strings and M theory,''
  Nucl.\ Phys.\  {\bf B544 } (1999)  469-502
  [hep-th/9807011].
  
\bibitem{ssi3}
I.~Antoniadis, G.~D'Appollonio, E.~Dudas and A.~Sagnotti,
  ``Partial breaking of supersymmetry, open strings and M theory,''
  Nucl.\ Phys.\  {\bf B553 } (1999)  133-154
  [hep-th/9812118].
  
\bibitem{Angelantonj:1999gm}
  C.~Angelantonj, I.~Antoniadis, K.~Forger,
  ``Nonsupersymmetric type I strings with zero vacuum energy,''
  Nucl.\ Phys.\  {\bf B555 } (1999)  116-134
  [hep-th/9904092].
   
\bibitem{magnetic1a}
C.~Bachas,
``A Way to break supersymmetry,''
arXiv:hep-th/9503030.

\bibitem{magnetic1b}
M.~Berkooz, M.~R.~Douglas and R.~G.~Leigh,
``Branes intersecting at angles,''
Nucl.\ Phys.\ B {\bf 480} (1996) 265 [arXiv:hep-th/9606139].

\bibitem{magnetic2a}
R.~Blumenhagen, L.~Goerlich, B.~Kors and D.~Lust,
``Noncommutative compactifications of type I strings on tori with  magnetic background flux,''
JHEP {\bf 0010} (2000) 006 [arXiv:hep-th/0007024].

\bibitem{magnetic2b}
 G.~Aldazabal, S.~Franco, L.~E.~Ibanez, R.~Rabadan, A.~M.~Uranga,
 ``Intersecting brane worlds,''
  JHEP {\bf 0102 } (2001)  047
  [hep-ph/0011132].

\bibitem{magnetic2c}  
    M.~Cvetic, G.~Shiu, A.~M.~Uranga,
  ``Three family supersymmetric standard-like models from intersecting brane worlds,''
  Phys.\ Rev.\ Lett.\  {\bf 87 } (2001)  201801
  [hep-th/0107143].
    
\bibitem{magnetic2d}   
F.~G.~Marchesano Buznego,
 ``Intersecting D-brane models,''
  [hep-th/0307252];
  
\bibitem{magnetic2e}    
R.~Blumenhagen, B.~Kors, D.~Lust, S.~Stieberger,
 ``Four-dimensional String Compactifications with D-Branes, Orientifolds and Fluxes,''
  Phys.\ Rept.\  {\bf 445 } (2007)  1-193
  [hep-th/0610327].
  
\bibitem{Larosa:2003mz}
  M.~Larosa, G.~Pradisi,
  ``Magnetized four-dimensional $\mathbb{Z}_2 \times  \mathbb{Z}_2$ orientifolds,''
  Nucl.\ Phys.\  {\bf B667 } (2003)  261-309.
  [hep-th/0305224].

\bibitem{bsb1a}
S.~Sugimoto,
 ``Anomaly cancellations in type I D9 - anti-D9 system and the USp(32) string theory,''
  Prog.\ Theor.\ Phys.\  {\bf 102 } (1999)  685-699.
  [hep-th/9905159]~;

\bibitem{bsb1b}
I.~Antoniadis, E.~Dudas and A.~Sagnotti,
``Brane supersymmetry breaking,''
Phys.\ Lett.\ B {\bf 464} (1999) 38 [arXiv:hep-th/9908023];

\bibitem{bsb1c}
C.~Angelantonj,
``Comments on open-string orbifolds with a non-vanishing $B_{ab}$,''
Nucl.\ Phys.\ B {\bf 566} (2000) 126 [arXiv:hep-th/9908064];

\bibitem{bsb1d}
G.~Aldazabal and A.~M.~Uranga,
``Tachyon-free non-supersymmetric type IIB orientifolds via  brane-antibrane systems,''
JHEP {\bf 9910} (1999) 024 [arXiv:hep-th/9908072].

\bibitem{Angelantonj:1999ms}
C.~Angelantonj, I.~Antoniadis, G.~D'Appollonio, E.~Dudas and
A.~Sagnotti,
``Type I vacua with brane supersymmetry breaking,''
Nucl.\ Phys.\ B {\bf 572} (2000) 36 [arXiv:hep-th/9911081].

\bibitem{Angelantonj:2003hr}
  C.~Angelantonj, I.~Antoniadis,
  ``Suppressing the cosmological constant in nonsupersymmetric type I strings,''
  Nucl.\ Phys.\  {\bf B676 } (2004)  129-148
  [hep-th/0307254].  
  
\bibitem{Angelantonj:2005hs}
  C.~Angelantonj, M.~Cardella, N.~Irges,
  ``Scherk-Schwarz breaking and intersecting branes,''
  Nucl.\ Phys.\  {\bf B725} (2005) 115-154
  [hep-th/0503179].

\bibitem{jihad}
E.~Dudas, J.~Mourad,
  ``Consistent gravitino couplings in nonsupersymmetric strings,''
  Phys.\ Lett.\  {\bf B514 } (2001)  173-182
  [hep-th/0012071]~;

\bibitem{pr}
G.~Pradisi, F.~Riccioni,
  ``Geometric couplings and brane supersymmetry breaking,''
  Nucl.\ Phys.\  {\bf B615 } (2001)  33-60
  [hep-th/0107090].

\bibitem{Sen:1998sm}
  A.~Sen,
  ``Tachyon condensation on the brane anti-brane system,''
  JHEP {\bf 9808 } (1998)  012.
  [hep-th/9805170].

\bibitem{Sen:2004nf}
  A.~Sen,
  ``Tachyon dynamics in open string theory,''
  Int.\ J.\ Mod.\ Phys.\  {\bf A20 } (2005)  5513-5656.
  [hep-th/0410103].

\bibitem{Angelantonj:2000xf}
  C.~Angelantonj, R.~Blumenhagen, M.~R.~Gaberdiel,
  ``Asymmetric orientifolds, brane supersymmetry breaking and nonBPS branes,''
  Nucl.\ Phys.\  {\bf B589 } (2000)  545-576
  [hep-th/0006033].

\bibitem{carloemilian}
  C.~Angelantonj, E.~Dudas,
  ``Metastable string vacua,''
  Phys.\ Lett.\  {\bf B651 } (2007)  239-245
  [arXiv:0704.2553 [hep-th]].

\bibitem{Dudas:2005jx}
  E.~Dudas, C.~Timirgaziu,
  ``Internal magnetic fields and supersymmetry in orientifolds,''
  Nucl.\ Phys.\  {\bf B716 } (2005)  65-87
  [hep-th/0502085].

\bibitem{Blumenhagen:2005tn}
  R.~Blumenhagen, M.~Cvetic, F.~Marchesano, G.~Shiu,
  ``Chiral D-brane models with frozen open string moduli,''
  JHEP {\bf 0503 } (2005)  050  [hep-th/0502095]  
  
\bibitem{ms}  
  F.~Marchesano, G.~Shiu,
``Building MSSM flux vacua,''
  JHEP {\bf 0411 } (2004)  041
  [hep-th/0409132], Phys.\ Rev.\  {\bf D71 } (2005)  011701
  [hep-th/0408059].

\bibitem{zavala}
S.~Forste, I.~Zavala,
 ``Oddness from Rigidness,''
  JHEP {\bf 0807 } (2008)  086
  [arXiv:0806.2328 [hep-th]],

\bibitem{honecker}
S.~Forste, G.~Honecker,
 ``Rigid D6-branes on $T^6/(\mathbb{Z}_2 \times \mathbb{Z}_{2M} \times \varOmega R)$ with discrete torsion,''
  JHEP {\bf 1101 } (2011)  091
  [arXiv:1010.6070 [hep-th]].

\bibitem{Blumenhagen:2002wn}
  R.~Blumenhagen, V.~Braun, B.~Kors, D.~Lust,
  ``Orientifolds of K3 and Calabi-Yau manifolds with intersecting D-branes,''
  JHEP {\bf 0207 } (2002)  026.
  [hep-th/0206038].

\bibitem{Bianchi:1990yu}
  M.~Bianchi, A.~Sagnotti,
  ``On the systematics of open string theories,''
  Phys.\ Lett.\  {\bf B247 } (1990)  517-524.

\bibitem{Gimon:1996rq}
  E.~G.~Gimon, J.~Polchinski,
  ``Consistency conditions for orientifolds and d manifolds,''
  Phys.\ Rev.\  {\bf D54 } (1996)  1667-1676.
  [hep-th/9601038].
  
\bibitem{Cremades:2002cs}
  D.~Cremades, L.~E.~Ibanez, F.~Marchesano,
 ``Intersecting brane models of particle physics and the Higgs mechanism,''
  JHEP {\bf 0207 } (2002)  022.
  [hep-th/0203160].
  
\bibitem{Angelantonj:1999xf}
  C.~Angelantonj, R.~Blumenhagen,
  ``Discrete deformations in type I vacua,''
  Phys.\ Lett.\  {\bf B473 } (2000)  86-93.
  [hep-th/9911190].
  
\bibitem{Vafa:1994rv}
  C.~Vafa, E.~Witten,
  ``On orbifolds with discrete torsion,''
  J.\ Geom.\ Phys.\  {\bf 15 } (1995)  189-214.
  [hep-th/9409188].

\bibitem{Angelantonj:2009yj}
  C.~Angelantonj, C.~Condeescu, E.~Dudas, M.~Lennek,
  ``Stringy Instanton Effects in Models with Rigid Magnetised D-branes,''
  Nucl.\ Phys.\  {\bf B818 } (2009)  52-94.
  [arXiv:0902.1694 [hep-th]].

\bibitem{Angelantonj:2002ct}
  C.~Angelantonj, A.~Sagnotti,
  ``Open strings,''
  Phys.\ Rept.\  {\bf 371 } (2002)  1-150.
  [hep-th/0204089].

 \bibitem{Dudas:2000bn}
  E.~Dudas,
``Theory and phenomenology of type I strings and M theory,''
  Class.\ Quant.\ Grav.\  {\bf 17 } (2000)  R41-R116.
  [hep-ph/0006190].
  
  \bibitem{Camara:2010zm}
  P.~G.~Camara, C.~Condeescu, E.~Dudas, M.~Lennek,
  ``Non-perturbative Vacuum Destabilization and D-brane Dynamics,''
  JHEP {\bf 1006 } (2010)  062.
  [arXiv:1003.5805 [hep-th]].

\bibitem{tsyetlin}
 A.A.~Tseytlin,
``Born-Infeld action, supersymmetry and string theory,''
  In  Shifman, M.A. (ed.): ``The many faces of the superworld'' 417-452  [hep-th/9908105].

\bibitem{karim}
 P.~Anastasopoulos, I.~Antoniadis, K.~Benakli, M.~D.~Goodsell, A.~Vichi,
 ``One-loop adjoint masses for non-supersymmetric intersecting branes,''
  [arXiv:1105.0591 [hep-th]].

\bibitem{Dudas:2004nd}
  E.~Dudas, G.~Pradisi, M.~Nicolosi, A.~Sagnotti,
  ``On tadpoles and vacuum redefinitions in string theory,''
  Nucl.\ Phys.\  {\bf B708 } (2005)  3-44.
  [hep-th/0410101].

 \bibitem{superfield1}
 N.~Marcus, A.~Sagnotti, W.~Siegel,
  ``Ten-dimensional Supersymmetric Yang-mills Theory In Terms Of Four-dimensional Superfields,''
  Nucl.\ Phys.\  {\bf B224 } (1983)  159.

  \bibitem{superfield2}
   N.~Arkani-Hamed, T.~Gregoire, J.~G.~Wacker,
  ``Higher dimensional supersymmetry in 4-D superspace,''
  JHEP {\bf 0203 } (2002)  055.
  [hep-th/0101233].

 \bibitem{nekrasov}
 N.~Nekrasov, A.~S.~Schwarz,
  ``Instantons on noncommutative $R^4$ and (2,0) superconformal six-dimensional theory,''
  Commun.\ Math.\ Phys.\  {\bf 198 } (1998)  689-703.
  [hep-th/9802068].

\bibitem{sw}
N.~Seiberg, E.~Witten,
  ``String theory and noncommutative geometry,''
  JHEP {\bf 9909 } (1999)  032.
  [hep-th/9908142].

\bibitem{Terashima}
  S.~Terashima,
  ``U(1) instanton in Born-Infeld action and noncommutative gauge theory,''
  Phys.\ Lett.\  B {\bf 477} (2000) 292
  [arXiv:hep-th/9911245].

\bibitem{mmms}
  M.~Marino, R.~Minasian, G.~W.~Moore and A.~Strominger,
  ``Nonlinear instantons from supersymmetric p-branes,''
  JHEP {\bf 0001} (2000) 005
  [arXiv:hep-th/9911206].

\bibitem{ncinst}
  M.~Billo, M.~Frau, S.~Sciuto, G.~Vallone and A.~Lerda,
  ``Non-commutative (D)-instantons,''
  JHEP {\bf 0605} (2006) 069
  [arXiv:hep-th/0511036].

\bibitem{bpst}
  A.~A.~Belavin, A.~M.~Polyakov, A.~S.~Schwartz and Yu.~S.~Tyupkin,
  ``Pseudoparticle solutions of the Yang-Mills equations,''
  Phys.\ Lett.\  B {\bf 59} (1975) 85.

  


\end{thebibliography}
\end{document}